\documentclass[10pt]{IEEEtran}
\usepackage[ruled,linesnumbered]{algorithm2e}
\usepackage{anyfontsize}
\usepackage{multirow,hhline}
\usepackage{amsmath}
\usepackage{amssymb}
\usepackage{graphicx}
\usepackage{cite}
\usepackage{dirtytalk}

\newcommand\norm[1]{\left\lVert#1\right\rVert}

\makeatletter
\newlength \figwidth
\if@twocolumn
\else
\fi
\makeatother

\title{PSD Estimation and Source Separation in a Noisy Reverberant Environment using a Spherical Microphone Array}

\author{
Abdullah Fahim, Prasanga N. Samarasinghe, Thushara D. Abhayapala\\
\thanks{This work is supported by the Australian Research Council (ARC) Discovery Project DP180102375.}
}

\begin{document}

\IEEEoverridecommandlockouts
\IEEEpubid{\makebox[\columnwidth]{\hfill 10.1109/TASLP.2018.2835723
\copyright2018
IEEE \hfill}}

\maketitle

%\thispagestyle{empty}
%\pagestyle{empty}

%%%%%%%%%%%%%%%%%%%%%%%%%%%%%%%%%%%%%%%%%%%%%%%%%%%%%%%%%%%%%%%%%%%%%%%%%%%%%%%%
\begin{abstract}
In this paper, we propose an efficient technique for estimating individual power spectral density (PSD) components, i.e., PSD of each desired sound source as well as of noise and reverberation, in a multi-source reverberant sound scene with coherent background noise. We formulate the problem in the spherical harmonics domain to take the advantage of the inherent orthogonality of the spherical harmonics basis functions and extract the PSD components from the cross-correlation between the different sound field modes. We also investigate an implementation issue that occurs at the nulls of the Bessel functions and offer an engineering solution. The performance evaluation takes place in a practical environment with a commercial microphone array in order to measure the robustness of the proposed algorithm against all the deviations incurred in practice. We also exhibit an application of the proposed PSD estimator through a source septation algorithm and compare the performance with a contemporary method in terms of different objective measures.
\end{abstract}
%
%%%%%%%%%%%%%%%%%%%%%%%%%%%%%%%%%%%%%%%%%%%%%%%%%%%%%%%%%%%%%%%%%%%%%%%%%%%%%%%%
\section{Introduction}
\IEEEPARstart{T}{he} power spectral density (PSD) of an audio signal carries useful information about the signal characteristics. Many spectral enhancement techniques, most commonly Wiener filter and spectral subtraction, use the knowledge of the PSD to suppress undesired signal components such as background noise\cite{benesty2005study}, late reverberation \cite{lebart2001new,naylor2010speech}, or both \cite{kuklasinski2016maximum}. Few other applications of PSD include computing direct to reverberation energy ratio (DRR) \cite{hioka2011estimating} or separating sound sources in a mixed acoustic environment \cite{hioka2013underdetermined}. Most of the existing spectral enhancement techniques, including the aforementioned ones, focused on estimating PSD components under strict assumptions such as a noiseless, free-field or a single-source scenario. In this work, we develop a new technique for estimating individual signal PSDs along with the PSD of the reverberation and coherent noise components in a multi-source noisy and reverberant environment using a spherical microphone array.\par
\subsection{Literature review}
Several techniques have been proposed in the literature for single-source reverberant environment to estimate both the source and reverberation PSDs. Lebart et al. \cite{lebart2001new} used a statistical model of room impulse responses (RIR) to estimate the reverberation PSD and used that in a spectral subtraction-based speech dereverberation technique. Braun et al. \cite{braun2013dereverberation} proposed a PSD estimator in using a reference signal under the strict statistical assumption of diffused reverberant field. Kuklasi{\'n}ski et al. \cite{kuklasinski2016maximum} developed a maximum likelihood-based method for estimating speech and late reverberation PSD in a single-source noisy reverberant environment assuming a known value of noise PSD. The spatial correlation between the received microphone signals were utilized in \cite{hioka2011estimating} to compute direct to reverberant energy ratio in a noiseless environment by estimating PSDs of the direct and reverberant components.\par
Saruwatari et al. \cite{saruwatari2000speech} proposed a method to suppress undesired signals in a multi-source environment using complementary beamformers. Hioka et al. \cite{hioka2013underdetermined} used a similar idea for PSD estimation and source separation which utilized multiple fixed beamformer to estimate source PSDs. While \cite{hioka2013underdetermined} offered a PSD estimation technique for an underdetermined system capable of working with a larger number of sources compared to \cite{saruwatari2000speech}, both the algorithms were developed for non-reverberant case and required the directions of each speech and noise sources. An improvement to \cite{hioka2013underdetermined} was suggested in \cite{niwa2016psd} where the property of an M-matrix was utilized to design the fixed beamformers, whereas the former chose the steering direction in an empirical manner. However, while \cite{niwa2016psd} eliminated the performance issue due to ill-conditioning of the de-mixing matrix, the method was developed under the assumption of a noiseless free-field environment.\par
\IEEEpubidadjcol % For aligning copyright
A common use of PSD is in post-filter design to be used in conjunction with a beamformer, e.g. multi-channel Wiener filter \cite{doclo2007frequency}, in order to enhance the desired speech signal in a mixed environment. Beamforming is a common speech enhancement technique used for decades \cite{johnson1992array,Bourgeois2010beamforming}. However, a post-filter at the beamformer output is known to enhance system performance by boosting interference rejection \cite{marro1998analysis}.\par
In the recent past, several methods have been proposed in the spherical harmonics domain to estimate PSD components of an acoustic signal \cite{samarasinghe2017estimating} or for interference rejection through beamforming \cite{beh2014adaptive,kumar2016near}. One of the major advantages of spherical harmonics domain representation of a signal \cite{williams1999fourier,abhayapala1999modal} is the inherent orthogonality of its basis functions. Hence, the spherical harmonics-based solutions are becoming popular in other fields of acoustics signal processing, such as source localization \cite{evers2014multiple}, speech dereverberation \cite{yamamoto2016spherical}, and noise suppression \cite{jarrett2014noise}. However, to the best of our knowledge, no methods have been developed in the harmonics domain to estimate individual PSD components in a multi-source mixed acoustic environment. A harmonics-based single-source PSD estimator was proposed by the authors of this paper \cite{samarasinghe2017estimating} for a noiseless reverberant environment in order to estimate DRR.\par
\subsection{Our approach and contribution}
We consider a multi-source reverberant environment with coherent background noise and propose a novel technique to estimate the individual PSDs of each desired source as well as the PSDs of the reverberant and noise components. The method is developed in the spherical harmonics domain to take the advantage of the inherent orthogonality of the spherical harmonics basis functions which ensures a well-posed solution without the requirement of any additional design criteria such as an M-matrix consideration as adopted in \cite{niwa2016psd}. Additionally, in contrast to the conventional beamformer-based methods \cite{hioka2013underdetermined} where only the autocorrelation coefficients of the beamformer output were used, we also incorporate the cross-correlation between the spherical harmonics coefficients in our solution. This latter approach was first used in \cite{hioka2011estimating} for estimating DRR in a single-source environment. The larger number of correlation coefficients makes the algorithm suitable for separating a larger number of sources compared to the conventional techniques. We also derive a harmonics-based novel closed form expression for the spatial correlation of a coherent noise field. We carry out detailed theoretical analysis, demonstrate the practical impact and offer an engineering solution to a Bessel-zero issue which, if not addressed in a correct way, significantly limits the performance of spherical array-based systems. Initial work on this approach was published in \cite{fahim2017psd} where a limited-scope implementation of the idea was presented under the assumption of a noiseless case. Furthermore, a detailed analysis on the implementation challenges and a benchmark in terms of perceived quality were not included in our prior work of \cite{fahim2017psd}.\par
It is worth mentioning that, the performance evaluation of the proposed algorithm took place in a practical environment with a commercially available microphone array, \lq{Eigenmike}\rq \cite{eigenmikeweb}, without any prior knowledge about the source characteristics. Hence, the evaluation process incorporated all the practical deviations like the source localization error, thermal noise at the microphones, non-ideal characteristics of the sound fields, etc. We validate the performance of the algorithm by carrying out $350$ experiments in $3$ different acoustic environments with varying number of speakers using independent mixed-gender speech signals. The improved objective measures in terms of perceptual evaluation of speech quality (PESQ) \cite{recommendation2001perceptual} and frequency-weighted segmental signal to noise ratio (FWSegSNR) \cite{hu2008evaluation} indicate that the proposed algorithm is robust against such practical distortions and produce a better estimation compared to the competing methods.\par
The paper is structured as follows. Section \ref{lab:problem-formulation} contains problem statement and defines the objective of the work. In Section \ref{lab:framework}, we develop a framework for PSD estimation based on the spatial correlation of sound field coefficients in a noisy reverberant environment. We use the framework in Section \ref{lab:psd-estimation} to formulate a PSD estimation technique, and discuss and offer solutions to a practical challenge which we term as Bessel-zero issue. In Section \ref{lab:practical-application}, we outline a practical application of the estimated PSDs. Finally in Section \ref{lab:experimental-results}, we evaluate and compare the performance of the proposed algorithm with other methods based on objective metrics and graphical aids.
\section{Problem formulation} \label{lab:problem-formulation}
Let us consider a microphone array consisting of $Q$ microphones to capture the sound field in a noisy reverberant room with $L$ distinct sound sources. The received signal at the $q^{th}$ microphone is given by
\begin{equation} \label{eq:basic-model-td}
p(\boldsymbol{x}_q, t) = \sum \limits_{\ell=1}^{L} h_{\ell}(\boldsymbol{x}_q, t) \ast s_{\ell}(t) + z(\boldsymbol{x}_q, t)
\end{equation}
where $q \in [1, Q]$, $\ell \in [1, L]$, $\boldsymbol{x}_q = (r_q, \theta_q, \phi_q)$ denotes the $q^{th}$ microphone position, $h_{\ell}(\boldsymbol{x}_q, t)$ is the RIR between the $\ell^{th}$ source and the $q^{th}$ microphone, $t$ is the discrete time index, $\ast$ denotes the convolution operation, $s_{\ell}(t)$ is the source excitation for the $\ell^{th}$ sound source, and $z(\boldsymbol{x}_q, t)$ is the coherent noise\footnote{Here the coherent noise refers to the colored background noise, different from the white thermal noise, which can be originated from any unknown noise source such as room air-conditioning system.} at the $q^{th}$ microphone position. The RIR can be decomposed into two parts
\begin{equation} \label{eq:rir-td}
h_{\ell}(\boldsymbol{x}_q, t) = h^{(d)}_{\ell}(\boldsymbol{x}_q, t) + h^{(r)}_{\ell}(\boldsymbol{x}_q, t)
\end{equation}
where $h^{(d)}_{\ell}(\boldsymbol{x}_q, t)$ and $h^{(r)}_{\ell}(\boldsymbol{x}_q, t)$ are the direct and reverberant path components, respectively. Substituting \eqref{eq:rir-td} into \eqref{eq:basic-model-td} and converting into frequency domain using short-time Fourier transform (STFT), we obtain
\begin{multline} \label{eq:basic-model-fd}
P(\boldsymbol{x}_q, \tau, k) = \sum \limits_{\ell=1}^{L} S_{\ell}(\tau, k) \bigg( H^{(d)}_{\ell}(\boldsymbol{x}_q, \tau, k) + H^{(r)}_{\ell}(\boldsymbol{x}_q, \tau, k) \bigg) \\
+ Z(\boldsymbol{x}_q, \tau, k)
\end{multline}
where $\{P, S, H, Z\}$ represent the corresponding signals of $\{p, s, h, z\}$ in the STFT domain, $\tau$ is the time frame index, $k = 2 \pi f / c$, $f$ denotes the frequency, and $c$ is the speed of sound propagation. In the subsequent sections, the time frame index $\tau$ is omitted for brevity.\par
Given the measured sound pressure $p(\boldsymbol{x}_q, t) \text{ } \forall q$, we aim to estimate the individual source PSDs, $E \Big\{ \lvert S_{\ell}(k) \rvert ^2 \Big \} \text{ } \forall \ell$, where $E\{\cdot\}$ represents the expected value over time. As an application of a common use of PSD, we further measure the signal enhancement of a beamformer output through a basic Wiener post filter driven by the estimated PSDs, and demonstrate an improved estimation technique for source separation.
\section{Framework for PSD estimation} \label{lab:framework}
In this section, we develop a spherical harmonics domain framework to establish the relationship between the sound field coefficients and the individual PSD components in a multi-source noisy and reverberant environment. We use this model in Section \ref{lab:psd-estimation} and \ref{lab:practical-application} to estimate individual PSD components and separate sound sources from a mixed recording.
\subsection{Spatial domain representation of room transfer function} \label{lab:rtf-modal}
We model the direct and reverberant path of room transfer function (RTF) in the spatial domain as
\begin{equation} \label{eq:rtf-modal-direct}
H^{(d)}_{\ell}(\boldsymbol{x}_q, k) = G^{(d)}_{\ell}(k) \text{ } e^{ik \text{ } \boldsymbol{\hat{y}}_{\ell} \cdot \boldsymbol{x}_q}
\end{equation}
\begin{equation} \label{eq:rtf-modal-reverb}
H^{(r)}_{\ell}(\boldsymbol{x}_q, k) = \int_{\boldsymbol{\hat{y}}} G^{(r)}_{\ell}(k, \boldsymbol{\hat{y}}) \text{ } e^{ik \text{ } \boldsymbol{\hat{y}} \cdot \boldsymbol{x}_q} \text{ } d\boldsymbol{\hat{y}}
\end{equation}
where $G^{(d)}_{\ell}(k)$ represents the direct path gain at the origin for the $\ell^{th}$ source, $i = \sqrt[]{-1}$, $\boldsymbol{\hat{y}_{\ell}}$ is a unit vector towards the direction of the $\ell^{th}$ source, and $G^{(r)}_{\ell}(k, \boldsymbol{\hat{y}})$ is the reflection gain at the origin along the direction of $\boldsymbol{\hat{y}}$ for the $\ell^{th}$ source. Hence, we obtain the spatial domain equivalent of \eqref{eq:basic-model-fd} by substituting the spatial domain RTF from \eqref{eq:rtf-modal-direct} and \eqref{eq:rtf-modal-reverb} as
\begin{multline} \label{eq:basic-model-sd}
P(\boldsymbol{x}_q, k) = \sum \limits_{\ell=1}^{L} S_{\ell}(k) \Bigg( G^{(d)}_{\ell}(k) \text{ } e^{ik \text{ } \boldsymbol{\hat{y}}_{\ell} \cdot \boldsymbol{x}_q} + \\
\int_{\boldsymbol{\hat{y}}} G^{(r)}_{\ell}(k, \boldsymbol{\hat{y}}) \text{ } e^{ik \text{ } \boldsymbol{\hat{y}} \cdot \boldsymbol{x}_q} \text{ } d\boldsymbol{\hat{y}} \Bigg) + Z(\boldsymbol{x}_q, k).
\end{multline}
\subsection{Spherical harmonics decomposition}
In this section, we derive the spherical harmonics expansion of \eqref{eq:basic-model-sd} using the existing mathematical models. Spherical harmonics are a set of orthonormal basis functions which can represent a function over a sphere. A spherical function $F(\boldsymbol{\hat{x}})$ can be expressed in the spherical harmonics domain as
\begin{equation}\label{eq: spherical-harmonics-general}
F(\boldsymbol{\hat{x}}) = \sum\limits_{nm}^{\infty} \text{ } a_{nm} Y_{nm}(\boldsymbol{\hat{x}})
\end{equation}
where $\boldsymbol{\hat{x}} = (1, \theta, \phi)$ is defined over a sphere, $\sum \limits_{nm}^{\infty} = \sum \limits_{n = 0}^{\infty} \sum \limits_{m = -n}^{n}$, $Y_{nm}(\cdot)$ denotes the spherical harmonic of order $n$ and degree $m$, and $a_{nm}$ indicates corresponding coefficient. The spherical harmonics inherently possess the orthonormal property, i.e.
\begin{equation}\label{eq: spherical-harmonics-orthonormal}
\int_{\boldsymbol{\hat{x}}} Y_{nm}(\boldsymbol{\hat{x}}) \text{ } Y^*_{n'm'}(\boldsymbol{\hat{x}})  \text{ } d\boldsymbol{\hat{x}} = \delta_{nn'} \delta_{mm'}
\end{equation}
where $\delta_{nn'}$ is the Kronecker delta function. Using \eqref{eq: spherical-harmonics-general} and \eqref{eq: spherical-harmonics-orthonormal}, sound field coefficients $a_{nm}$ can be calculated as
\begin{equation} \label{eq:alpha-theory}
a_{nm} = \int_{\boldsymbol{\hat{x}}} F(\boldsymbol{\hat{x}}) \text{ } Y_{nm}^*(\boldsymbol{\hat{x}}) \text{ } d\boldsymbol{\hat{x}}
\end{equation}
where $(\cdot)^*$ denotes the complex conjugate operation. As the realization of \eqref{eq:alpha-theory} requires an impractical continuous microphone array over a sphere, several array structures have been proposed in the literature to estimate $a_{nm}$, such as a spherical open/rigid array \cite{abhayapala2002theory,meyer2002highly}, multiple circular arrays \cite{abhayapala2010spherical}, or a planar array with differential microphones \cite{chen2015theory}. The subsequent theory is developed for a spherical microphone array, however, this is equally applicable for any array geometry given that they can produce $a_{nm}$ (e.g. \cite{samarasinghe2017planar}). \par
The spherical harmonics decomposition of the $3$D incident sound field of \eqref{eq:basic-model-sd} is given by \cite{williams1999fourier}
\begin{equation}\label{eq: spherical-harmonics-decomposition}
P(\boldsymbol{x}_q, k) = \sum\limits_{nm}^{\infty} \text{ } \underbrace{\alpha_{nm} (k) b_n(kr)}_{a_{nm} (kr)} Y_{nm}(\boldsymbol{\hat{x}}_q)
\end{equation}
where $r$ is the array radius, $\boldsymbol{\hat{x}}_q = \boldsymbol{x}_q/r$ is a unit vector towards the direction of the $q^{th}$ microphone, and $\alpha_{nm}(k)$ is the array-independent sound field coefficient. The function $b_n(\cdot)$ depends on the array configuration and is defined as
\begin{equation} \label{eq:bn}
b_n(\xi) =
\begin{cases}
j_n(\xi) & \text{for an open array} \\
j_n(\xi) - \frac{j'_n(\xi)}{h'_n(\xi)} h_n(\xi) & \text{for a rigid spherical array}
\end{cases}
\end{equation}
where $\xi \in \mathbb{R}$, $j_n(\cdot)$ and $h_n(\cdot)$ denote the $n^{th}$ order spherical Bessel and Hankel functions of the first kind, respectively, and $(\cdot)'$ refers to the corresponding first derivative term. Eq. \eqref{eq: spherical-harmonics-decomposition} can be truncated at the sound field order $N = \lceil k e r / 2 \rceil$ due to the high-pass nature of the higher order Bessel functions \cite{jones2002dimensionality,ward2001reproduction}, where $e \approx 2.7183$ and $\lceil \cdot \rceil$ denoting the ceiling operation. We can estimate the sound field coefficients $\alpha_{nm}(k)$ with a spherical microphone array as \cite{abhayapala2002theory,meyer2002highly}
\begin{equation} \label{eq:alpha}
\alpha_{nm}(k) \approx \frac{1}{b_n(kr)} \sum \limits_{q=1}^{Q} w_q \text{ } P(\boldsymbol{x}_q, k) \text{ } Y_{nm}^*(\boldsymbol{\hat{x}}_q)
\end{equation}
where $Q \geq (N+1)^2$ is imposed to avoid spatial aliasing and $w_q$ are suitable microphone weights that enforce the orthonormal property of the spherical harmonics with a limited number of sampling points, i.e.
\begin{equation}\label{eq: spherical-harmonics-orthonormal-band-limited}
\sum \limits_{q=1}^{Q} w_q \text{ } Y_{nm}(\boldsymbol{\hat{x}}_q) \text{ } Y^*_{n'm'}(\boldsymbol{\hat{x}}_q) \approx \delta_{nn'} \delta_{mm'}.
\end{equation}
Similarly, the spherical harmonics decomposition of the coherent noise component $Z(\boldsymbol{x}_q, k)$  of \eqref{eq:basic-model-sd} is
\begin{equation}\label{eq:coherent-noise-sh}
Z(\boldsymbol{x}_q, k) = \sum\limits_{nm}^{\infty} \text{ } \eta_{nm} (k) b_n(kr) Y_{nm}(\boldsymbol{\hat{x}}_q)
\end{equation}
where $\eta_{nm}(k)$ is the sound field coefficient due to the coherent noise sources. Finally, the spherical harmonics expansion of the Green's function is given by \cite[pp. 27--33]{colton2012inverse}
\begin{equation} \label{eq:far-field-source}
e^{ik \text{ } \boldsymbol{\hat{y}}_{\ell} \cdot \boldsymbol{x}_q} = \sum \limits_{nm}^{\infty} \text{ } 4 \pi i^{n} \text{ } Y^*_{nm}(\boldsymbol{\hat{y}}_{\ell}) \text{ } b_n(kr) \text{ } Y_{nm}(\boldsymbol{\hat{x}}_q).
\end{equation}
Using \eqref{eq: spherical-harmonics-decomposition}, \eqref{eq:coherent-noise-sh} and \eqref{eq:far-field-source}  in \eqref{eq:basic-model-sd}, we obtain the harmonics-domain representation of a noisy reverberant sound field by
\begin{multline} \label{eq:basic-model-sh}
\sum\limits_{nm}^{\infty} \text{ } \alpha_{nm}(k) b_n(kr) Y_{nm}(\boldsymbol{\hat{x}}_q) = \sum\limits_{nm}^{\infty} \text{ } \Bigg[ 4 \pi i^{n} \text{ } \sum \limits_{\ell=1}^{L} S_{\ell}(k) \\
\Bigg( G^{(d)}_{\ell}(k) \text{ } Y^*_{nm}(\boldsymbol{\hat{y}}_{\ell}) + \int_{\boldsymbol{\hat{y}}} G^{(r)}_{\ell}(k, \boldsymbol{\hat{y}}) \text{ } Y^*_{nm}(\boldsymbol{\hat{y}}) \text{ } d\boldsymbol{\hat{y}} \Bigg)\\
+ \eta_{nm}(k) \Bigg] b_n(kr) Y_{nm}(\boldsymbol{\hat{x}}_q).
\end{multline}
Hence, the expression for the combined sound field coefficients is obtained from \eqref{eq:basic-model-sh} as
\begin{align}
\alpha_{nm} (k) & = 4 \pi i^{n} \text{ } \sum \limits_{\ell=1}^{L} S_{\ell}(k)  \Bigg( G^{(d)}_{\ell}(k) \text{ } Y^*_{nm}(\boldsymbol{\hat{y}}_{\ell}) + \nonumber \\
& \int_{\boldsymbol{\hat{y}}} G^{(r)}_{\ell}(k, \boldsymbol{\hat{y}}) \text{ } Y^*_{nm}(\boldsymbol{\hat{y}}) \text{ } d\boldsymbol{\hat{y}} \Bigg) + \eta_{nm}(k) \label{eq:alpha-model} \\
&= \lambda_{nm}(k) + \eta_{nm}(k) \label{eq:alpha-model-shortcut}
\end{align}
where $\lambda_{nm}(k)$ is defined as the sound field coefficients related to the direct and reverberant components of the sound signals.\par
It is important to note that we consider a far-field sound propagation model in \eqref{eq:rtf-modal-direct} and \eqref{eq:rtf-modal-reverb}. For a near-field sound propagation, the corresponding Green's function and its spherical harmonics expansion is defined as \cite[pp. 31]{colton2012inverse}
\begin{multline} \label{eq:near-field-source}
\frac{e^{ik \norm{\boldsymbol{x}_q - \boldsymbol{y}_{\ell}}}}{4 \pi \norm{\boldsymbol{x}_q - \boldsymbol{y}_{\ell}}} = \sum \limits_{nm}^{\infty} \text{ } i k \text{ } h_n(k r_{\ell}) \text{ } Y^*_{nm}(\boldsymbol{\hat{y}}_{\ell}) \times \\
b_n(kr) \text{ } Y_{nm}(\boldsymbol{\hat{x}}_q)
\end{multline}
where $\boldsymbol{y}_{\ell} = (r_{\ell}, \boldsymbol{\hat{y}}_{\ell})$ is the position vector of $\ell^{th}$ source and $\norm{\cdot}$ denotes the Euclidean norm. In this work, we use the far-field assumption for mathematical tractability, however, the model is equally applicable for a near-field sound propagation.
\subsection{Spatial correlation of the sound field coefficients}
In this section, we propose novel techniques to develop closed form expressions of the spatial correlation between the harmonics coefficients of reverberant and noise fields in a multi-source environment. From \eqref{eq:alpha-model}, the spatial correlation between $\alpha_{nm}(k)$ and $\alpha_{n'm'}(k)$ is
\begin{multline} \label{eq:cross-alpha-0}
E \Big\{ \alpha_{nm}(k) \alpha^*_{n'm'}(k) \Big \} = E \Big\{ \lambda_{nm}(k) \lambda^*_{n'm'}(k) \Big \} + \\
E \Big \{ \eta_{nm}(k) \eta^*_{n'm'}(k) \Big \}
\end{multline}
where we assume uncorrelated speech and noise sources, i.e.
\begin{equation} \label{eq:uncorrelated-noise-speech}
E \Big\{ \lambda_{nm}(k) \text{ } \eta^*_{n'm'}(k) \Big \} = 0.
\end{equation}
\subsubsection{Spatial correlation of the direct and reverberant components}
From \eqref{eq:alpha-model} and \eqref{eq:alpha-model-shortcut}, the spatial cross-correlation between the direct and reverberant path coefficients is
\begin{multline} \label{eq:cross-alpha-1}
E \Big\{ \lambda_{nm}(k) \lambda^*_{n'm'}(k) \Big \} = C_{nn'} \sum \limits_{\ell=1}^{L} \sum \limits_{\ell'=1}^{L}  \text{ } E\{S_{\ell}(k) \text{ } S^*_{\ell'}(k)\} \times \\
E \Bigg\{ \Bigg( G_{\ell}^{(d)}(k) \text{ } Y^*_{nm}(\boldsymbol{\hat{y}}_{\ell}) + \int_{\boldsymbol{\hat{y}}} G^{(r)}_{\ell}(k, \boldsymbol{\hat{y}}) \text{ } Y^*_{nm}(\boldsymbol{\hat{y}}) \text{ } d\boldsymbol{\hat{y}} \Bigg) \times \\
\Bigg( G_{\ell'}^{(d)*}(k) \text{ } Y_{n'm'}(\boldsymbol{\hat{y}}_{\ell'}) + \int_{\boldsymbol{\hat{y}'}} G_{\ell'}^{(r)*}(k, \boldsymbol{\hat{y}'}) \text{ } Y_{n'm'}(\boldsymbol{\hat{y}'}) \text{ } d\boldsymbol{\hat{y}'} \Bigg) \Bigg\}
\end{multline}
where $C_{nn'} = 16 \pi^2 i^{n-n'}$.
Due to the autonomous behavior of the reflective surfaces in a room (i.e., the reflection gains from the reflective surfaces are independent from the direct path gain), the cross-correlation between the direct and reverberant gains is negligible, i.e.,
\begin{equation} \label{eq:uncorrelated-reverb-direct}
E \Big \{ G_{\ell}^{(d)}(k) \text{ } G_{\ell}^{(r)*}(k, \boldsymbol{\hat{y}}) \Big \} = 0.
\end{equation}
Furthermore, we assume that the sources are uncorrelated with each other, and so are the reverberant path gains from different directions, i.e.
\begin{equation} \label{eq:uncorrelated-direct-sound}
E \Big\{ S_\ell(k) \text{ } S^*_{\ell'}(k) \Big\} = E \Big\{ \lvert S_\ell(k) \rvert ^2 \Big \} \text{ } \delta_{\ell \ell'}
\end{equation}
\begin{equation} \label{eq:uncorrelated-reverb-sound}
E \Big\{ G_{\ell}^{(r)}(k, \boldsymbol{\hat{y}}) \text{ } G_{\ell}^{(r)*}(k, \boldsymbol{\hat{y}'}) \Big \} = E \Big\{ \lvert G_{\ell}^{(r)}(k, \boldsymbol{\hat{y}}) \rvert ^2 \Big \} \text{ } \delta_{\boldsymbol{\hat{y}} \boldsymbol{\hat{y}'}}
\end{equation}
where $\lvert \cdot \rvert$ denotes absolute value. Using \eqref{eq:uncorrelated-reverb-direct}, we eliminate the cross terms of the right hand side of \eqref{eq:cross-alpha-1} and deduce
\begin{multline} \label{eq:cross-alpha-2}
E\Big\{ \lambda_{nm}(k) \lambda^*_{n'm'}(k) \Big\} = C_{nn'} \sum \limits_{\ell=1}^{L} \sum \limits_{\ell'=1}^{L}  \text{ } E \Big\{S_{\ell}(k) \text{ } S^*_{\ell'}(k) \Big\} \\
\Bigg( E\left\{ G_{\ell}^{(d)}(k) G_{\ell'}^{(d)*}(k) \right\} \text{ } Y^*_{nm}(\boldsymbol{\hat{y}}_{\ell}) Y_{n'm'}(\boldsymbol{\hat{y}}_{\ell'}) + \\
\int_{\boldsymbol{\hat{y}}} \! \int_{\boldsymbol{\hat{y}'}} \!\!\!\! E \Big\{ G^{(r)}_{\ell}(k, \boldsymbol{\hat{y}}) G_{\ell'}^{(r)*}(k, \boldsymbol{\hat{y}'}) \Big\}  Y^*_{nm}(\boldsymbol{\hat{y}}) Y_{n'm'}(\boldsymbol{\hat{y}'}) d\boldsymbol{\hat{y}} d\boldsymbol{\hat{y}'} \Bigg).
\end{multline}
Defining $\Phi_\ell(k)= \left( E \Big\{\lvert S_{\ell}(k) \rvert ^2 \Big\} \text{ } E \Big\{ \lvert G_{\ell}^{(d)}(k) \rvert ^2 \Big\} \right)$ as the PSD of the $\ell^{th}$ source at the origin, we use \eqref{eq:uncorrelated-direct-sound} and \eqref{eq:uncorrelated-reverb-sound} in \eqref{eq:cross-alpha-2} to obtain
\begin{multline} \label{eq:cross-alpha-3}
\!\!\!\! E\Big\{ \lambda_{nm}(k) \lambda^*_{n'm'}(k) \Big\} = C_{nn'} \!\! \sum \limits_{\ell=1}^{L} \!\! \Bigg( \Phi_\ell(k) \text{ } Y^*_{nm}(\boldsymbol{\hat{y}_{\ell}}) \text{ } Y_{n'm'}(\boldsymbol{\hat{y}}_{\ell}) \\
 + E \Big\{ \lvert S_{\ell}(k) \rvert ^2 \Big\} \int_{\boldsymbol{\hat{y}}} E \Big\{ \lvert G^{(r)}_{\ell}(k, \boldsymbol{\hat{y}}) \rvert ^2 \Big\} Y^*_{nm}(\boldsymbol{\hat{y}}) Y_{n'm'}(\boldsymbol{\hat{y}}) d\boldsymbol{\hat{y}} \Bigg).
\end{multline}
Since $\lvert G^{(r)}_{\ell}(k, \boldsymbol{\hat{y}}) \rvert ^2$ is defined over a sphere, we can represent it using the spherical harmonics decomposition as
\begin{equation} \label{eq:spherical-power}
E \Big\{ \lvert G^{(r)}_{\ell}(k, \boldsymbol{\hat{y}}) \rvert ^2 \Big\} = \sum \limits_{vu}^{V} \text{ } E \Big\{ \gamma^{(\ell)}_{vu}(k) \Big\} \text{ } Y_{vu}(\boldsymbol{\hat{y}})
\end{equation}
where $\gamma^{(\ell)}_{vu}(k)$ is the coefficient of the power of a reverberant sound field due to $\ell^{th}$ source and $V$ is a non-negative integer defining corresponding order. Substituting the value of $E\{ \lvert G^{(r)}_{\ell}(k, \boldsymbol{\hat{y}}) \rvert ^2 \}$ from \eqref{eq:spherical-power} into \eqref{eq:cross-alpha-3}, we derive
\begin{multline}\label{eq:cross-alpha-4}
\!\!\!\! E \Big\{ \lambda_{nm}(k) \lambda^*_{n'm'}(k) \Big\} = C_{nn'} \!\! \sum \limits_{\ell=1}^{L} \Bigg( \Phi_\ell(k) Y^*_{nm}(\boldsymbol{\hat{y}}_{\ell}) Y_{n'm'}(\boldsymbol{\hat{y}}_{\ell}) \\
+ E \Big\{\lvert S_{\ell}(k) \rvert ^2 \! \Big\} \! \sum \limits_{vu}^{V} \! E \Big\{ \gamma^{(\ell)}_{vu}(k) \Big\} \!\! \int_{\boldsymbol{\hat{y}}} \!\!\!\! Y_{vu}(\boldsymbol{\hat{y}}) Y^*_{nm}(\boldsymbol{\hat{y}}) Y_{n'm'}(\boldsymbol{\hat{y}}) d\boldsymbol{\hat{y}} \!\! \Bigg).
\end{multline}
Using the definition of Wigner constants $W_{v, n, n'}^{u, m, m'}$ from Appendix \ref{app:integral-sh}, we rewrite \eqref{eq:cross-alpha-4} as
\begin{multline}\label{eq:cross-alpha-5}
E \Big\{ \lambda_{nm}(k) \lambda^*_{n'm'}(k) \Big\} = \sum \limits_{\ell=1}^{L} \Phi_\ell(k) \text{ } C_{nn'} \text{ } Y^*_{nm}(\boldsymbol{\hat{y}}_{\ell}) \times \\
Y_{n'm'}(\boldsymbol{\hat{y}}_{\ell}) + \sum \limits_{vu}^{V} \text{ } \Gamma_{vu}(k) \text{ } C_{nn'} \text{ } W_{v, n, n'}^{u, m, m'}
\end{multline}
where
\begin{equation} \label{eq:Gamma-rev}
\Gamma_{vu}(k) = \left( \sum \limits_{\ell=1}^{L} E\{\lvert S_{\ell}(k) \rvert ^2\} \text{ } E \Big\{ \gamma^{(\ell)}_{vu}(k) \Big\} \right)
\end{equation}
is the total reverberant power for order $v$ and degree $u$. Please note that the spatial correlation model developed in \cite{samarasinghe2017estimating} was derived for a single source case, i.e. $L = 1$, and did not include background noise in the model.
\subsubsection{Spatial correlation model for coherent noise}
In a similar way \eqref{eq:alpha-theory} is derived, we obtain the expression for $\eta_{nm}$ from \eqref{eq:coherent-noise-sh} as
\begin{equation} \label{eq:noise-alpha}
\eta_{nm}(k) = \frac{1}{b_n(kr)} \int_{\boldsymbol{\hat{x}}} Z(\boldsymbol{x}, k) \text{ } Y_{nm}^*(\boldsymbol{\hat{x}}) \text{ } d\boldsymbol{\hat{x}}
\end{equation}
where $\boldsymbol{x} = (r, \boldsymbol{\hat{x}})$. Hence, we deduce
\begin{multline} \label{eq:noise-alpha-correaltion}
E \Big\{ \eta_{nm}(k) \eta^*_{n'm'}(k) \Big\} = \frac{1}{\lvert b_n(kr) \rvert ^2} \times \\
\int_{\boldsymbol{\hat{x}}} \int_{\boldsymbol{\hat{x}}'} E \Big\{ Z(\boldsymbol{x}, k) Z^*(\boldsymbol{x}', k) \Big\} Y_{nm}^*(\boldsymbol{\hat{x}}) Y_{n'm'}(\boldsymbol{\hat{x}}') \text{ } d\boldsymbol{\hat{x}} \text{ } d\boldsymbol{\hat{x}}'.
\end{multline}
The spatial correlation of the coherent noise is given by \cite{teal2002spatial}
\begin{multline} \label{eq:noise-spatial-correlation}
\!\!\!\! E \Big\{ Z(\boldsymbol{x}, k) Z^*(\boldsymbol{x}', k) \Big\} = E \Big\{ \lvert Z(\boldsymbol{x}, k) \rvert ^2 \Big\} \sum \limits_{nm}^{N} j_n(k \norm{\boldsymbol{x} - \boldsymbol{x}'}) \\
4 \pi i^n \text{ } Y_{nm} \left( \frac{\boldsymbol{x}' - \boldsymbol{x}}{\norm{\boldsymbol{x} - \boldsymbol{x}'}} \right) \int_{\boldsymbol{\hat{y}}} \frac{E \left\{ \lvert A(\boldsymbol{\hat{y}}) \rvert ^2 \right\} }{\int_{\boldsymbol{\hat{y}}} \lvert A(\boldsymbol{\hat{y}}) \rvert ^2 d\boldsymbol{\hat{y}}} \text{ } Y^*_{nm}(\boldsymbol{\hat{y}}) \text{ } d\boldsymbol{\hat{y}}
\end{multline}
where $A(\boldsymbol{\hat{y}})$ is the complex gain of the noise sources from $\boldsymbol{\hat{y}}$ direction. In a reverberant room, the noise field can be assumed to be diffused \cite{mccowan2003microphone}, hence \eqref{eq:noise-spatial-correlation} reduces to
\begin{equation} \label{eq:noise-spatial-correlation-diffuse}
E \Big\{ Z(\boldsymbol{x}, k) Z^*(\boldsymbol{x}', k) \Big\} = \Phi_{z_x}(k) \text{ } j_0(k \norm{\boldsymbol{x} - \boldsymbol{x}'})
\end{equation}
where $\Phi_{z_x}(k)$ is the PSD of the noise field at $\boldsymbol{x}$. Furthermore, for the sake of simplicity, we assume that the noise field is spatially white within the small area of a spherical microphone array (e.g., a commercially available spherical microphone array \lq{Eigenmike}\rq \cite{eigenmikeweb} has a radius of $4.2$ cm), i.e. $\Phi_{z_x}(k) = \Phi_{z}(k) \text{ } \forall \boldsymbol{x}$. Hence, from \eqref{eq:noise-alpha-correaltion} and \eqref{eq:noise-spatial-correlation-diffuse}, we get
\begin{multline} \label{eq:noise-alpha-correaltion-2}
E \Big\{ \eta_{nm}(k) \eta^*_{n'm'}(k) \Big\} = \Phi_{z}(k) \frac{1}{\lvert b_n(kr) \rvert ^2} \\
\int_{\boldsymbol{\hat{x}}} \int_{\boldsymbol{\hat{x}}'} j_0(k \norm{\boldsymbol{x} - \boldsymbol{x}'}) \text{ } Y_{nm}^*(\boldsymbol{\hat{x}}) Y_{n'm'}(\boldsymbol{\hat{x}}') \text{ } d\boldsymbol{\hat{x}} \text{ } d\boldsymbol{\hat{x}}'.
\end{multline}
\subsubsection{The combined model} \label{sec:combined-model}
Finally, from \eqref{eq:cross-alpha-5} and \eqref{eq:noise-alpha-correaltion-2}, we obtain the complete model of the spatial correlation in a noisy reverberant environment as
\begin{multline}\label{eq:final-model-rev-theroem}
E \Big\{ \alpha_{nm}(k) \alpha^*_{n'm'}(k) \Big\} = \sum \limits_{\ell=1}^{L} \Phi_\ell(k) \text{ } \Upsilon_{nm}^{n'm'}(\boldsymbol{\hat{y}}_{\ell}) + \\
\sum \limits_{vu}^{V} \text{ } \Gamma_{vu}(k) \text{ } \Psi_{n,n',v}^{m,m',u} + \Phi_{z}(k) \text{ } \Omega_{nm}^{n'm'}(k)
\end{multline}
where
\begin{equation}\label{eq:final-model-direct-coeff}
\Upsilon_{nm}^{n'm'}(\boldsymbol{\hat{y}}_{\ell}) = C_{nn'} \text{ } Y^*_{nm}(\boldsymbol{\hat{y}}_{\ell}) \text{ } Y_{n'm'}(\boldsymbol{\hat{y}}_{\ell})
\end{equation}
\begin{equation}\label{eq:final-model-reverb-coeff}
\Psi_{n,n',v}^{m,m',u} = C_{nn'} \text{ } W_{n,n',v}^{m,m',u}
\end{equation}
\begin{multline}\label{eq:final-model-noise-coeff}
\Omega_{nm}^{n'm'}(k) = \frac{1}{\lvert b_n(kr) \rvert ^2} \int_{\boldsymbol{\hat{x}}} \int_{\boldsymbol{\hat{x}}'} j_0(k \norm{\boldsymbol{x} - \boldsymbol{x}'}) \times \\
Y_{nm}^*(\boldsymbol{\hat{x}}) Y_{n'm'}(\boldsymbol{\hat{x}}') \text{ } d\boldsymbol{\hat{x}} \text{ } d\boldsymbol{\hat{x}}'.
\end{multline}
The integrals of \eqref{eq:final-model-noise-coeff} can be evaluated using a numerical computing tool. An approximation of \eqref{eq:final-model-noise-coeff} can be made through the finite summations as
\begin{multline} \label{eq:final-model-noise-coeff1}
\Omega_{nm}^{n'm'}(k) \approx \frac{1}{\lvert b_n(kr) \rvert ^2} \sum \limits_{q=1}^{Q'} \sum \limits_{q'=1}^{Q'} w_q \text{ } w_{q'}^* \text{ } j_0(k \norm{\boldsymbol{x}_q - \boldsymbol{x}_{q'}}) \times \\
Y_{nm}^*(\boldsymbol{\hat{x}}_q) Y_{n'm'}(\boldsymbol{\hat{x}}_{q'})
\end{multline}
where $\boldsymbol{\hat{x}_q}$ and $w_q$ are chosen such a way that the orthonormal property of the spherical harmonics holds. Also, a closed-form expression for \eqref{eq:final-model-noise-coeff} is derived in Appendix \ref{app:closed-form-noise} with the help of the addition theorem of the spherical Bessel functions \cite{chew1995waves} as
\begin{equation} \label{eq:final-model-noise-coeff2}
\Omega_{nm}^{n'm'} \! (k) \! = \! \frac{(4 \pi)^{\frac{3}{2}} i^{(n-n'+2m+2m')} j_{n}(k r) j_{n'}(k r) W_{n, n', 0}^{-m, -m', 0}}{\lvert b_n(kr) \rvert ^2}.
\end{equation}
The spatial correlation model of \eqref{eq:final-model-rev-theroem} is developed considering a far-field sound propagation. Following the discussion of Section \ref{lab:rtf-modal}, it is evident from \eqref{eq:far-field-source} and \eqref{eq:near-field-source} that a near-field source consideration for the direct path signals changes the direct path coefficient $\Upsilon_{nm}^{n'm'}(\boldsymbol{\hat{y}_\ell})$ of \eqref{eq:final-model-rev-theroem} as
\begin{equation} \label{eq:near-field-change}
\Upsilon_{nm}^{n'm'}(\boldsymbol{\hat{y}}_{\ell}, k) = k^2 h_n(k r_{\ell}) h^*_{n'}(k r_{\ell}) \text{ } Y^*_{nm}(\boldsymbol{\hat{y}}_{\ell}) \text{ } Y_{n'm'}(\boldsymbol{\hat{y}}_{\ell}).
\end{equation}
Hence, to design a system with near-field sources, we require the additional knowledge of the source distance $r_{\ell}$.
\section{PSD estimation} \label{lab:psd-estimation}
In this section, we reformulate \eqref{eq:final-model-rev-theroem} into a matrix form and solve it in the least square sense to estimate the source, reverberant and noise PSDs. We also discuss an implementation issue and offer engineering solutions to the problem.
\subsection{Source PSDs}
Defining
\begin{equation} \label{eq:Lambda-corr}
\Lambda_{nm}^{n'm'} = E \Big\{ \alpha_{nm}(k) \alpha^*_{n'm'}(k) \Big\},
\end{equation}
we can write \eqref{eq:final-model-rev-theroem} in a matrix form by considering the cross-correlation between all the available modes as
\begin{equation} \label{eq:final-model-rev-matrix}
\boldsymbol{\Lambda} = \boldsymbol{T} \text{ } \boldsymbol{\Theta}
\end{equation}
where
\begin{equation} \label{eq:Lambda}
\boldsymbol{\Lambda} = [\Lambda_{00}^{00} \quad \Lambda_{00}^{1-1} \dots \Lambda_{00}^{NN} \quad \Lambda_{1-1}^{00} \dots \Lambda_{NN}^{NN}]^T_{1 \times (N+1)^4}
\end{equation}
\begin{equation} \label{eq:T}
\boldsymbol{T} =
\underbrace{
\begingroup % keep the change local
\setlength\arraycolsep{2pt}
\begin{bmatrix}
    \Upsilon_{00}^{00}(\boldsymbol{\hat{y}_1}) & \dots  &  {\Upsilon_{00}^{00}(\boldsymbol{\hat{y}_L})}  & \Psi_{0,0,0}^{0,0,0} & \dots & \Psi_{0,0,V}^{0,0,V} & \Omega_{00}^{00} \\
    \vdots & \vdots & \vdots & \vdots & \vdots & \vdots & \vdots \\
    \vdots & \vdots & \vdots & \vdots & \vdots & \vdots & \vdots \\
    \Upsilon_{NN}^{NN}(\boldsymbol{\hat{y}_1}) & \dots  &  {\Upsilon_{NN}^{NN}(\boldsymbol{\hat{y}_L})}  & \Psi_{N,N,0}^{N,N,0} & \dots & \Psi_{N,N,V}^{N,N,V} & \Omega_{NN}^{NN} \\
  \end{bmatrix}
  \endgroup
}_{(N+1)^4 \times (L+\{V+1\}^2+1)}
\end{equation}
\begin{equation} \label{eq:Psi}
\boldsymbol{\Theta} = [\Phi_1 \dots \Phi_{L} \quad \Gamma_{00} \dots \Gamma_{VV} \quad \Phi_{z}]^T_{1 \times (L+\{V+1\}^2+1)}
\end{equation}
where $(\cdot)^T$ denotes transpose operation. Note that, the frequency dependency is omitted in \eqref{eq:final-model-rev-matrix}-\eqref{eq:Psi} to simplify the notation. We estimate the source, reverberant, and noise PSDs by
\begin{equation} \label{eq:solution}
\boldsymbol{\hat{\Theta}} = \boldsymbol{T}^{\dagger} \text{ } \boldsymbol{\Lambda}
\end{equation}
where $^{\dagger}$ indicates the pseudo-inversion of a matrix. In a practical implementation, a half-wave rectification or similar measure is required on \eqref{eq:solution} to avoid negative PSDs. The terms $\Phi_{\ell}$ and $\Phi_{z}$ in the vector $\boldsymbol{\hat{\Theta}}$ of \eqref{eq:solution} represent the estimated source and noise PSDs at the origin, respectively. It is worth noting that, \eqref{eq:solution} can readily be used for estimating source PSDs in a non-reverberant or noiseless environment by respectively discarding the $\Psi$ and $\Omega$ terms from the translation matrix $\boldsymbol{T}$ in \eqref{eq:T}.
\subsection{Reverberant PSD}
The total reverberation PSD at the origin due to all the sound sources is
\begin{equation} \label{eq:reverb-power}
\Phi_r(k) = \sum \limits_{\ell=1}^{L} E \Big\{ \lvert S_{\ell}(k) \rvert ^2 \Big\} \int_{\boldsymbol{\hat{y}}} E \Big\{ \lvert G^{(\ell)}_r(k, \boldsymbol{\hat{y}}) \rvert ^2 \Big\} \text{ } d\boldsymbol{\hat{y}}.
\end{equation}
Using \eqref{eq:spherical-power}, the definition of $\Gamma_{vu}(k)$ in \eqref{eq:Gamma-rev}, and the symmetrical property of the spherical harmonics, \eqref{eq:reverb-power} can be written as
\begin{align} \label{eq:reverb-power-2}
\Phi_r(k) & = \sum \limits_{vu}^{V} \text{ } \Gamma_{vu}(k) \text{ } \int_{\boldsymbol{\hat{y}}} Y_{vu}(\boldsymbol{\hat{y}}) \text{ } d\boldsymbol{\hat{y}} \nonumber \\
& = \sum \limits_{vu}^{V} \text{ } \Gamma_{vu}(k) \text{ } \frac{\delta(v) \delta(u)}{\sqrt[]{4\pi}} = \frac{\Gamma_{00}(k)}{\sqrt[]{4\pi}}
\end{align}
where $\delta(\cdot)$ is the Dirac delta function. PSD estimation process for a single frequency bin is shown in Algorithm \ref{algo:psd-estimator}.
\begin{algorithm} \label{algo:psd-estimator}
  \SetAlgoLined
  \KwData{$\boldsymbol{x}_q, P(\boldsymbol{x}_q, k) \text{ } \forall q$}
  Find $\alpha_{nm}$ using \eqref{eq:alpha}. $w_q$ is manufacture defined;\\
  Get $\hat{y}_{\ell} \text{ } \forall {\ell}$ using any suitable DOA estimation technique;\\
  Calculate $\Upsilon_{nm}^{n'm'}$, $\Psi_{n,n',v}^{m,m',u}$, and $\Omega_{nm}^{n'm'}$ from \eqref{eq:final-model-direct-coeff}, \eqref{eq:final-model-reverb-coeff} and \eqref{eq:final-model-noise-coeff2}, respectively;\\
  Get the expected value $\Lambda_{nm}^{n'm'}$ using \eqref{eq:expected-value};\\
  Solve \eqref{eq:solution} for $\boldsymbol{\Theta}$ using the defintions from \eqref{eq:Lambda} - \eqref{eq:Psi}.
  \caption{Algorithm to estimate PSD components}
\end{algorithm}
\subsection{Bessel-zero issue} \label{sec:bessel-zero-issue}
One of the challenges in calculating the $\boldsymbol{\Lambda}$ vector is the Bessel-zero issue. We define Bessel-zero issue as the case when $\lvert b_n(kr) \rvert$ of \eqref{eq:alpha} takes a near-zero value and thus causes noise amplification and induces error in $\alpha_{nm}$ estimation. This situation arises in 3 distinct scenarios:
\subsubsection{At low frequencies} \label{sec:bessel-zero-issue-low-f}
\begin{figure}[!t]
\begin{minipage}[b]{0.49\linewidth}
  \centering
  \centerline{\includegraphics[width=\linewidth]{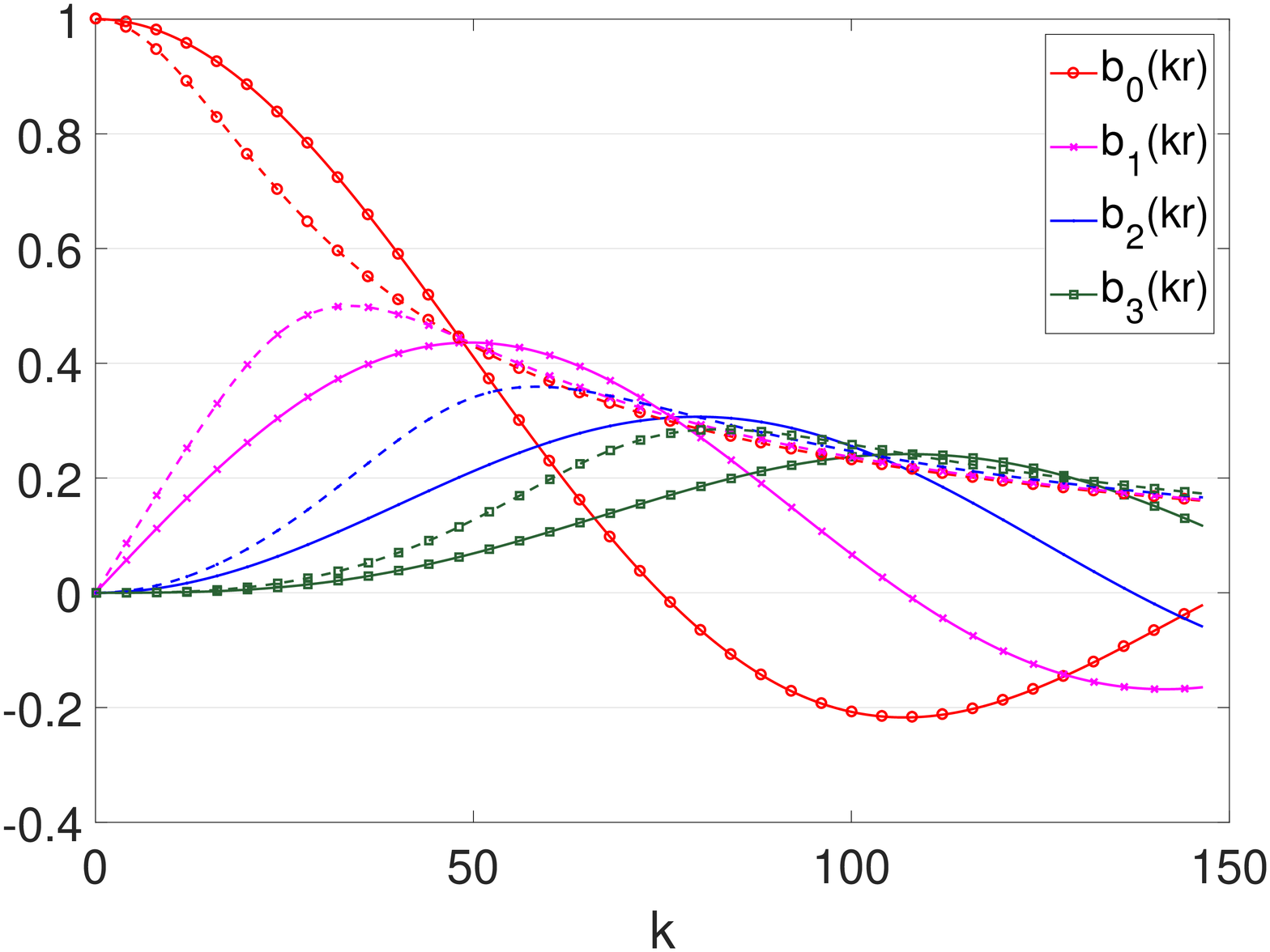}}
  \centerline{(a) Before modification}\medskip
\end{minipage}
\hfill
\begin{minipage}[b]{.49\linewidth}
  \centering
  \centerline{\includegraphics[width=\linewidth]{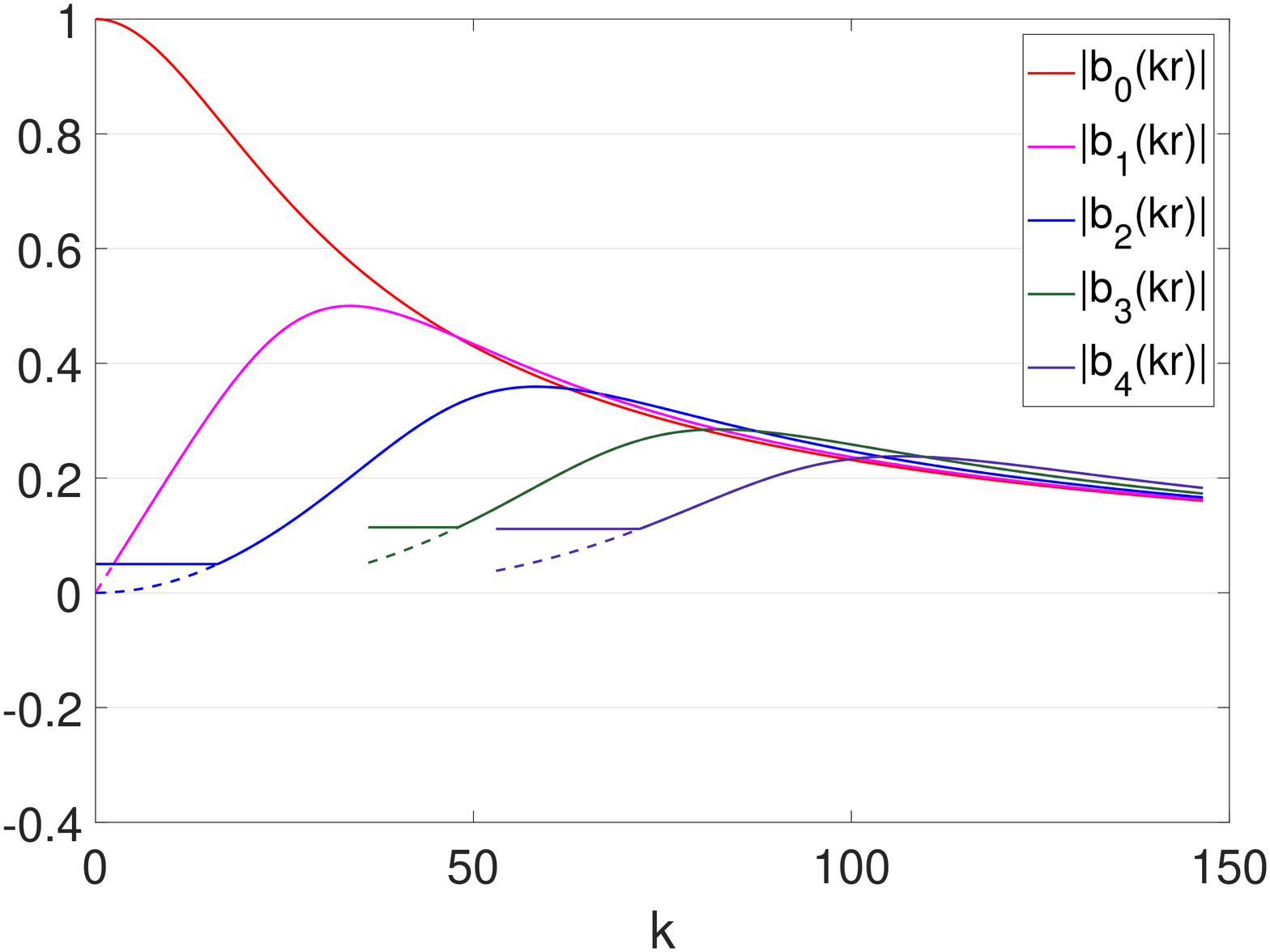}}
  \centerline{(b) After modification}\medskip
\end{minipage}
\caption{The unaltered Bessel functions with the modified version to alleviate the Bessel-zero issue. (a) Plots unaltered $b_n(kr)$ as a function of $k$. The complex values are plotted as magnitudes. Solid and dashed lines denote open and rigid arrays, respectively. (b) Shows $\lvert b_n(kr) \rvert$ after modification. Dashed extension denotes the original value.}
\label{fig:bessel-functions}
\end{figure}

To avoid underdetermined solutions as well as to improve the estimation accuracy of \eqref{eq:solution} by incorporating extra spatial modes, we force a minimum value of the sound field order $N$ at the lower frequency bins. For example, with $V=3$, $L=4$ and $f=500$ Hz, the calculated sound field order is $N=1$ and the dimension of $\boldsymbol{T}$ of \eqref{eq:final-model-rev-matrix} becomes $[16 \times 21]$ which results in an underdetermined system. In another scenario, if we choose a smaller value of $V=1$, though we can avoid an underdetermined system, the availability of a fewer spatial modes affects the estimation accuracy. Hence, we impose a lower boundary on $N$ for all frequency bins such that $N = \text{Max} \{ N, N_\text{min} \}$ where $\text{Max} \{ \cdot \}$ denotes the maximum value and $N_\text{min}$ is the lower limit of $N$. For this work, we choose $N_\text{min} = 2$ in an empirical manner. This, however, results in the aforementioned Bessel-zero issue for $n \in [1, N_\text{min}]$ at the lower frequencies as shown Fig. \ref{fig:bessel-functions}(a). To avoid this issue, we impose a lower boundary on $b_n(kr)$ as well such that
\begin{equation} \label{b_n-mod-1}
\lvert b_n(kr) \rvert = \text{Max} \Big \{ \lvert b_n(kr) \rvert, b_{n_{\text{min}}} \Big \} \text{, } n \in [1, N_\text{min}]
\end{equation}
where $b_{n_{\text{min}}}$ is a pre-defined floor value for $\lvert b_n(kr) \rvert$.
\subsubsection{At the mode activation boundary} \label{sec:bessel-zero-issue-mode-act}
This scenario appears at the first few frequency bins after a higher order mode ($N > N_\text{min}$) becomes active. As an example, for $r = 4.2$ cm, $3^{rd}$ order modes are activated approximately at $k_3^a=35$ and $k_3^b = 48$, where $k_3^a$ and $k_3^b$ are defined as the values of $k$ when we consider $N = \lceil k e r / 2 \rceil$ and $N = \lceil k r \rceil$, respectively. In the proposed algorithm, the $3^{rd}$ order modes are introduced at $k = k_3^a$ and we observe from Fig. \ref{fig:bessel-functions}(a) that the value of $\lvert b_3(kr) \rvert$ is close to zero for the first few frequency bins after the activation of the $3^{rd}$ order modes. To overcome this, we introduce another lower boundary criterion on $\lvert b_n(kr) \rvert$ as
\begin{equation} \label{b_n-mod-2}
\lvert b_n(kr) \rvert = \text{Max} \Big \{ \lvert b_n(kr) \rvert, \lvert b_n(k_n^b r) \rvert \Big \} \text{, } n > N_\text{min}.
\end{equation}
It is important to note that, the modifications proposed in \eqref{b_n-mod-1} and \eqref{b_n-mod-2} only affect the higher order modes at each frequency bin whereas the lower-order modes remain unchanged. Hence the distortion resulted from these modifications is expected to have less adverse impact than the Bessel-zero issue.
\subsubsection{Zero-crossing at a particular frequency} \label{sec:bessel-zero-issue-high-f}
Another case of a Bessel-zero issue occurs when the Bessel functions cross the zero-line on the y-axis at higher frequencies. This is more prominent with the open array configuration as shown in Fig. \ref{fig:bessel-functions}(a). The use of a rigid microphone array in the experiment is a way to avoid this issue which we followed in our experiments. Also note that, the modifications we propose for the previous two scenarios also take care of this zero crossing issue of the Bessel functions for an open array, when $N > 0$.\par
Fig. \ref{fig:bessel-functions}(b) plots the magnitudes of $b_n(kr)$ after the modification for different values of $k$. The impact of the Bessel-zero issue and the improvement after the proposed modifications are discussed in the result section.
\section{A practical application of PSD estimation in source separation} \label{lab:practical-application}
\begin{figure}[!ht]
\centering
\includegraphics[width=\linewidth]{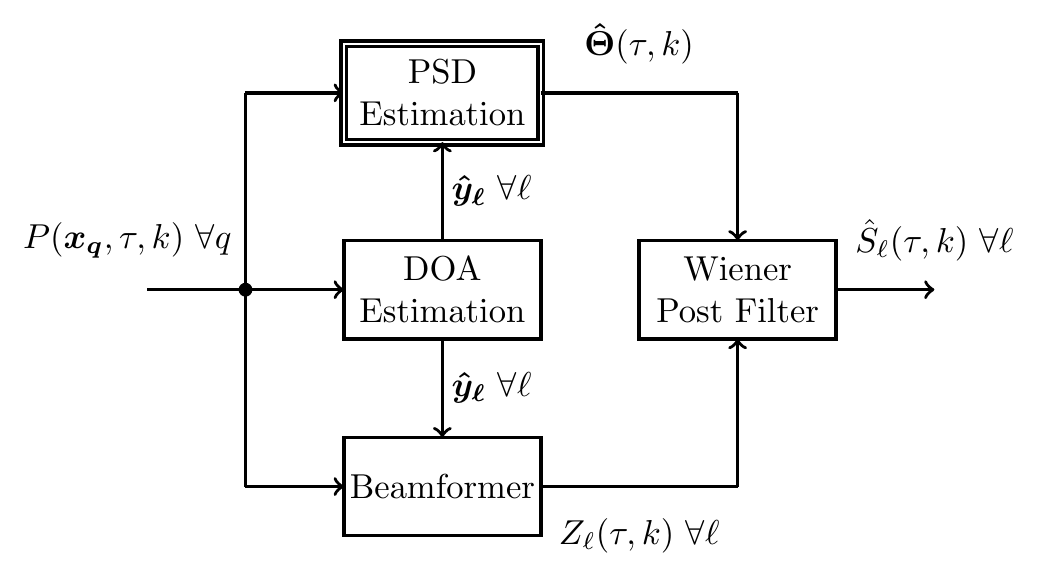}
\caption{The block diagram of the source separation as an application to the proposed PSD estimation method. The double-bordered block represents the proposed PSD estimation block.}
\label{fig:separation}
\end{figure}

An accurate knowledge of the PSDs of different signal components is desirable in many audio processing applications such as spectral subtraction, and Wiener filtering. In this section, we demonstrate a practical use of the proposed method in separating the speech signals from multiple concurrent speakers in a noisy reverberant room. This example uses the signal PSDs, estimated through the proposed algorithm, to significantly enhance the output of a conventional beamformer using a Wiener post filter. The performance of the Wiener filter largely depends on the estimation accuracy of the source and interferer PSDs, which is where the importance of a PSD estimation algorithm lies. A block digram of the complete source separation technique is shown in Fig. \ref{fig:separation} and explained in the subsequent sections.
\subsection{Estimation of the direction of arrival}
The proposed algorithm (and also any beamforming technique) requires the knowledge of the direction of arrival (DOA) of the desired speech signals as \textit{a priori}. If the source positions are unknown, any suitable localization technique, e.g., multiple signal classification, commonly known as MUSIC \cite{schmidt1986multiple}, can be used to estimate the DOA of the source signals. In the experiment where we measured the performance of the proposed algorithm, we used a frequency-smoothed approach of the MUSIC algorithm implemented in the spherical harmonics domain \cite{khaykin2009coherent}.
\subsection{Choice of Beamformer}
There are several beamforming techniques available in the literature such as delay and sum (DS), maximum directivity (MD), or minimum variance distortionless response (MVDR) etc. The choice of the beamforming technique depends on the use case in practice. In this work where the undesired signal includes the correlated reverberant component of the desired signal, an MVDR beamformer can result desired signal cancellation if the undesired PSD components at each microphone position are unknown. Hence, a simple delay and sum beamformer or a maximum directivity beamformer is more appropriate for the current case whose output, when steered towards $\ell^{th}$ far-field source, is given by \cite{meyer2002highly,rafaely2008spherical}
\begin{equation} \label{eq:bf-output}
\hat{S}_{bf}^{(\ell)}(k) = \sum \limits_{nm}^{N} \text{ } d_n(kr) \text{ } \alpha_{nm}(k) Y_{nm}(\theta_{\ell}, \phi_{\ell})
\end{equation}
where
\begin{equation} \label{eq:dn}
d_n(kr) = \begin{cases}
\frac{i^{-n}}{(N+1)^2} & \text{ for an MD beamformer} \\
\frac{4 \pi \lvert b_n(kr) \rvert ^2}{i^{n}} & \text{ for a DS beamformer}
\end{cases}.
\end{equation}
\subsection{Wiener post-filter}
Regardless of the choice of the beamformer, a post filter is found to enhance the beamformer output in most of the cases \cite{zelinski1988microphone,marro1998analysis,mccowan2003microphone}.  Hence, at the last stage, we apply a Wiener post filter at the beamformer output where the estimated PSD values is used. The transfer function of a Wiener filter for the $\ell^{th}$ source is given by
\begin{equation} \label{eq:wf-transfer-function}
H_{w}^{(\ell)}(k) = \frac{\Phi_{\ell}(k)}{\sum \limits_{\ell'=1}^{L} \Phi_{\ell'}(k) + \Phi_r(k) + \Phi_{z}(k)}.
\end{equation}
where all the PSD components are already estimated by the proposed algorithm and available in the vector $\boldsymbol{\hat{\Theta}}$. Hence, the $\ell^{th}$ source signal is estimated by
\begin{equation} \label{eq:final-estimation}
\hat{S}_{\ell}(k) = \hat{S}_{bf}^{(\ell)}(k) \text{ } H_{w}^{(\ell)}(k).
\end{equation}
\section{Experimental Results} \label{lab:experimental-results}
In this section, we demonstrate and discuss the experimental results based on practical recordings in a noisy and reverberant room using $4$, $6$, and $8$ speech sources.

\subsection{Experimental setup}
We evaluated the performance in $7$ distinct scenarios under $3$ different reverberant environments\footnote{8-speaker case was not tested in Room B \& C due to logistical issues.} as shown in Table \ref{table:experimen-cases}. The reverberation time $T_{60}$ and DRR in Table \ref{table:experimen-cases} were calculated based on the methods used in \cite{eaton2015ace}. All the experiments included background noise from the air-conditioning system and the vibration of the electrical equipments in the lab. We created separate training and evaluation datasets that consisted of $320$ male and female voices from the TIMIT database \cite{garofolo1993timit}. The training dataset was used to set the parameters such as $V$, $N_{min}$ etc. whereas the algorithm performance was measured using the evaluation dataset. Each of the $7$ scenarios was evaluated $50$ times with different mixtures of mixed-gender speech signals making it a total of $350$ unique experiments. We used the far-field assumption in each case for computational tractability. We measured the performance with the true and estimated DOA\footnote{The true and estimated DOAs for $L={4,6}$ are listed in Appendix \ref{app:source-directions}.} (denoted as \say{Proposed-GT} and \say{Proposed-EST}, respectively), where the latter was found with a MUSIC-based algorithm. We compared the performance with multiple beamformer-based method of \cite{hioka2013underdetermined} (denoted as \say{MBF}) and a conventional DS beamformer (denoted as \say{BF}) as, to the best of our knowledge, no similar harmonics-based technique has been proposed in the literature. Note that, for the fairness of comparison, we used all $32$ microphones of Eigenmike for all the competing methods. Furthermore, as the experiments were designed with the practical recordings instead of a simulation-based approach, the robustness of the proposed algorithm against the realistic thermal noise at the microphones was already evaluated through the process.\par
\begin{table}[ht]
\caption{Experimental environments. $d_{sm}$ denotes source to microphone distance.}
\label{table:experimen-cases}
\renewcommand{\arraystretch}{1.2}
\resizebox{\linewidth}{!}{%
\begin{tabular}{| c | c | c | c | c | c |}
\hline
 & Room size (m) & $T_{60}$ (ms) & DRR & $d_{sm}$ & \# Speakers \\
\hline
Room A & $6.5 \times 4.5 \times 2.75$ & $230$ & $10.9$ dB & $1$ m & $4, 6, 8$ \\
\hline
Room B & $6.5 \times 4.5 \times 2.75$ & $230$ & $2.5$ dB & $2$ m & $4, 6$ \\
\hline
Room C & $11 \times 7.5 \times 2.75$ & $640$ & $-0.6$ dB & $2.8$ m & $4, 6$ \\
\hline
\end{tabular}
}
\end{table}
The Eigenmike consists of $32$ pressure microphones distributed on the surface of a sphere with a radius of $4.2$ cm. The mixed sound was recorded at $48$ kHz sampling rate, but downsampled to $16$ kHz for computational efficiency. The recorded mixed signals were then converted to the frequency domain with a $8$ ms Hanning window, $50\%$ frame overlap, and a $128$-point fast Fourier transform (FFT). All the subsequent processing were performed in the STFT domain with the truncated sound field order $N = 4$, $N_{\text{min}} = 2$, and $b_{n_{\text{min}}} = 0.05$, unless mentioned otherwise. The noise PSD was assumed to have significant power up to $1$ kHz whereas all other PSD components were estimated for the whole frequency band. The expected value $\Lambda_{nm}^{n'm'}(k)$ of \eqref{eq:Lambda-corr} was computed using an exponentially weighted moving average as
\begin{equation} \label{eq:expected-value}
\Lambda_{nm}^{n'm'}(\tau, k) = \beta \Lambda_{nm}^{n'm'}(\tau-1, k) + (1 - \beta)  \alpha_{nm}(\tau, k) \alpha^*_{n'm'}(\tau, k)
\end{equation}
where $\beta \in [0, 1]$ is a smoothing factor, we chose $\beta = 0.8$.
\subsection{Selection of $V$}
$V$ represents the order of the power of a reverberation sound field. The exact harmonics analysis of the power of a reverberation sound field is a difficult task and depends on the structure, orientation, and characteristics of the reflective surfaces. Hence, unlike a free-field sound propagation, a reverberant field cannot be analytically decomposed into linear combination of Bessel functions which enables the truncation of a non-reverberant sound field order [31, 32]. Theoretically, $V$ extends to infinity, however, we need to consider several limiting factors such as
\begin{itemize}
\item Avoid an underdetermined system of equations in (43) which impose a limit on $V$ as
\begin{equation} \label{eq:V-limit}
V \leq \sqrt[]{(N+1)^2 - L - 1} - 1.
\end{equation}
\item Save computational complexity by choosing the minimum required value of $V$.
\end{itemize}
It is also important to note that the nature of the reverberation field plays an important role in determining $V$. As an example, for a perfectly diffused reverberant room with spatially-uniform reverberant power, only $0^{th}$ order ($V = 0$) mode is enough. On the other hand, a room with strong directional characteristics requires the higher orders to be considered. Hence, $V$ should be tuned separately for each reverberation environment to obtain an acceptable performance. In our experiments, we chose $V = 0, 4, 8$ for Room A, B, and C, respectively, based on the performance with the training dataset.
\subsection{Evaluation metrics}
We evaluate the performance through visual comparisons of the true and estimated PSDs of the sound sources. In addition to that, we also introduce an objective performance index to measure the full-band normalized PSD estimation error as
\begin{equation}
\Phi_{\text{err}_{\ell'}} = 10 \log_{10} \Bigg( \frac{1}{F} \sum \limits_{\forall k} \frac{E \big\{ \lvert \Phi_{\ell'}(\tau, k) - \hat{\Phi}_{\ell'}(\tau, k) \rvert \big\} }{E \big\{ \lvert \Phi_{\ell'}(\tau, k) \rvert \big\} } \Bigg)
\end{equation}
where $F$ is the total STFT frequency bands. To demonstrate an application of the source PSD estimation, we also evaluate the performance of a source separation algorithm based on a conventional delay and sum beamformer and a Wiener post-filter driven by the estimated PSDs. The overall speech qualities of the separated sources are measured by PESQ and FWSegSNR. For the purpose of a relative comparison, we analyze the original speech with reconstructed signals at the beamformer output and at the output of the Wiener filter driven by PSDs estimated through the proposed algorithm and the method proposed in \cite{hioka2013underdetermined}.
\begin{figure}[!ht]
\centering
\begin{minipage}[b]{0.49\linewidth}
  \centering
  \centerline{\includegraphics[width=\linewidth]{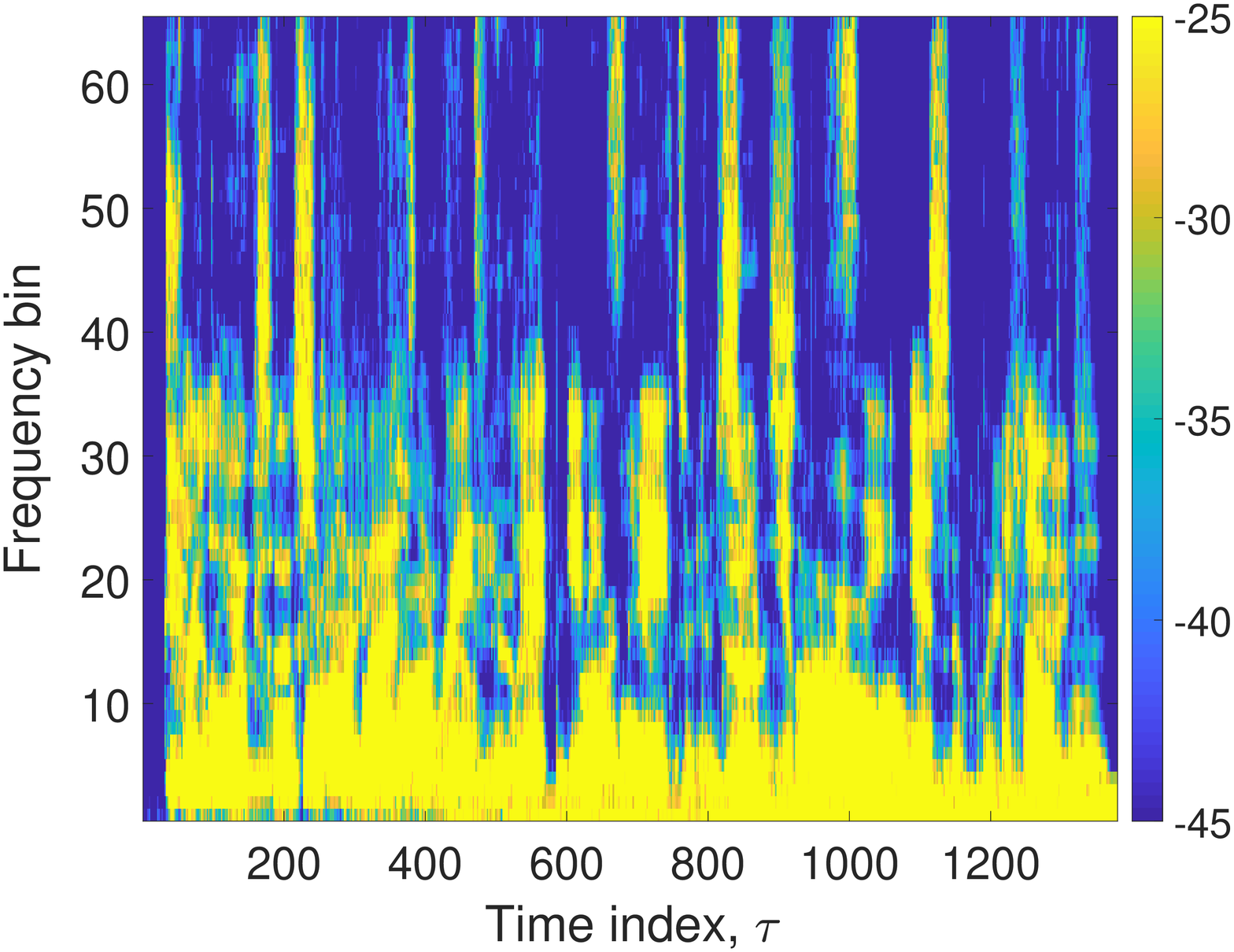}}
  \centerline{(a) Mixed signal PSD}\medskip
\end{minipage}
\begin{minipage}[b]{.49\linewidth}
  \centering
  \centerline{\includegraphics[width=\linewidth]{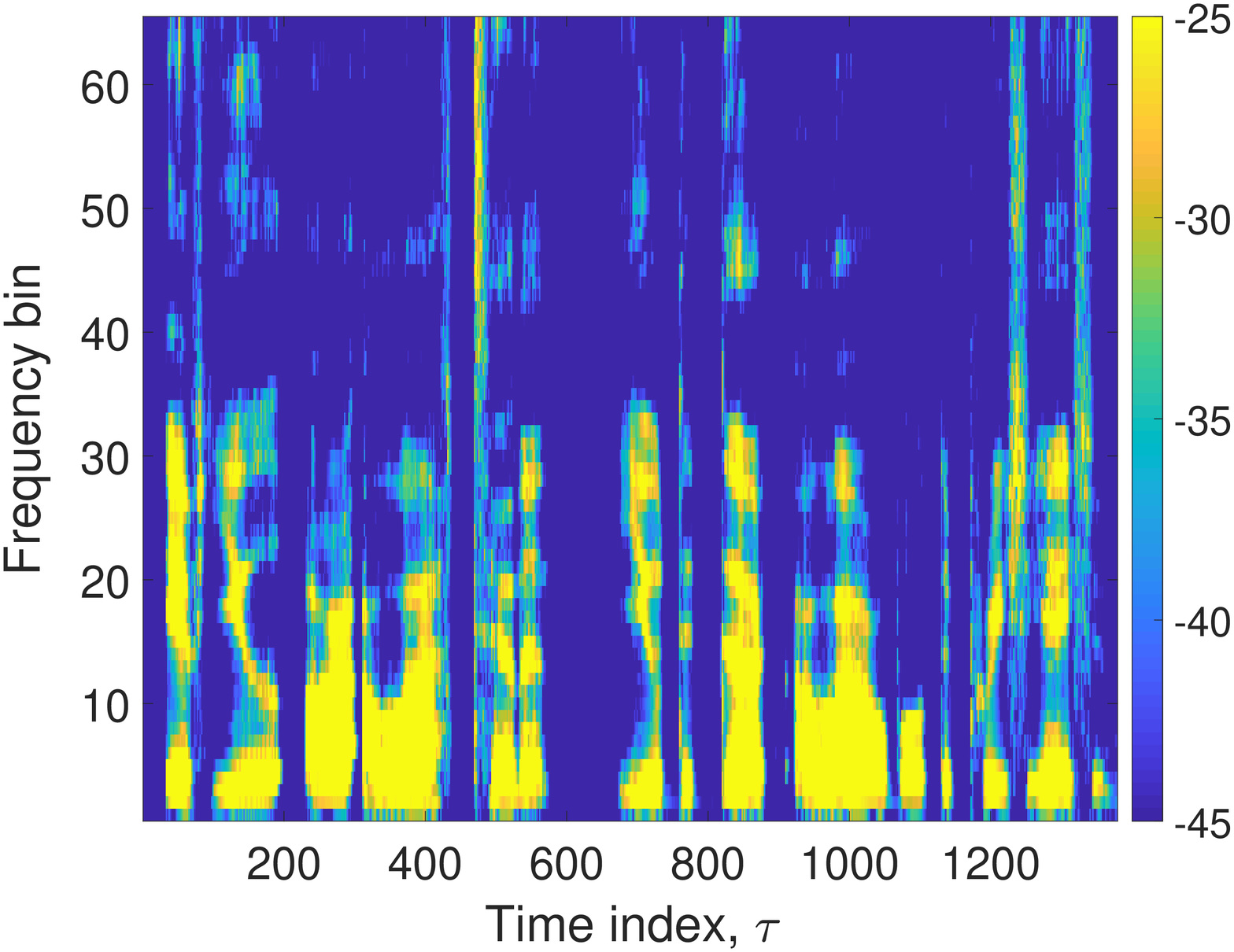}}
  \centerline{(b) Reference signal PSD}\medskip
\end{minipage}
\begin{minipage}[b]{0.49\linewidth}
  \centering
  \centerline{\includegraphics[width=\linewidth]{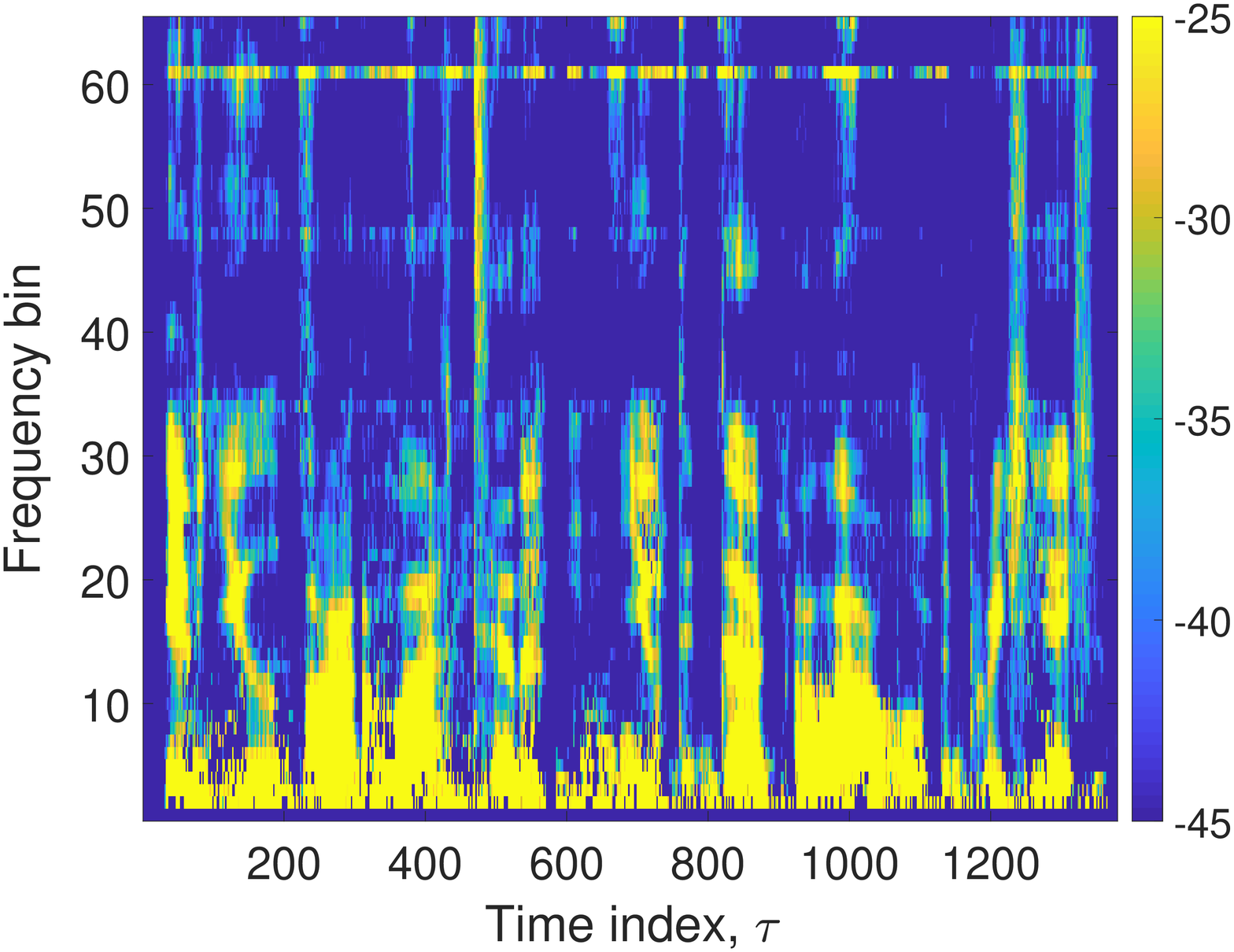}}
  \centerline{(c) Unmodified (open array)}\medskip
\end{minipage}
\begin{minipage}[b]{.49\linewidth}
  \centering
  \centerline{\includegraphics[width=\linewidth]{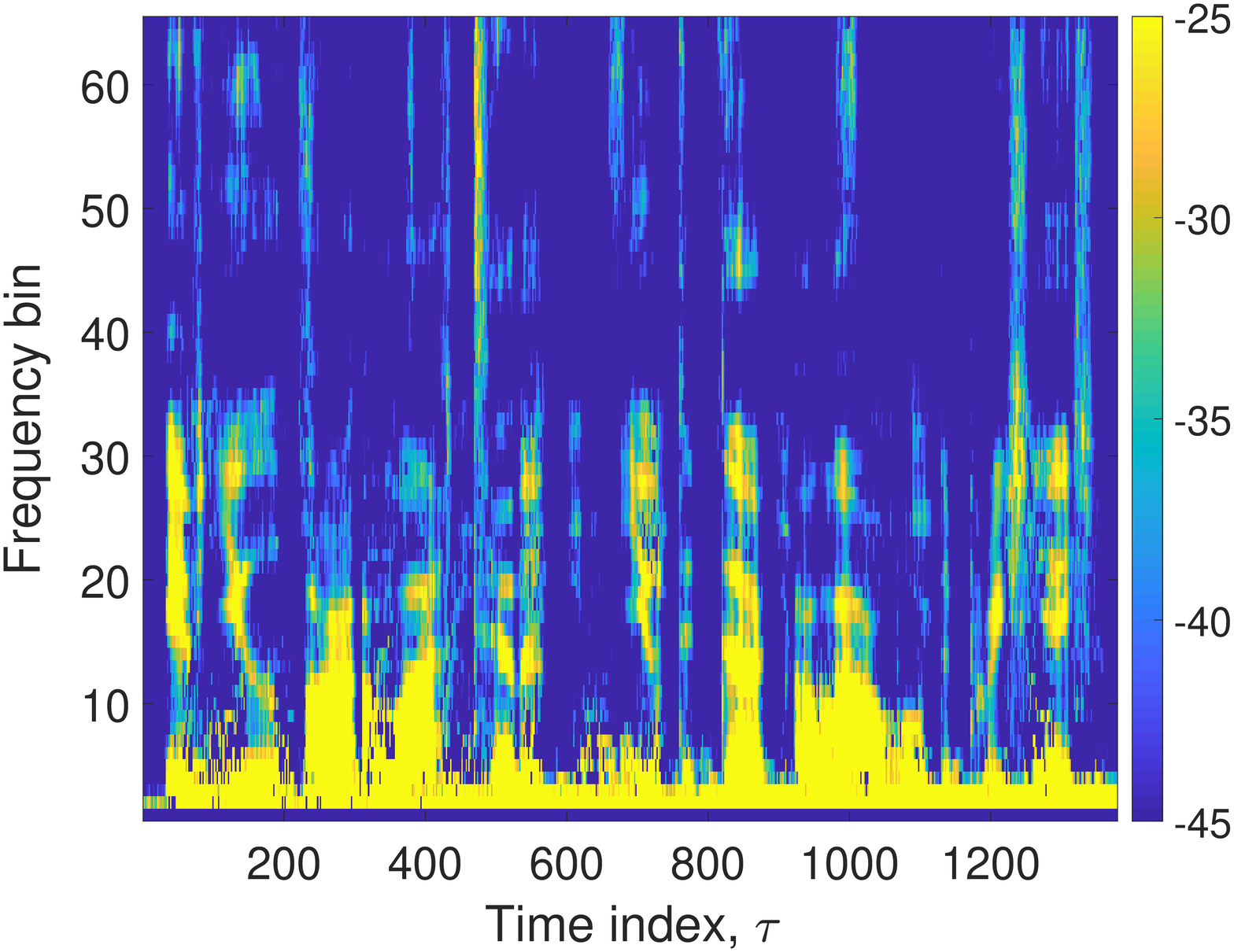}}
  \centerline{(d) Unmodified (rigid array)}\medskip
\end{minipage}
\begin{minipage}[b]{0.49\linewidth}
  \centering
  \centerline{\includegraphics[width=\linewidth]{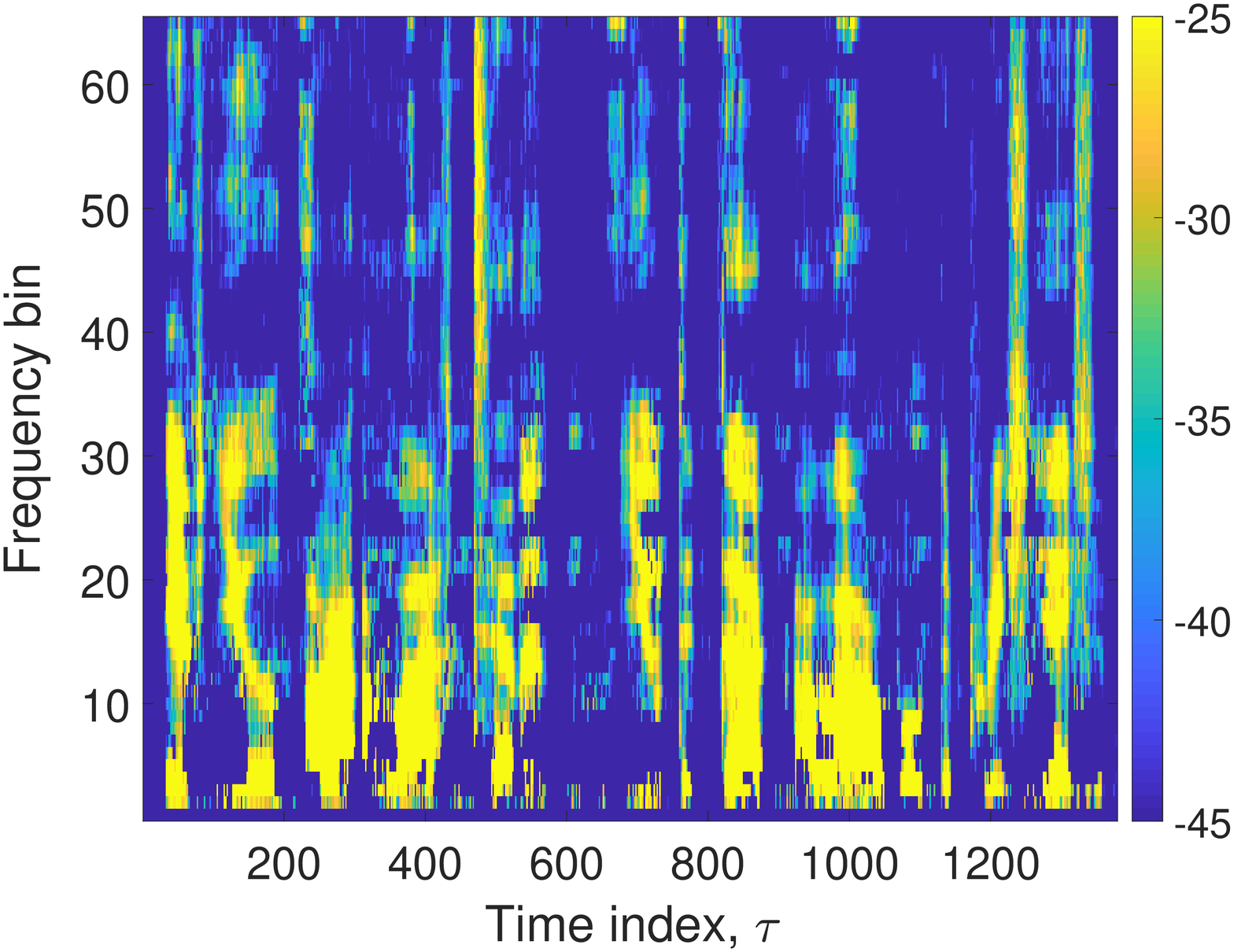}}
  \centerline{(e) Corrected (open array)}\medskip
\end{minipage}
\begin{minipage}[b]{0.49\linewidth}
  \centering
  \centerline{\includegraphics[width=\linewidth]{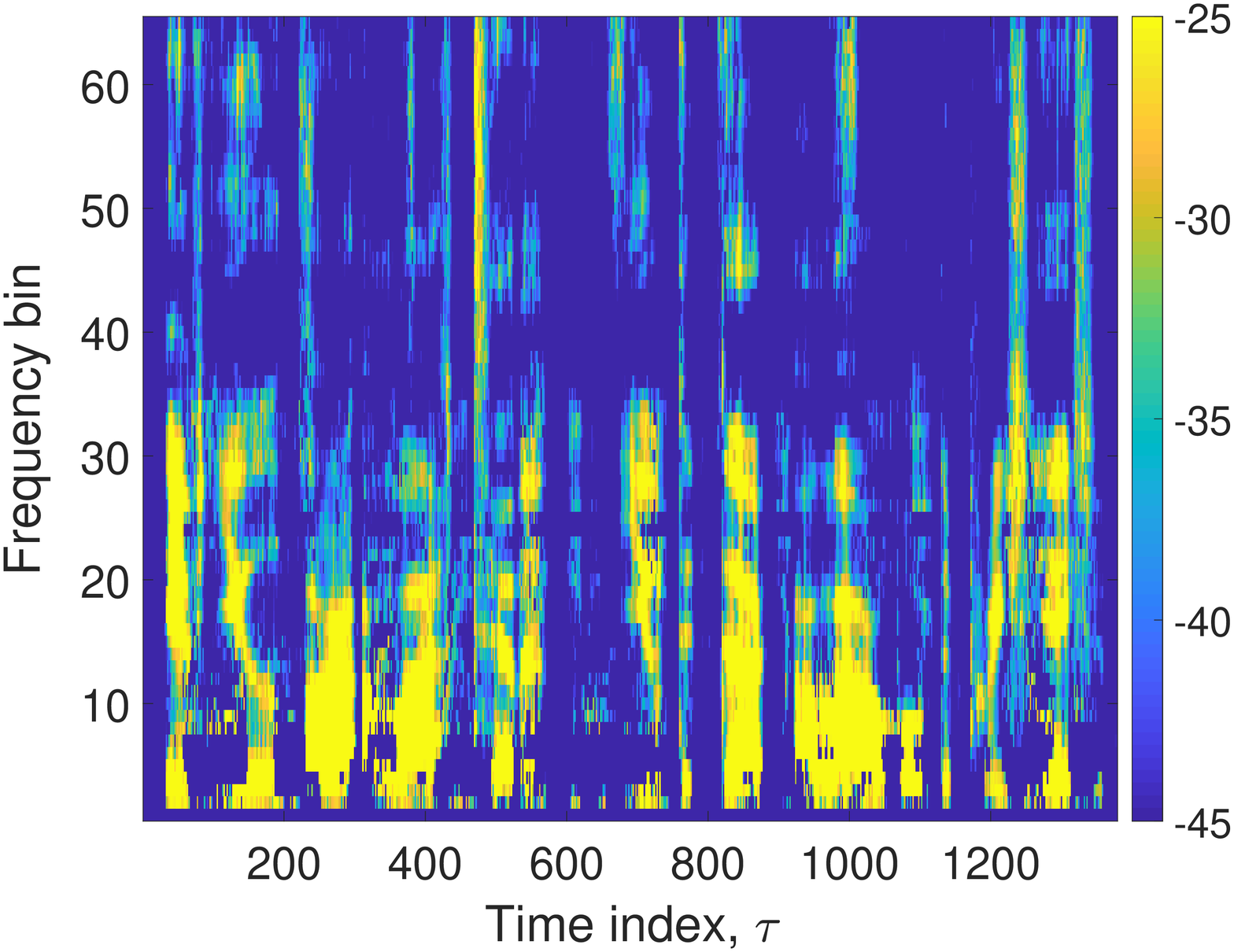}}
  \centerline{(f) Corrected (rigid array)}\medskip
\end{minipage}
\caption{The log-spectrograms of the PSDs in a simulated environment to demonstrate the Bessel-zero issue: (a) received signal at the first microphone, (b) true PSD, (c) and (d) estimated PSD without Bessel-zero correction using an open and a rigid array, respectively, and (e) and (f) estimated PSD of with Bessel-zero correction using an open and a rigid array, respectively.}
\label{fig:bessel-zero-correction-psd}
\end{figure}
\subsection{Visualization of Bessel-zero issue through simulation}
In this section, we discuss the practical impact of the Bessel-zero issue, described in Section \ref{sec:bessel-zero-issue}, on PSD estimation. For this section only, we used a simulated environment to generate the mixed signal, as we required the recordings of both open and rigid arrays to gain a better insight. The simulated environment had identical setup with the practical environment used in the experiments. Fig. \ref{fig:bessel-zero-correction-psd}(a) and (b) respectively show the PSDs of the mixed signal and the true PSD of the source signal for S-$01$. When we estimated the PSD for speaker $1$ using an open array and without the proposed Bessel-zero correction (Section \ref{sec:bessel-zero-issue}), it resulted in Fig. \ref{fig:bessel-zero-correction-psd}(c) where the spectral distortion is easily visible at the higher frequencies in the form of isolated horizontal bars (Section \ref{sec:bessel-zero-issue-high-f}) and some random distortions at the lower frequency range (Section \ref{sec:bessel-zero-issue-low-f}). The Bessel-zero issue described in Section \ref{sec:bessel-zero-issue-mode-act} is not prominent here as the impact depends on the spatial location and the relative power of the new incoming mode.\par
We also tried to solve the Bessel-zero issue by replacing the open array with a rigid array and the result is shown in Fig. \ref{fig:bessel-zero-correction-psd}(d). As expected, the rigid array removed the isolated distortions at the higher frequencies, but failed to act on the random distortions at the lower frequency range. It can also be observed that the rigid array resulted an inferior performance in terms of low-frequency noise suppression. As an alternative solution, we used the previous recording from the open array, but this time with the Bessel-zero correction as outlined in Section. \ref{sec:bessel-zero-issue}. The results, shown in Fig. \ref{fig:bessel-zero-correction-psd}(e), provided a better estimation this time by removing most of the Bessel-zero induced distortions. However, few distortions in the form of isolated spectral dots remained at the higher frequencies which were eventually removed when we integrated the proposed solution with a rigid array, as shown in Fig. \ref{fig:bessel-zero-correction-psd}(f).\par
Hence, we conclude that, irrespective of the array type, the most part of the Bessel-zero issue can be overcome through the proposed correction. However, for a better estimation accuracy, it is recommended to integrate the solution with a rigid microphone array. It must be noted that the Bessel-zero corrections can result in some spectral distortion especially at the lower frequency range due to a less number of active modes. However, the gain achieved through these corrections proved to be more significant compared to the resulting distortion, as evident from the source separation performances in Section \ref{sec:source-separation-application}.
\begin{figure}[!ht]
\begin{minipage}[b]{\linewidth}
  \centering
  \centerline{\includegraphics[width=\linewidth]{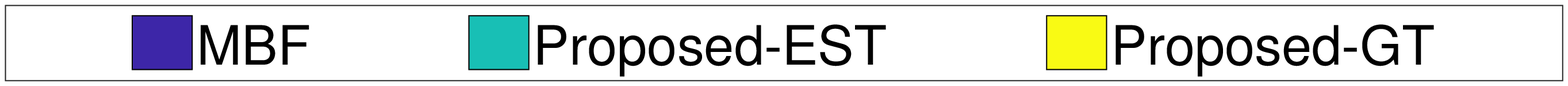}}
\end{minipage}
\begin{minipage}[b]{0.325\linewidth}
  \centering
  \centerline{\includegraphics[width=\linewidth]{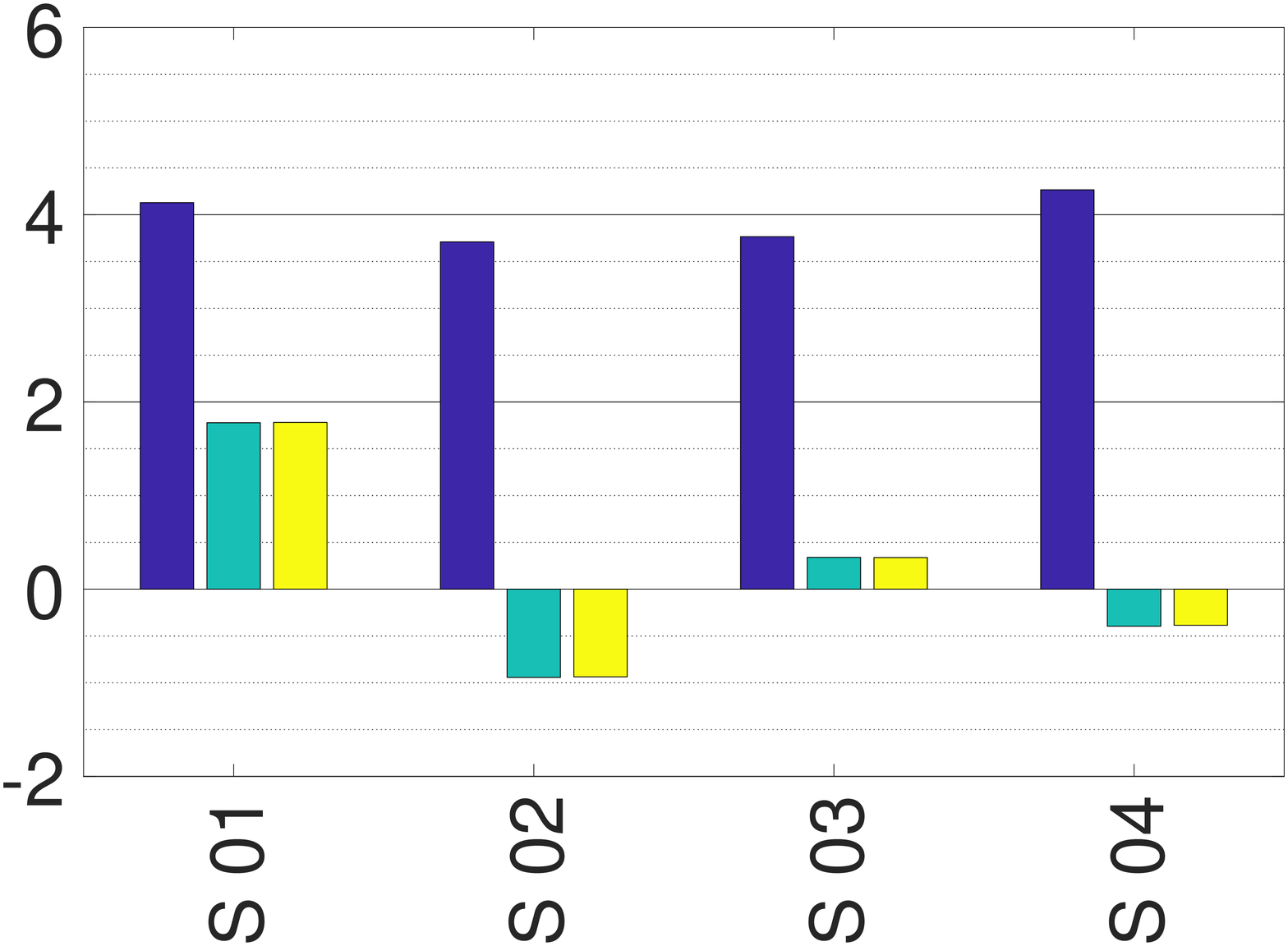}}
  \centerline{{\fontsize{9}{10}\selectfont(a) Room A ($L = 4$)}}\medskip
\end{minipage}
\begin{minipage}[b]{0.325\linewidth}
  \centering
  \centerline{\includegraphics[width=\linewidth]{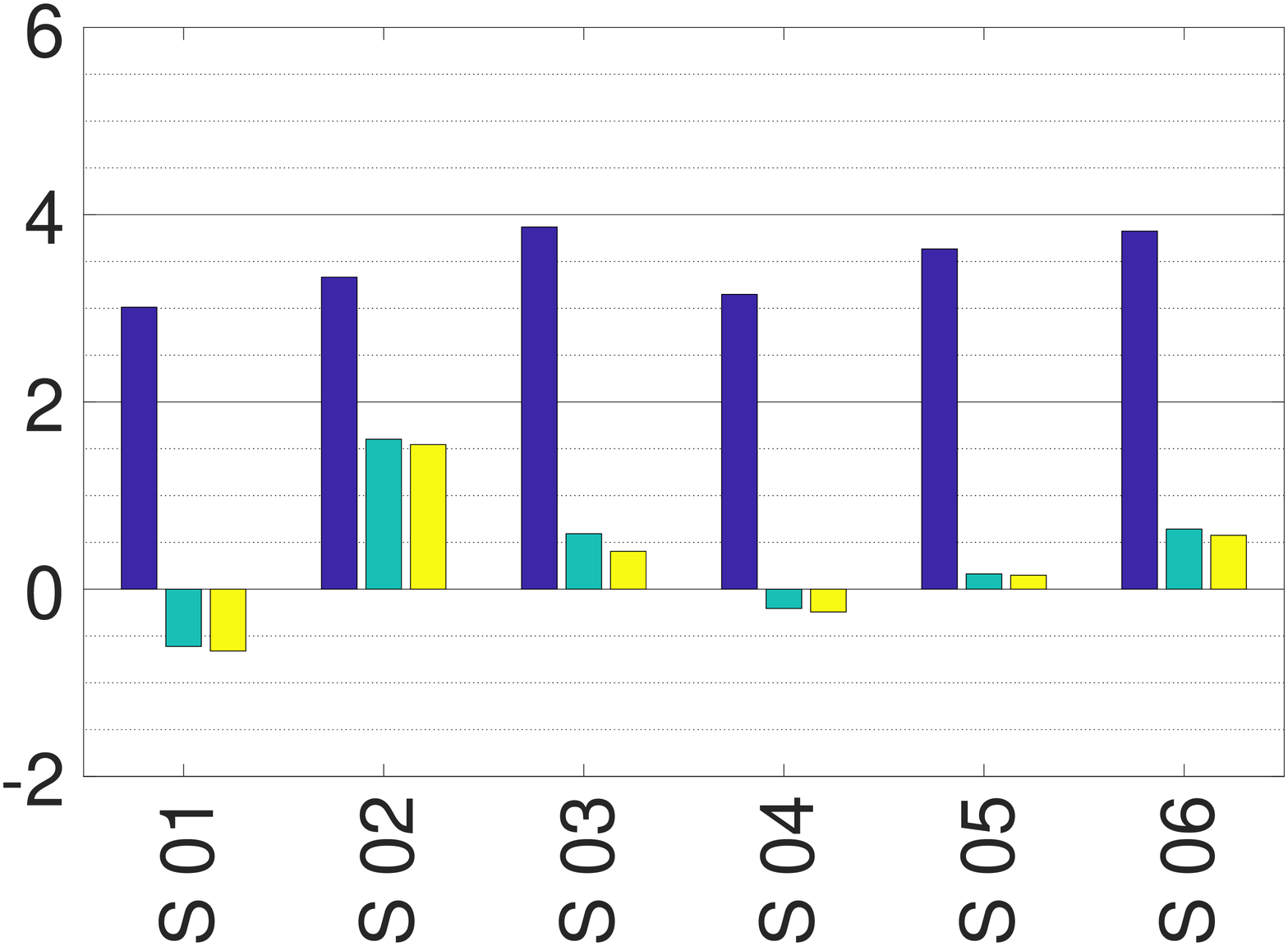}}
  \centerline{{\fontsize{9}{10}\selectfont(b) Room A ($L = 6$)}}\medskip
\end{minipage}
\begin{minipage}[b]{0.325\linewidth}
  \centering
  \centerline{\includegraphics[width=\linewidth]{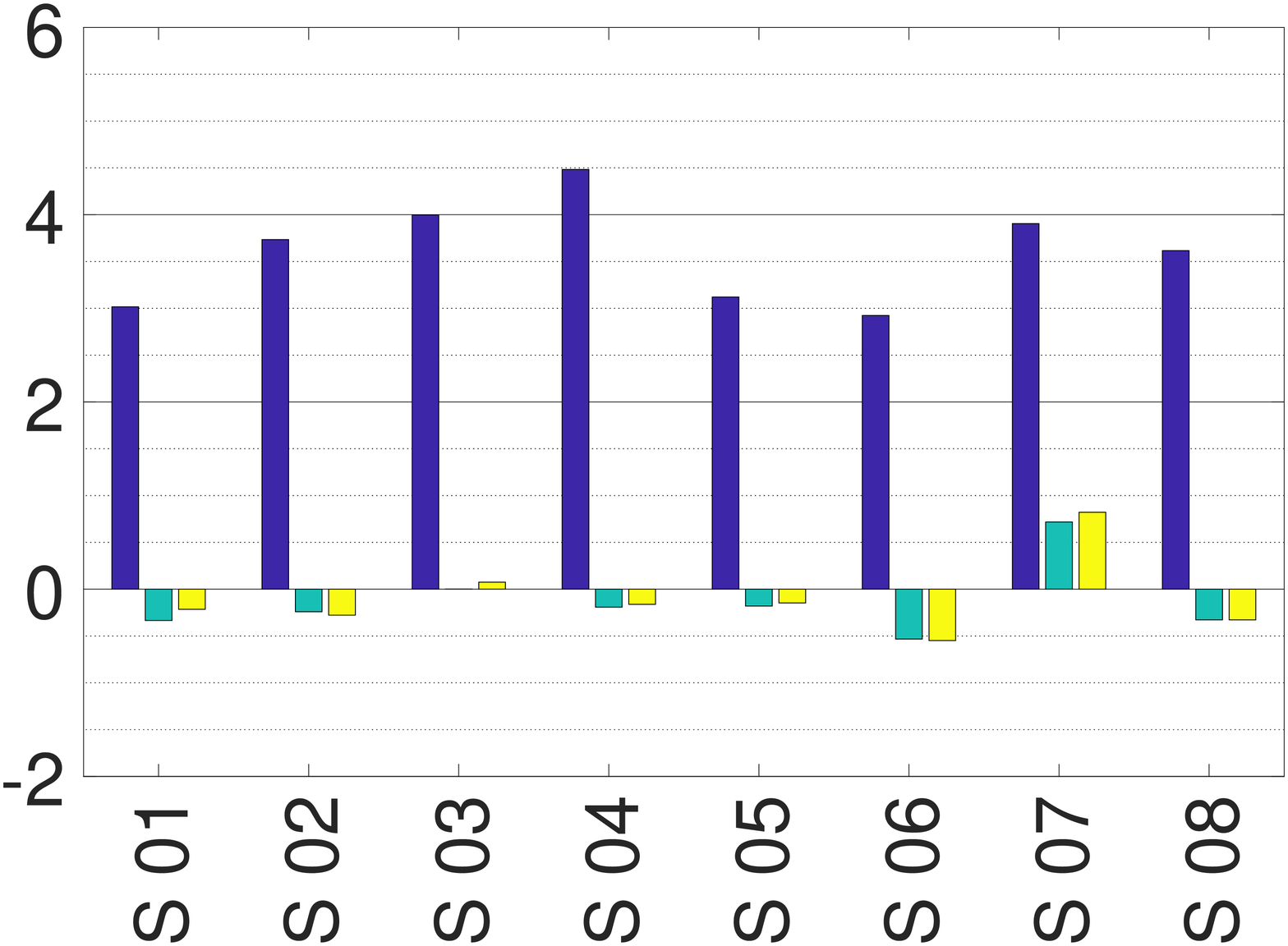}}
  \centerline{{\fontsize{9}{10}\selectfont(c) Room A ($L = 8$)}}\medskip
\end{minipage}
\begin{minipage}[b]{0.325\linewidth}
  \centering
  \centerline{\includegraphics[width=\linewidth]{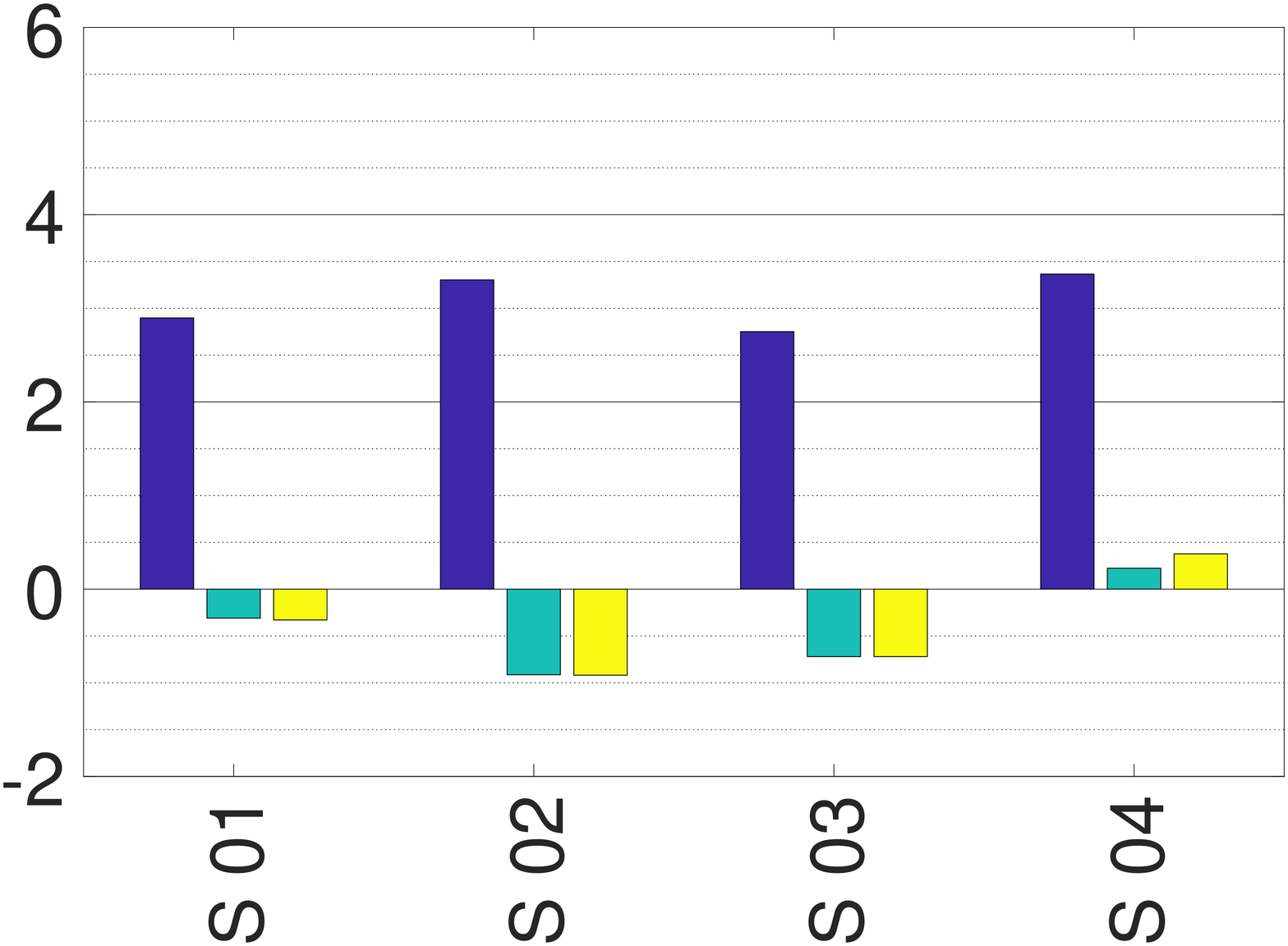}}
  \centerline{{\fontsize{9}{10}\selectfont(d) Room B ($L = 4$)}}\medskip
\end{minipage}
\begin{minipage}[b]{0.325\linewidth}
  \centering
  \centerline{\includegraphics[width=\linewidth]{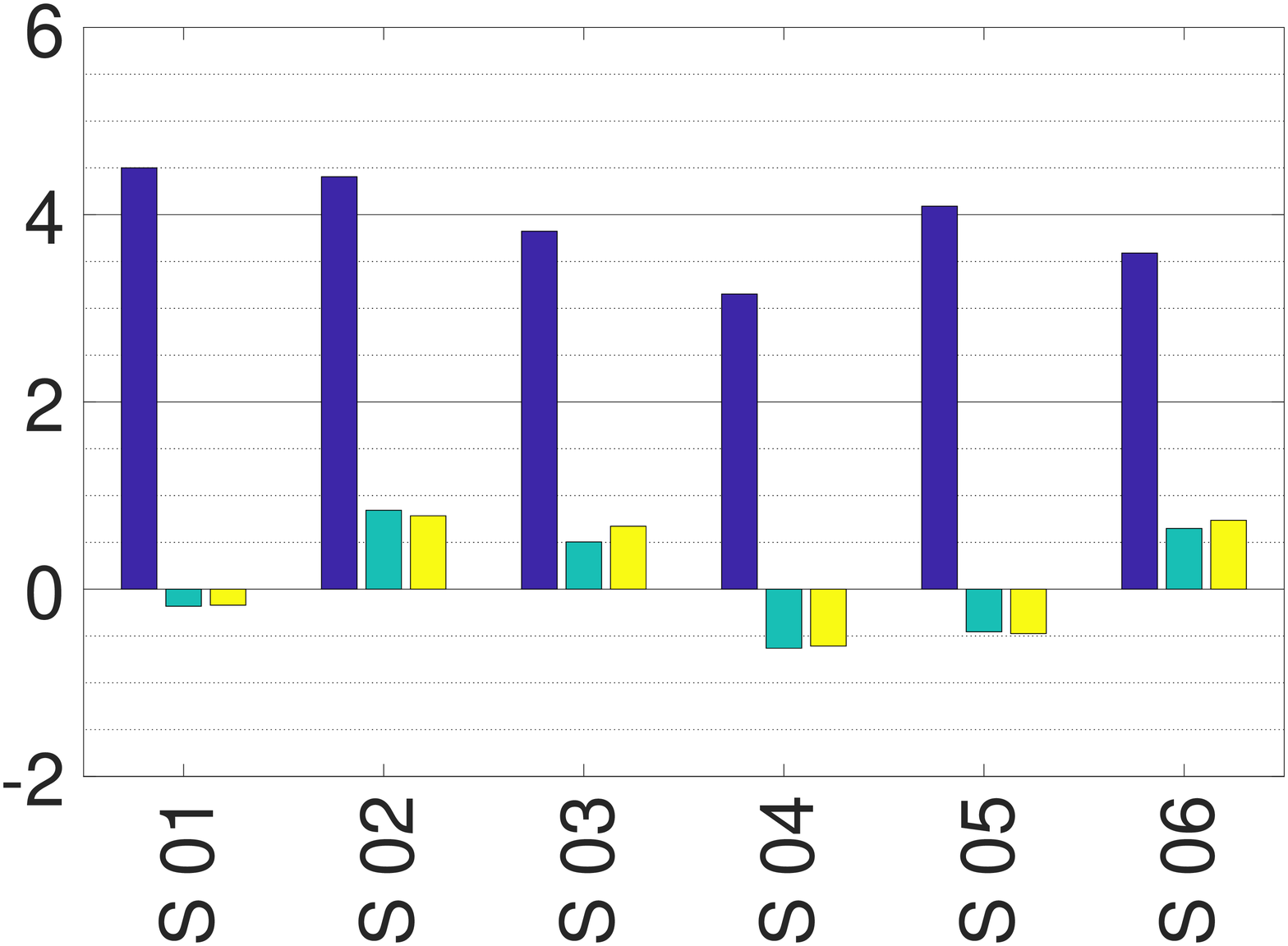}}
  \centerline{{\fontsize{9}{10}\selectfont(e) Room B ($L = 6$)}}\medskip
\end{minipage}
\hfill \\
\begin{minipage}[b]{0.325\linewidth}
  \centering
  \centerline{\includegraphics[width=\linewidth]{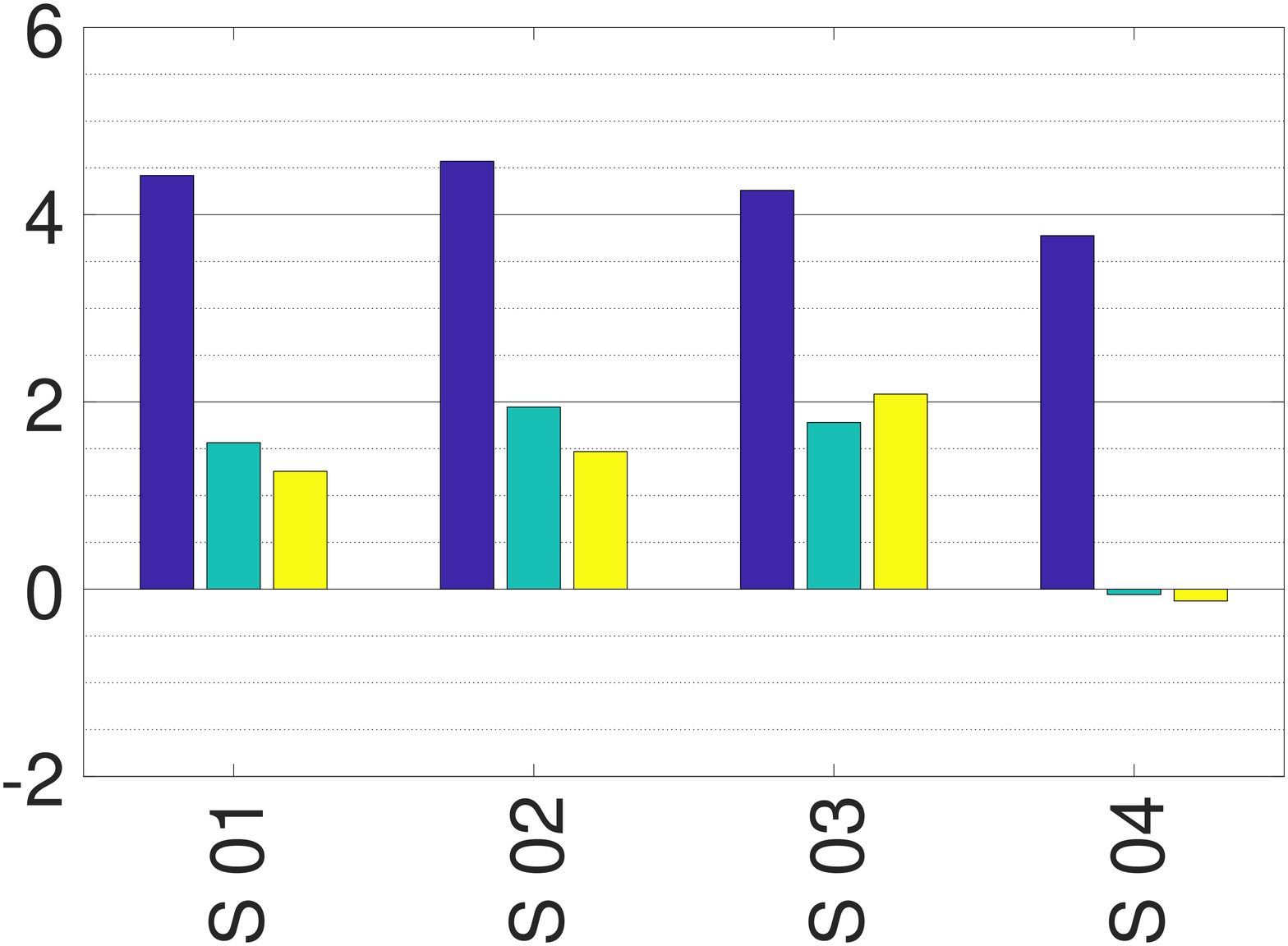}}
  \centerline{{\fontsize{9}{10}\selectfont(f) Room C ($L = 4$)}}\medskip
\end{minipage}
\begin{minipage}[b]{0.325\linewidth}
  \centering
  \centerline{\includegraphics[width=\linewidth]{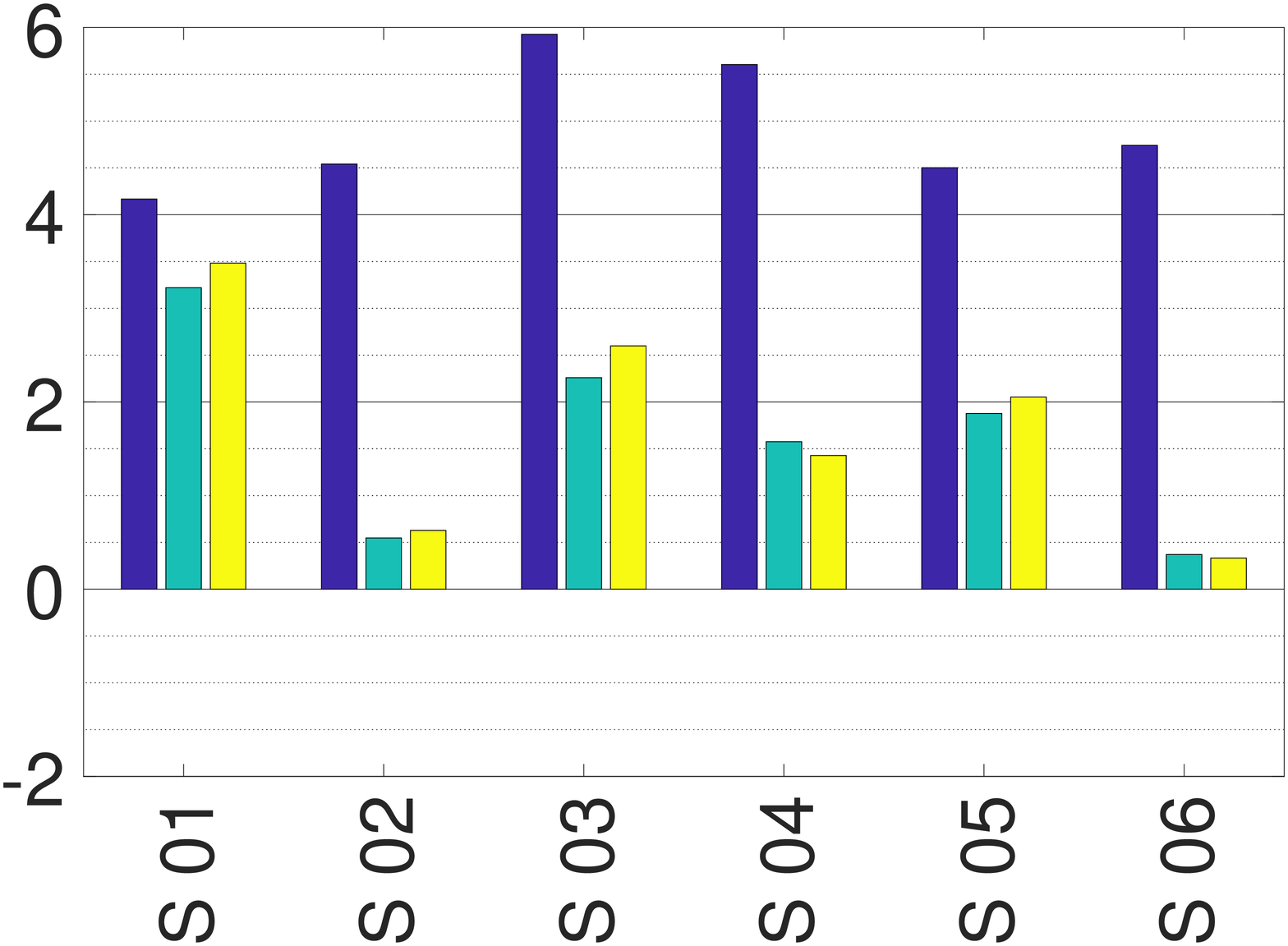}}
  \centerline{{\fontsize{9}{10}\selectfont(g) Room C ($L = 6$)}}\medskip
\end{minipage}
\caption{Full-band normalized PSD estimation error $\Phi_{\text{err}_{\ell'}}$ in $3$ distinct reverberant environments (Table \ref{table:experimen-cases}) for different number of sources.}
\label{fig:mse}
\end{figure}
\begin{figure}[!ht]
\begin{minipage}[b]{0.325\linewidth}
  \centering
  \centerline{\includegraphics[width=\linewidth]{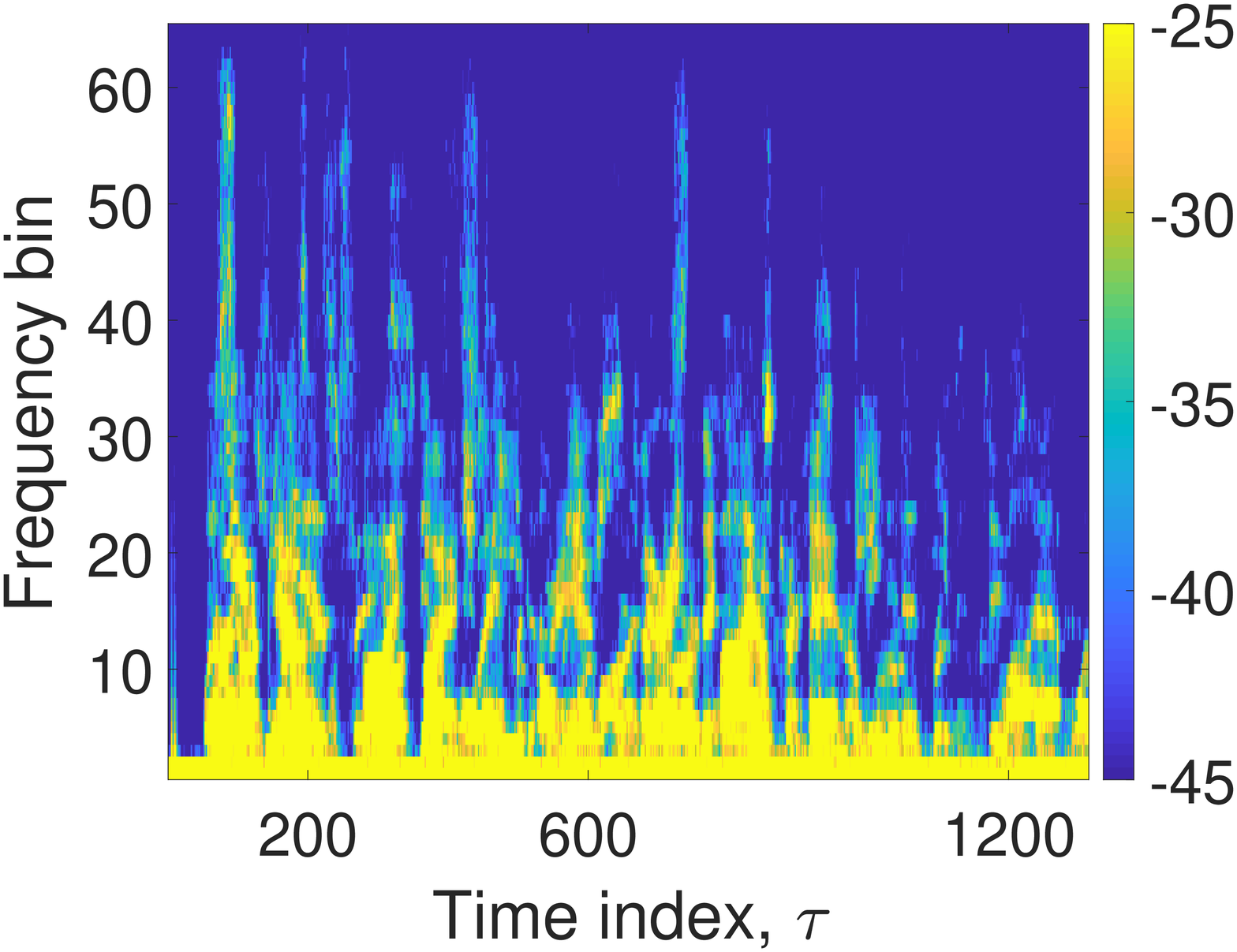}}
  \centerline{(a) Mixed}\medskip
\end{minipage}
\begin{minipage}[b]{0.325\linewidth}
  \centering
  \centerline{\includegraphics[width=\linewidth]{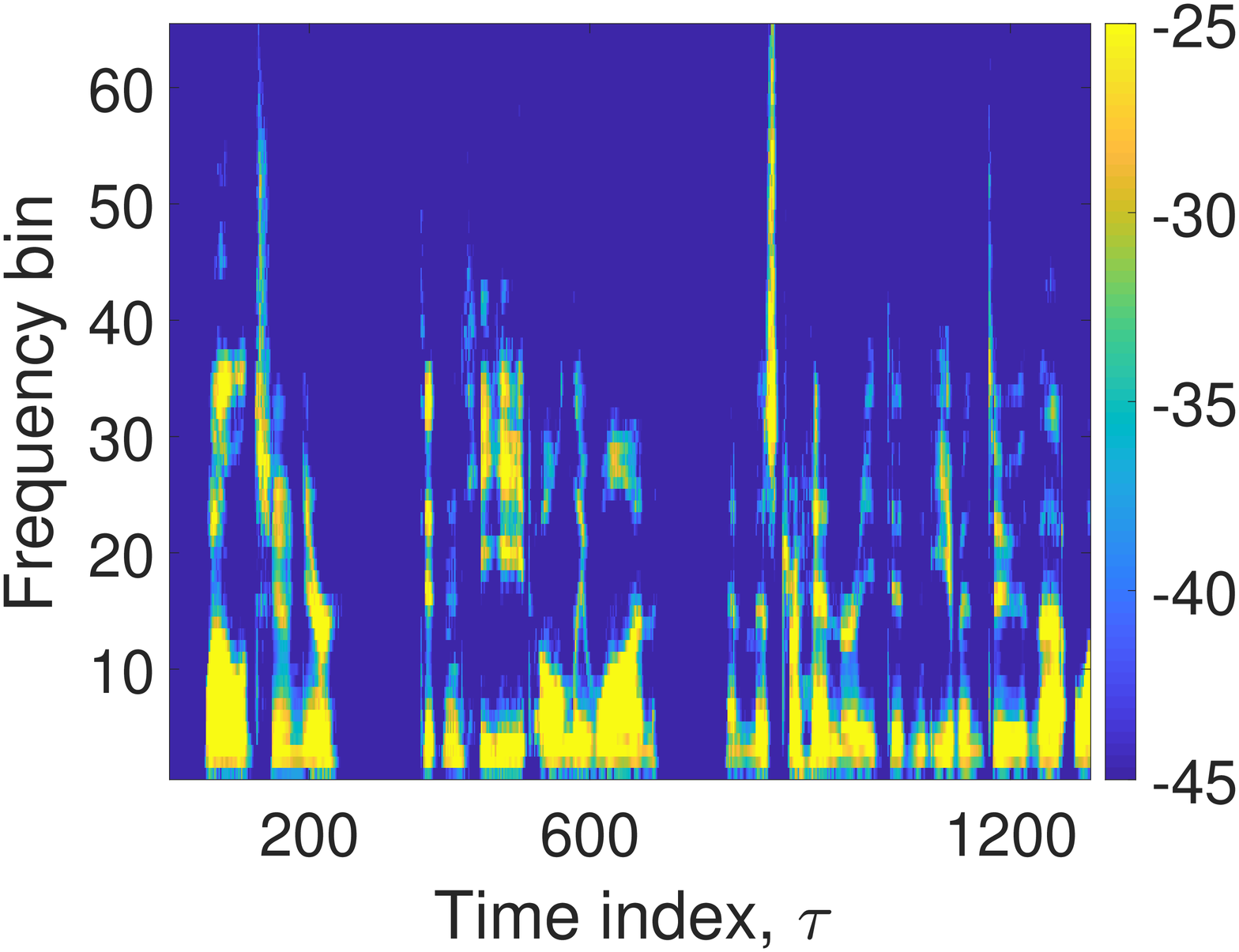}}
  \centerline{(b) Oracle (S-$03$)}\medskip
\end{minipage}
\begin{minipage}[b]{0.325\linewidth}
  \centering
  \centerline{\includegraphics[width=\linewidth]{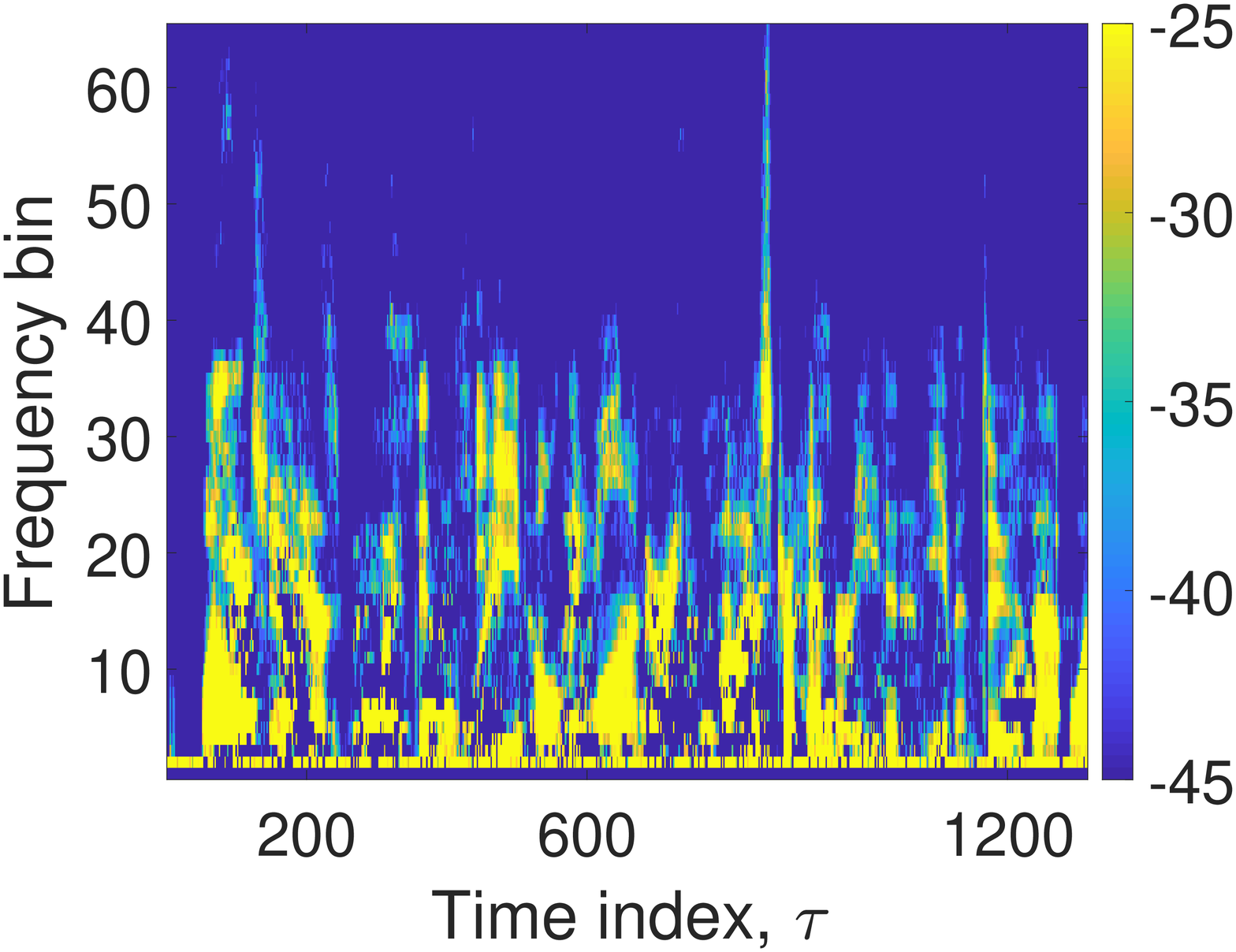}}
  \centerline{(c) Estimated (S-$03$)}\medskip
\end{minipage}
\begin{minipage}[b]{0.325\linewidth}
  \centering
  \centerline{\includegraphics[width=\linewidth]{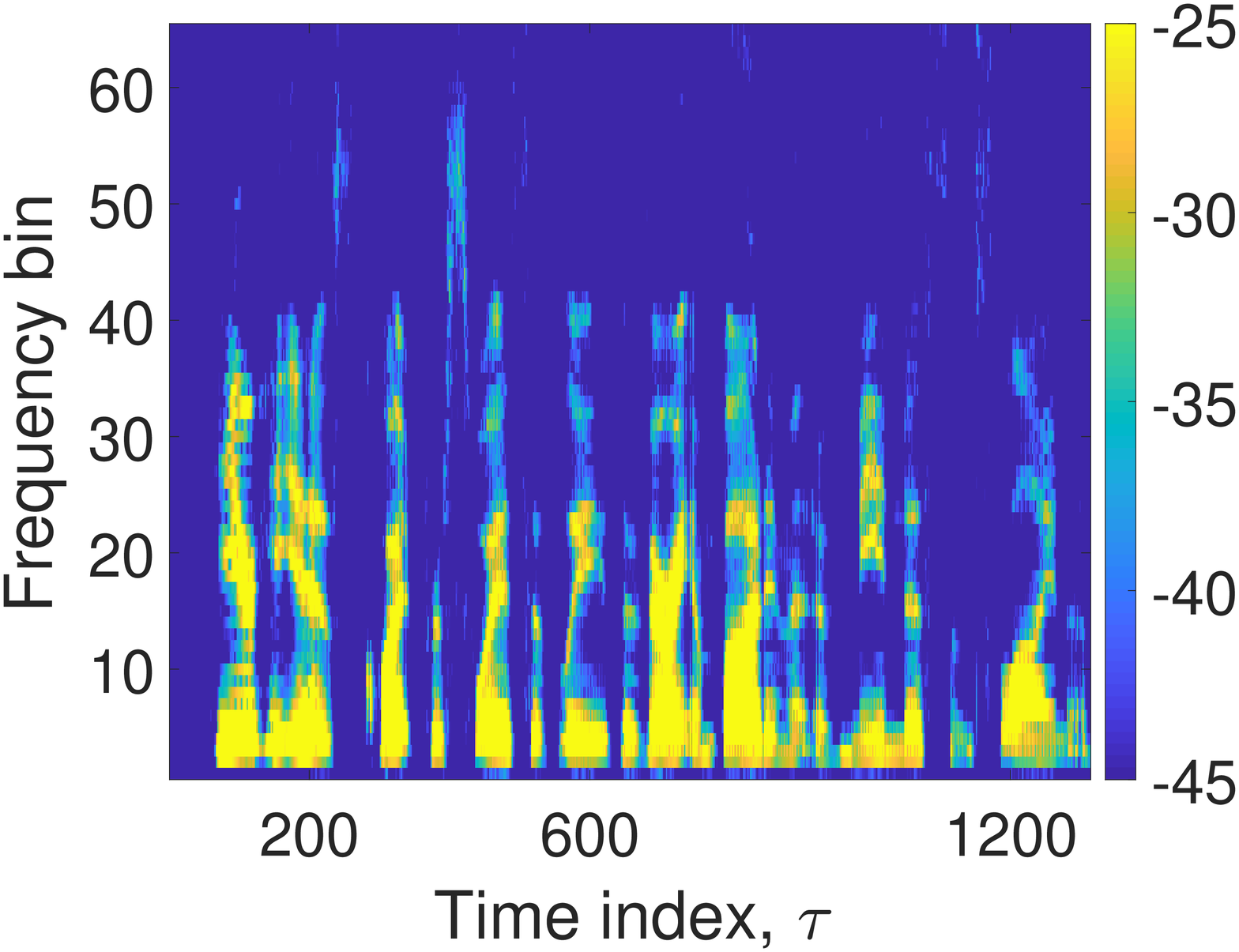}}
  \centerline{(d) Oracle (S-$04$)}\medskip
\end{minipage}
\begin{minipage}[b]{0.325\linewidth}
  \centering
  \centerline{\includegraphics[width=\linewidth]{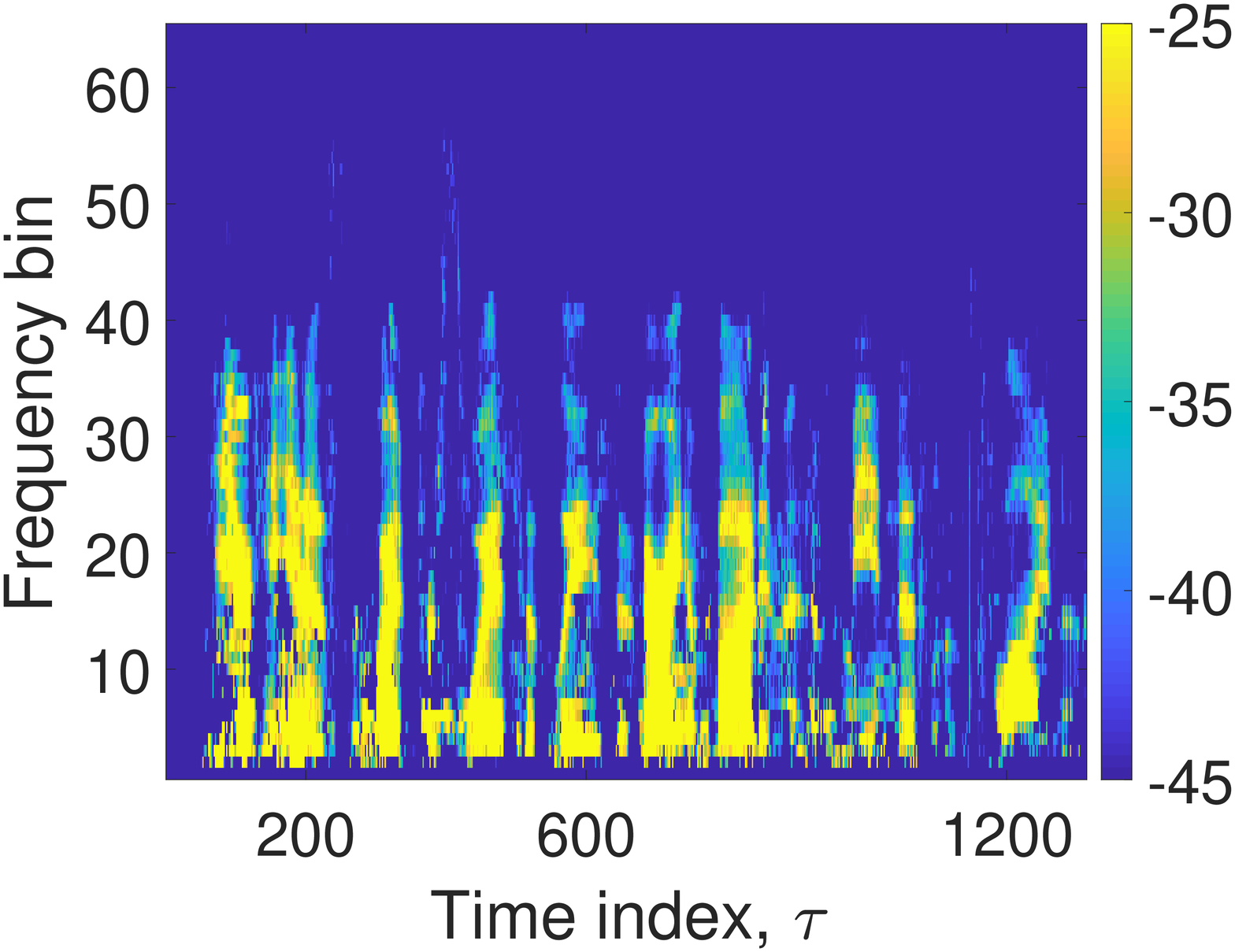}}
  \centerline{(e) Estimated (S-$04$)}\medskip
\end{minipage}
\caption{The log-spectrograms of the estimated PSDs for a $4$-source setup in Room A. The received signal PSD at microphone $1$ and the true PSDs are included for reference.}
\label{fig:estimate-psd-4-src}
\end{figure}
\subsection{Evaluation of PSD estimation accuracy}
\label{sec:psd-estimation-accuracy}
Fig. \ref{fig:mse} shows the normalized PSD estimation error in all $7$ scenarios where we observe improved PSD estimation for each individual source. In case of Room B (Fig. \ref{fig:mse}(d) \& (e)) where the source to microphone distance is close to the critical distance, the relative improvement offered by the proposed algorithm is significant which emphasizes on the importance of the use of cross-correlation coefficients in highly reverberant environments. We also observe notable improvements in Room C (Fig. \ref{fig:mse}(f) \& (g)) where we have a weaker direct path compared to the reverberant path (DRR $< 0$ dB). However, the performance in Room C was affected due to the non-uniform reflective surfaces (e.g. glass and brick walls) which resulted in relatively strong directional characteristics. This could be improved if the order $V$ was allowed to be increased which was not possible due to \eqref{eq:V-limit}. Finally, with the DOA estimation accuracy within $4$ degree (Table \ref{table:doa}), no major performance deviation was observed for true and estimated DOA consideration.\par
Fig. \ref{fig:estimate-psd-4-src} shows the original and the estimated PSDs in Room A for S-$03$ and S-$04$ along with the mixed signal PSD for $4$-speaker case. From Fig. \ref{fig:estimate-psd-4-src} we observe that S-$04$ estimation exhibits a very good resemblance to the original signal whereas S-$03$ are affected by few spectral distortions. This is due to the relative difference in signal strength in terms of signal to noise ratio (SNR) and signal to interference ratio (SIR) as S-$03$ possessed the lowest values of SNR and SIR among all the sources. We also notice the presence of very low frequency background noise in the estimated PSD of S-$03$. This can be a result of the spatial position of S-$03$ in addition to the aforementioned SNR and SIR issues. This problem can be resolved by pre-filtering the input signal with a high pass filter (HPF) to remove the signal below $200$ Hz. Furthermore, few random spectral distortions are observed in some frequencies which are mainly contributed by the practical limitations such as source and microphone positioning error, Bessel-zero correction, the deviation of the speaker and microphone characteristics from the ideal scenario, the finite correlation between the sources and the reverberation components due to limited STFT window length and imperfect room characteristics, measurement inaccuracies etc.
\begin{figure}[!ht]
\centering
\begin{minipage}[b]{0.43\linewidth}
  \centering
  \centerline{\includegraphics[width=\linewidth]{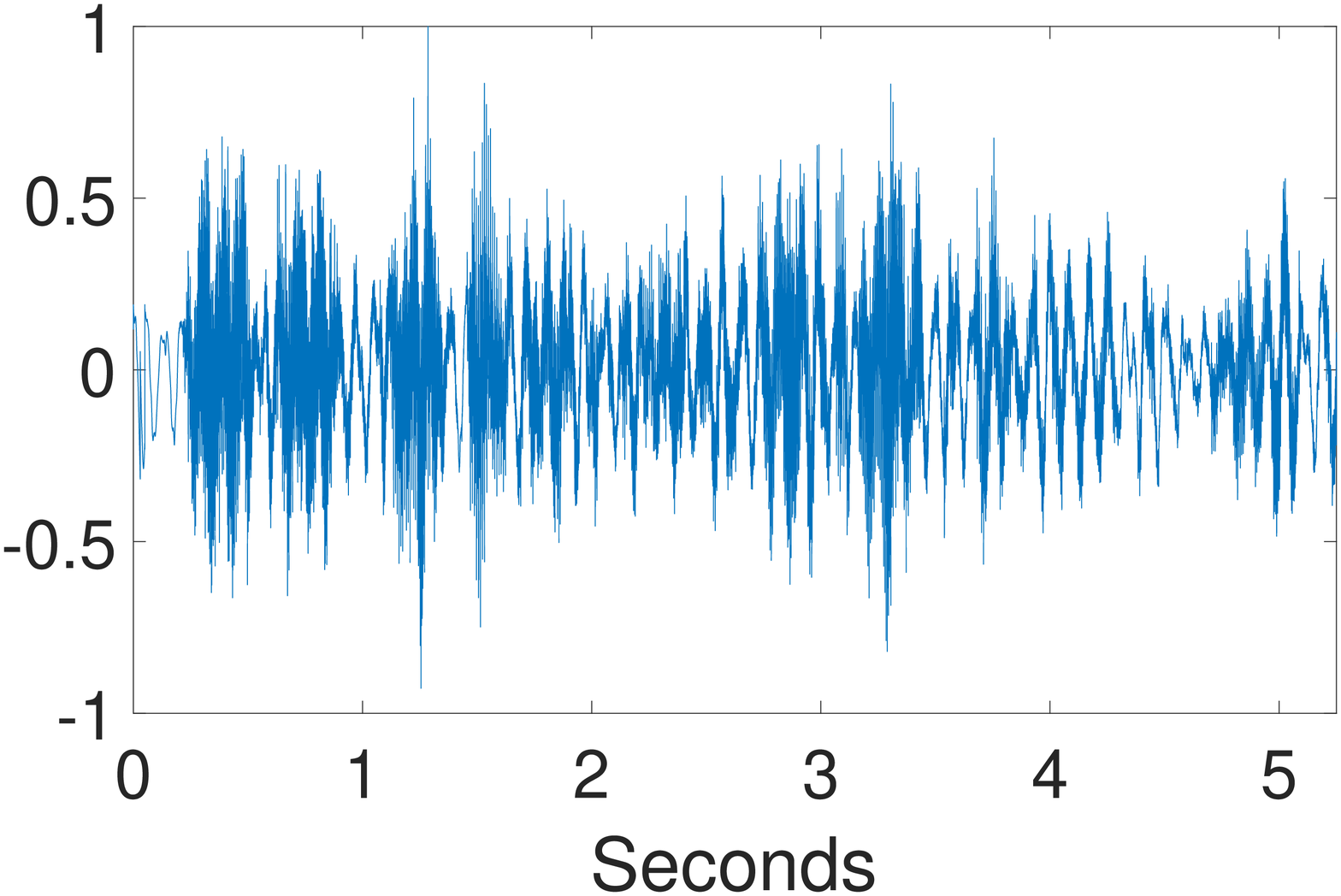}}
  \centerline{(a) Mixed signal}\medskip
\end{minipage}
\begin{minipage}[b]{0.43\linewidth}
  \centering
  \centerline{\includegraphics[width=\linewidth]{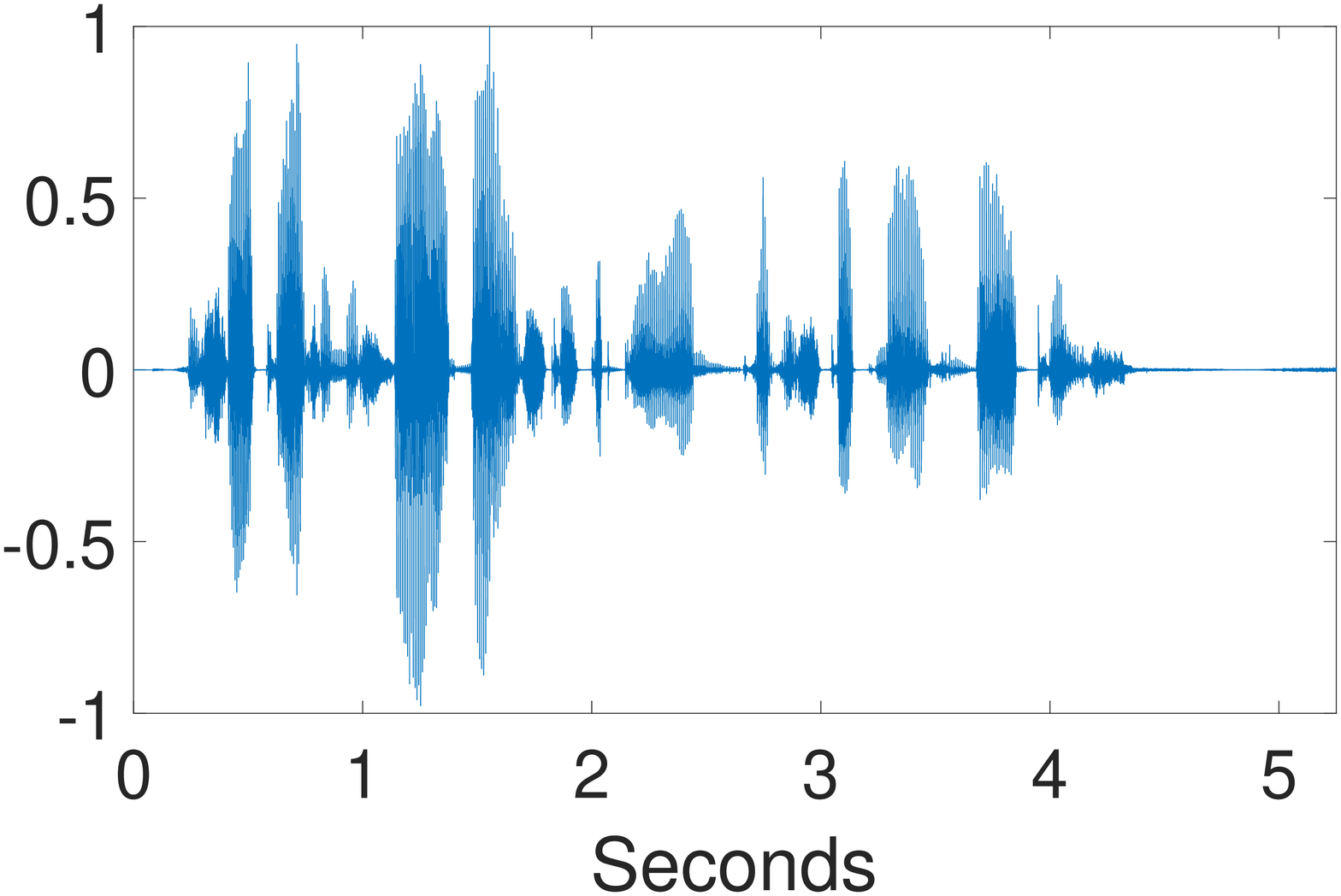}}
  \centerline{(b) True signal}\medskip
\end{minipage}\\
\begin{minipage}[b]{0.43\linewidth}
  \centering
  \centerline{\includegraphics[width=\linewidth]{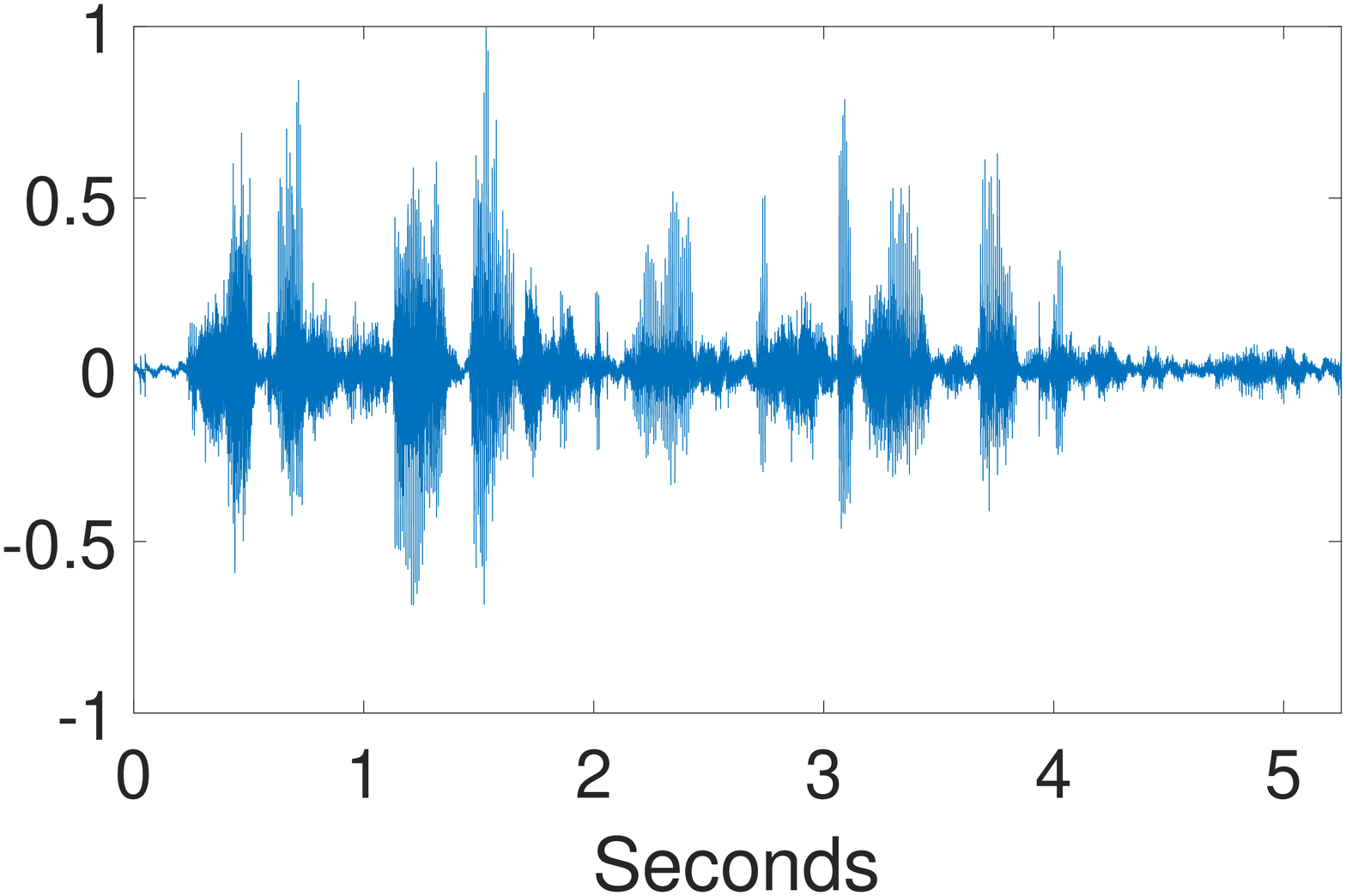}}
  \centerline{(c) Beamformer output}\medskip
\end{minipage}
\begin{minipage}[b]{0.43\linewidth}
  \centering
  \centerline{\includegraphics[width=\linewidth]{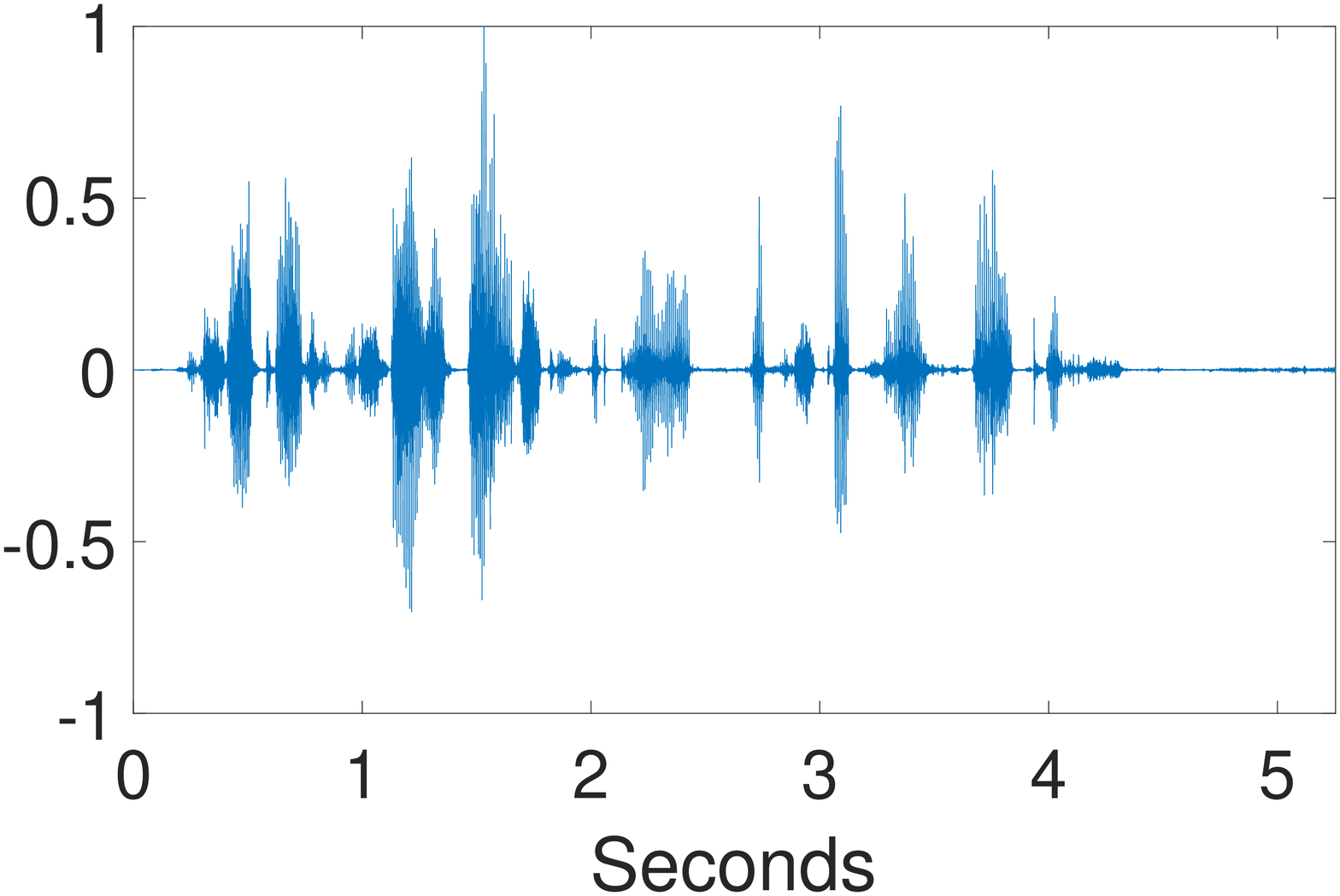}}
  \centerline{(d) Proposed method}\medskip
\end{minipage}
\caption{The estimated waveforms of speaker $1$ in Room A. The waveform at the beamformer output along with the original and the mixed signal waveforms are shown for reference.}
\label{fig:estimate-wave-4-src}
\end{figure}
\subsection{Evaluation of source separation as an application}
\label{sec:source-separation-application}
In this section, we evaluate the performance of the source separation algorithm as described in Section \ref{lab:practical-application} to demonstrate the application of the proposed PSD estimation algorithm. The Wiener filter in Fig. \ref{fig:separation} uses the estimated PSDs from the proposed algorithm. Therefore, the results in this section can be considered as an extension of the evaluation of PSD estimation accuracy. Fig. \ref{fig:estimate-wave-4-src} plots the time domain waveforms of speaker $1$ in Room A at different nodes. It is obvious from the plot that while the beamformer only partially restored the original signal waveform, the Wiener post-filter significantly improved the estimation accuracy. This indicates that the estimated PSDs were accurate enough to drive the Wiener filter for a better signal estimation.\par
\begin{figure}
\begin{minipage}[b]{\linewidth}
  \centering
  \centerline{\includegraphics[width=\linewidth]{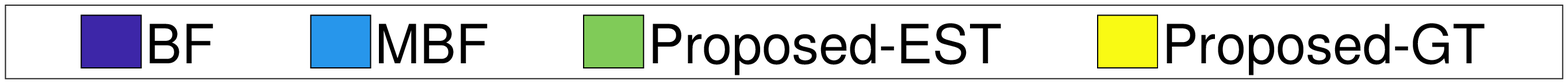}}
\end{minipage}
\begin{minipage}[b]{0.325\linewidth}
  \centering
  \centerline{\includegraphics[width=\linewidth]{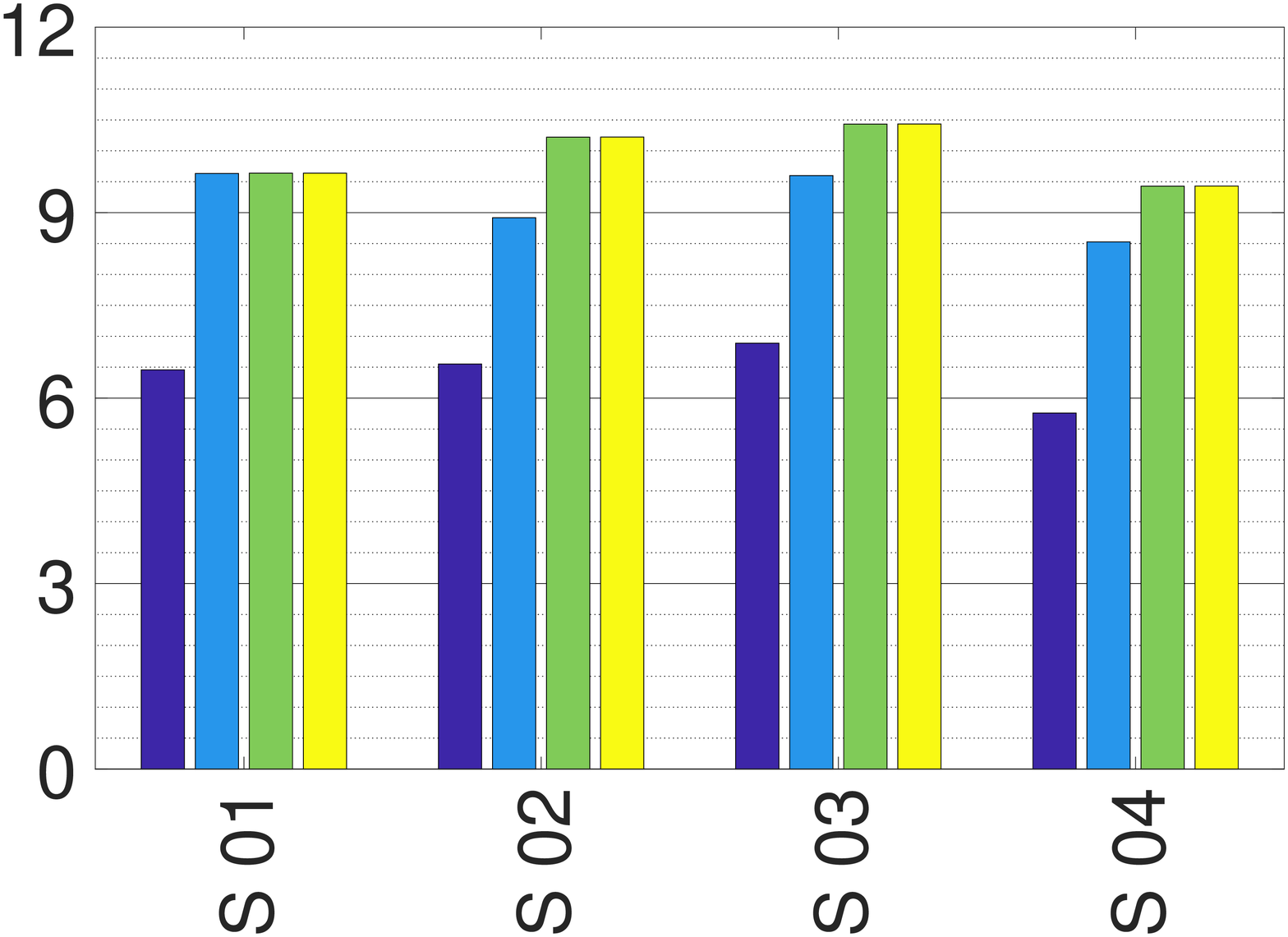}}
  \centerline{{\fontsize{9}{10}\selectfont(a) Room A ($L = 4$)}}\medskip
\end{minipage}
\begin{minipage}[b]{0.325\linewidth}
  \centering
  \centerline{\includegraphics[width=\linewidth]{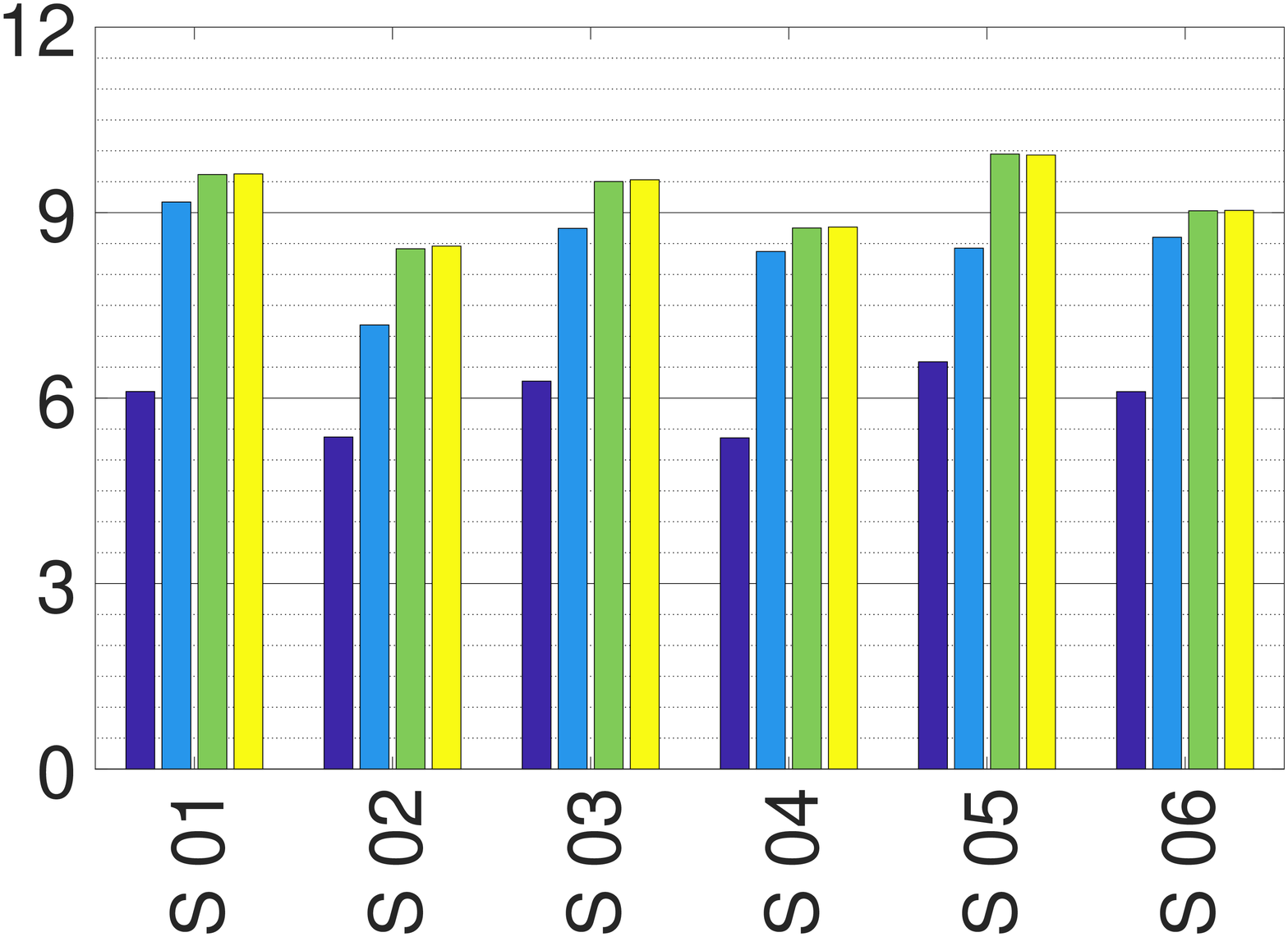}}
  \centerline{{\fontsize{9}{10}\selectfont(b) Room A ($L = 6$)}}\medskip
\end{minipage}
\begin{minipage}[b]{0.325\linewidth}
  \centering
  \centerline{\includegraphics[width=\linewidth]{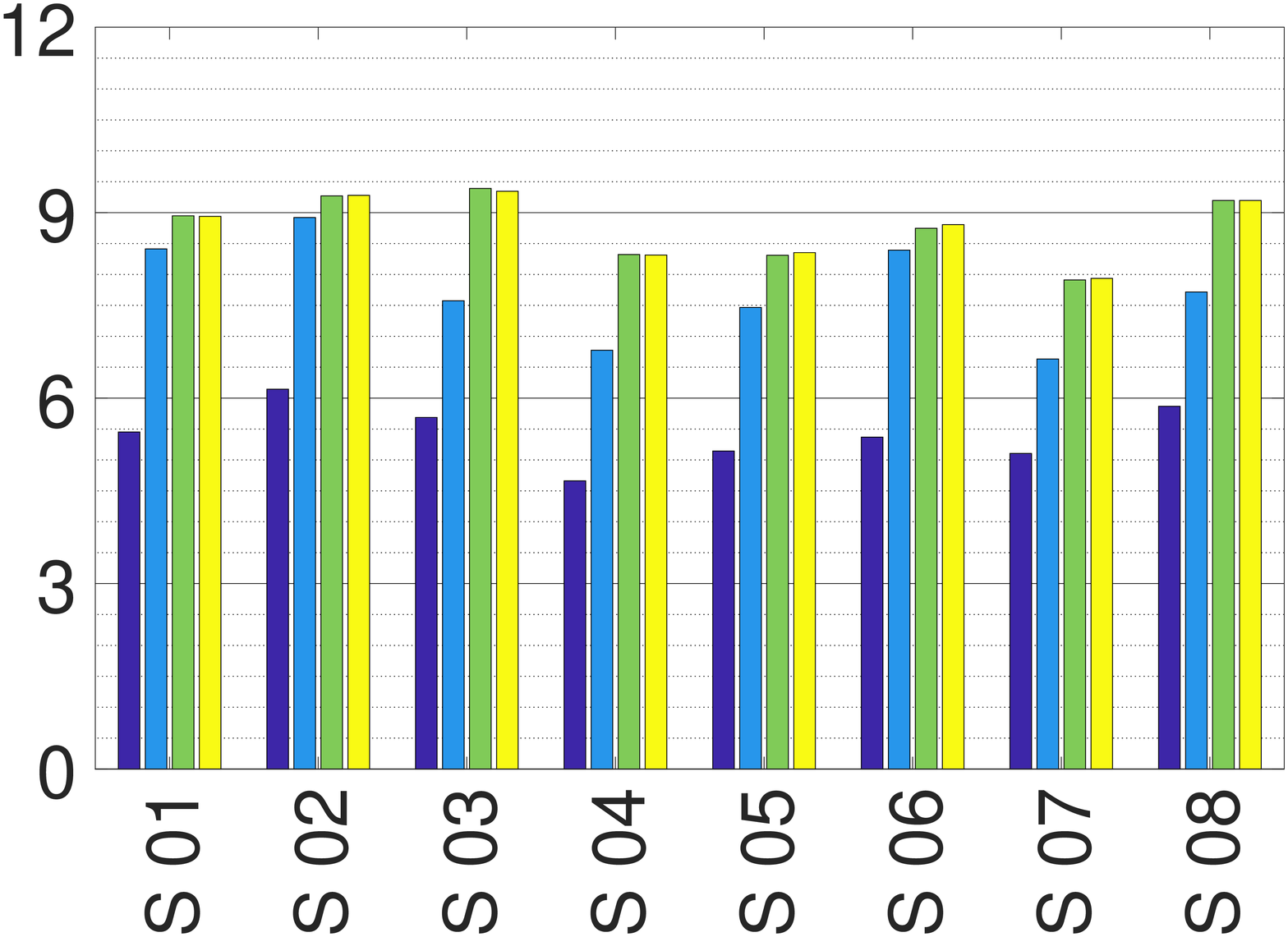}}
  \centerline{{\fontsize{9}{10}\selectfont(c) Room A ($L = 8$)}}\medskip
\end{minipage}\\
\begin{minipage}[b]{0.325\linewidth}
  \centering
  \centerline{\includegraphics[width=\linewidth]{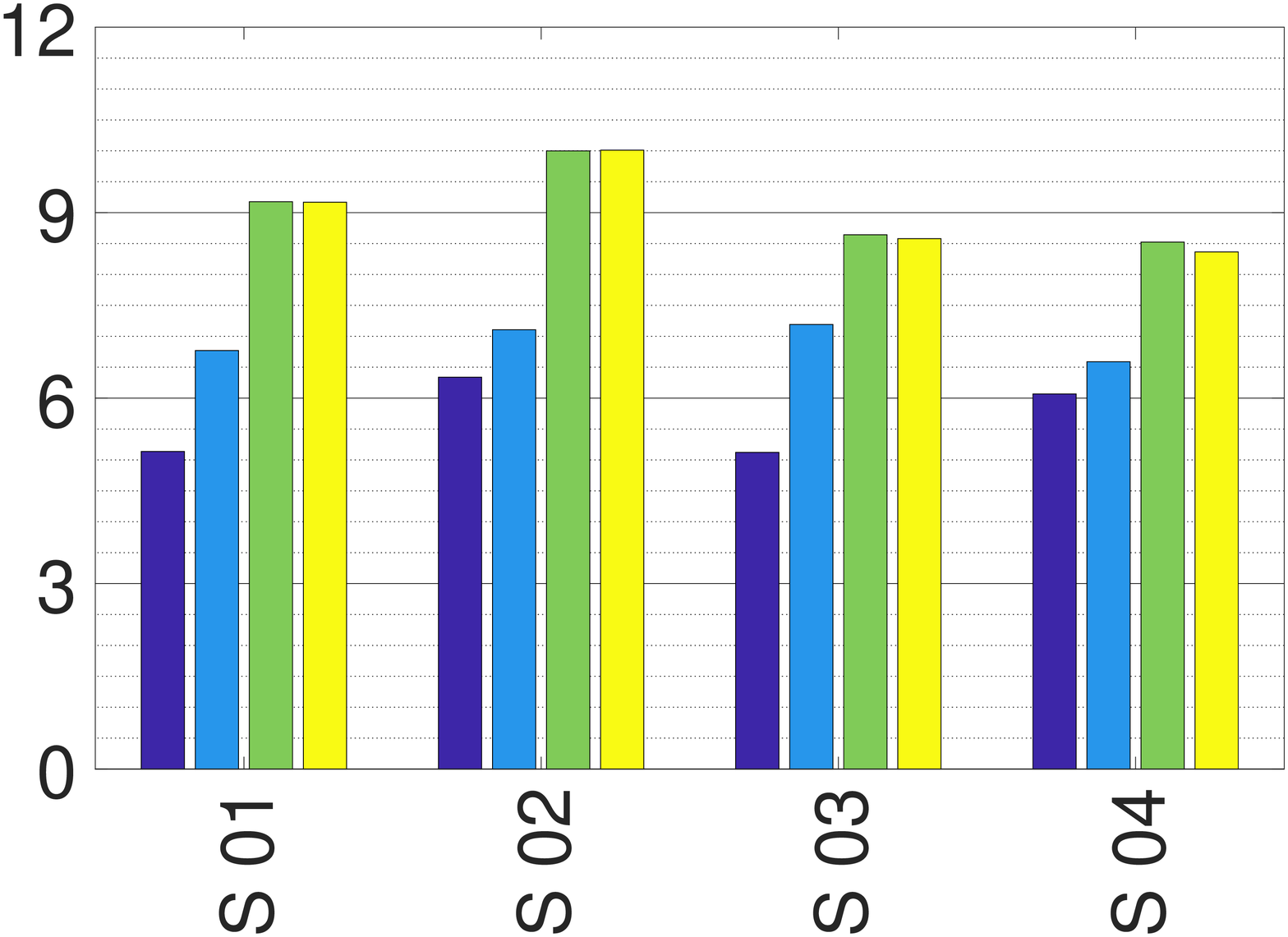}}
  \centerline{{\fontsize{9}{10}\selectfont(d) Room B ($L = 4$)}}\medskip
\end{minipage}
\begin{minipage}[b]{0.325\linewidth}
  \centering
  \centerline{\includegraphics[width=\linewidth]{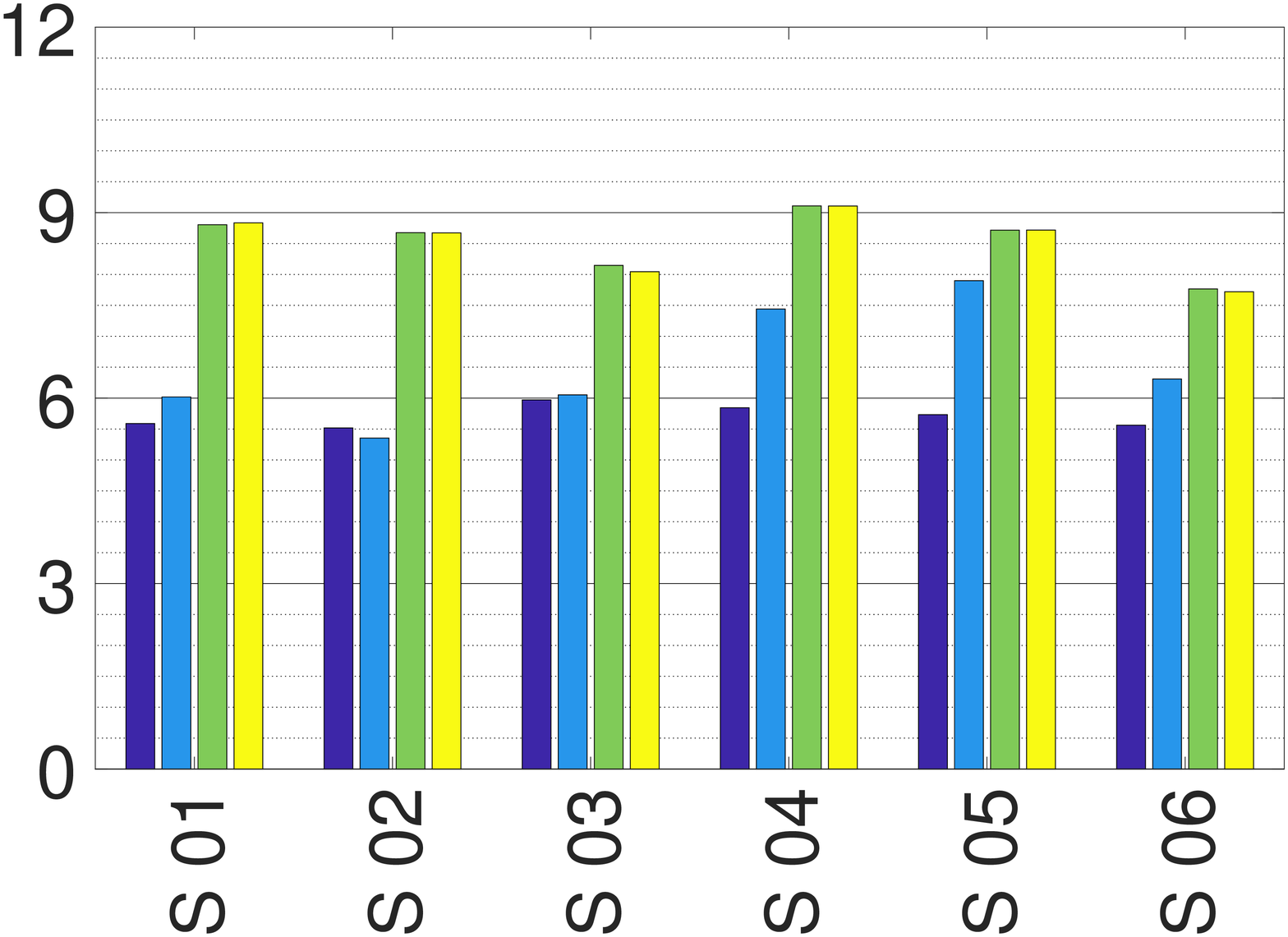}}
  \centerline{{\fontsize{9}{10}\selectfont(e) Room B ($L = 6$)}}\medskip
\end{minipage}
\hfill \\
\begin{minipage}[b]{0.325\linewidth}
  \centering
  \centerline{\includegraphics[width=\linewidth]{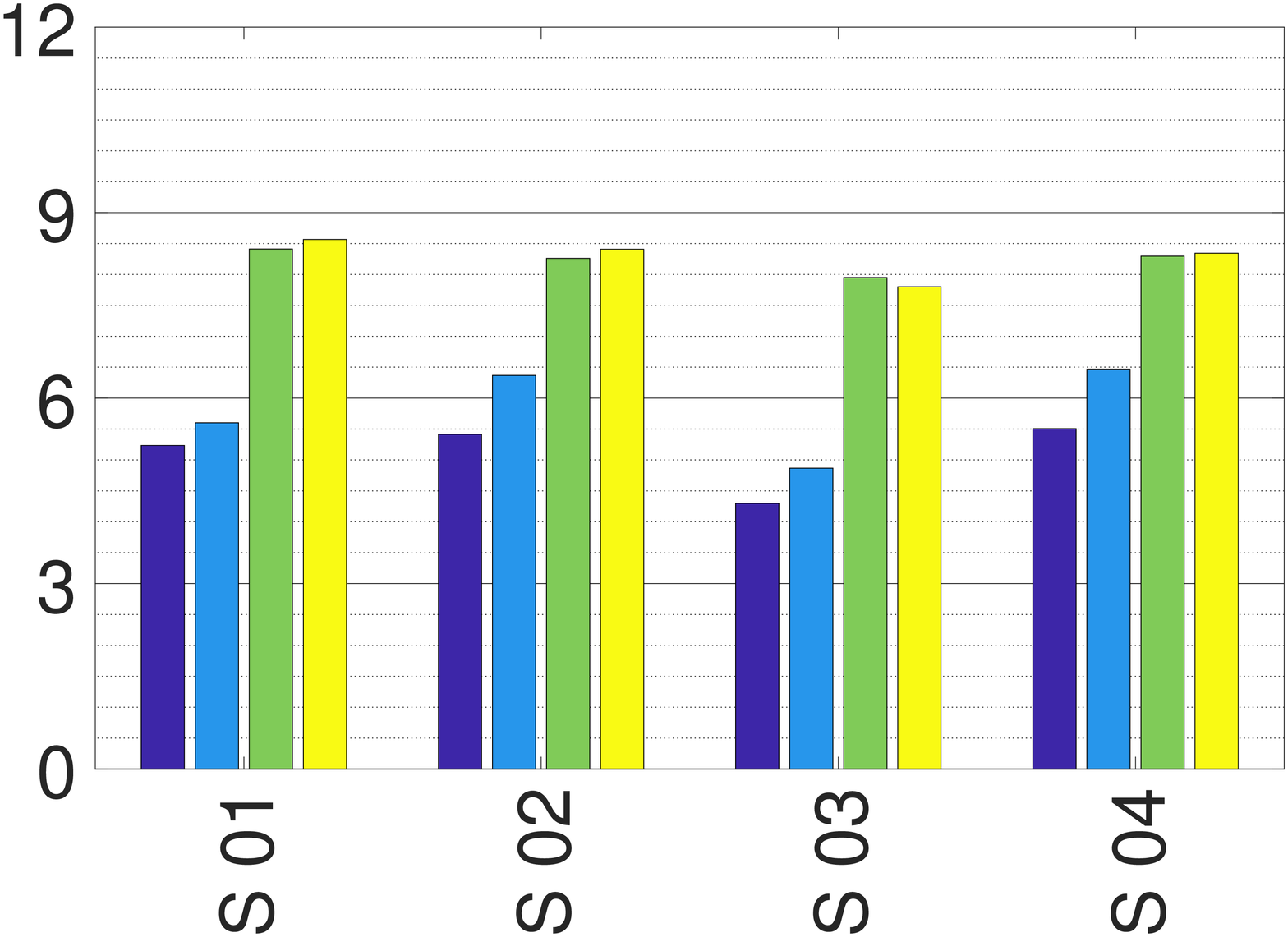}}
  \centerline{{\fontsize{9}{10}\selectfont(f) Room C ($L = 4$)}}\medskip
\end{minipage}
\begin{minipage}[b]{0.325\linewidth}
  \centering
  \centerline{\includegraphics[width=\linewidth]{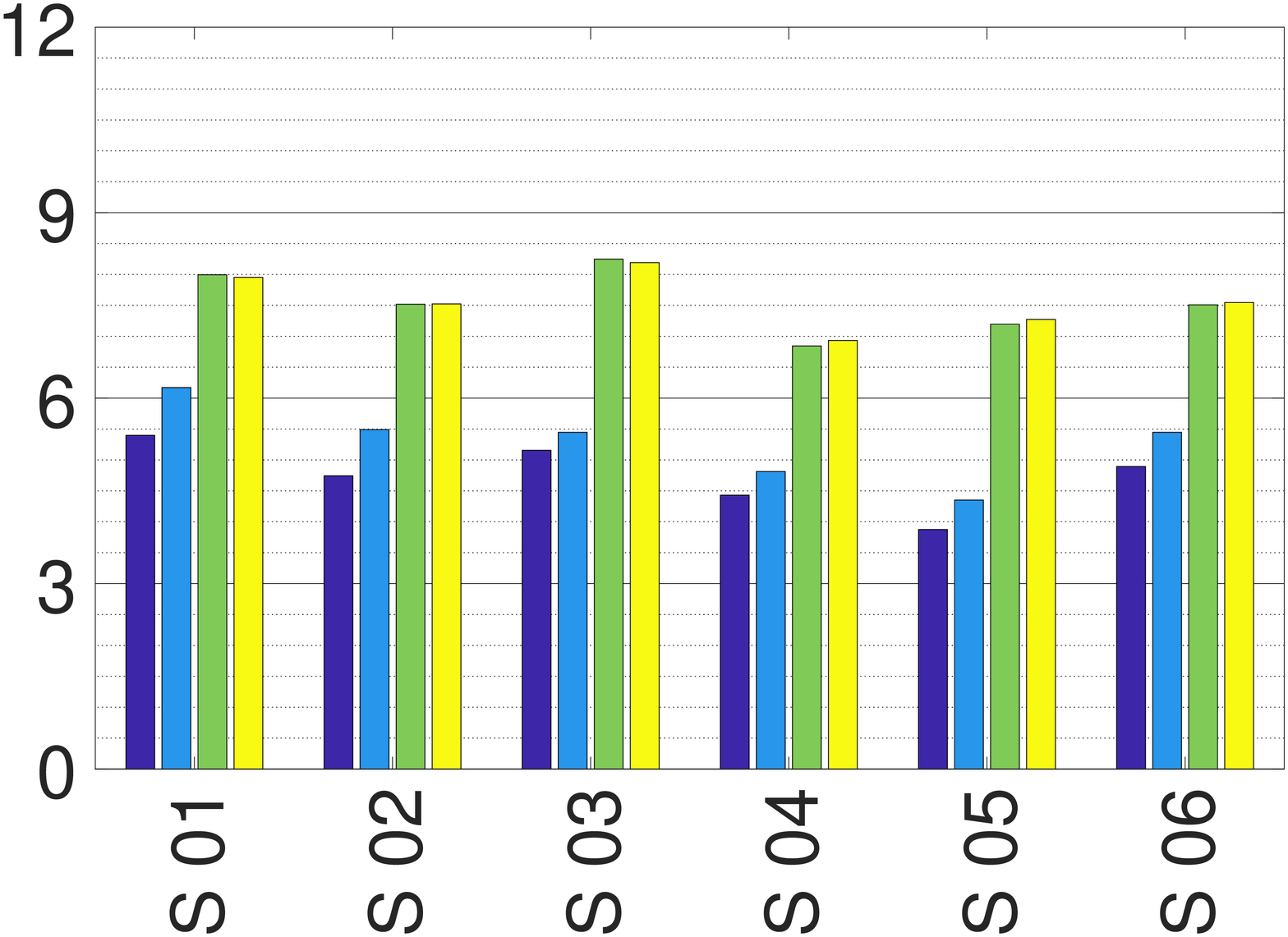}}
  \centerline{{\fontsize{9}{10}\selectfont(g) Room C ($L = 6$)}}\medskip
\end{minipage}
\hfill
\caption{FWSegSNR (dB) in $3$ distinct reverberant environments (Table \ref{table:experimen-cases}) for different number of sources.}
\label{fig:snr}
\end{figure}
\begin{figure}
\begin{minipage}[b]{\linewidth}
  \centering
  \centerline{\includegraphics[width=\linewidth]{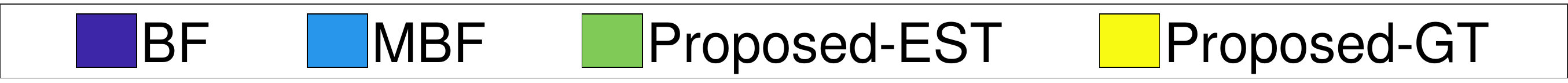}}
\end{minipage}
\begin{minipage}[b]{0.325\linewidth}
  \centering
  \centerline{\includegraphics[width=\linewidth]{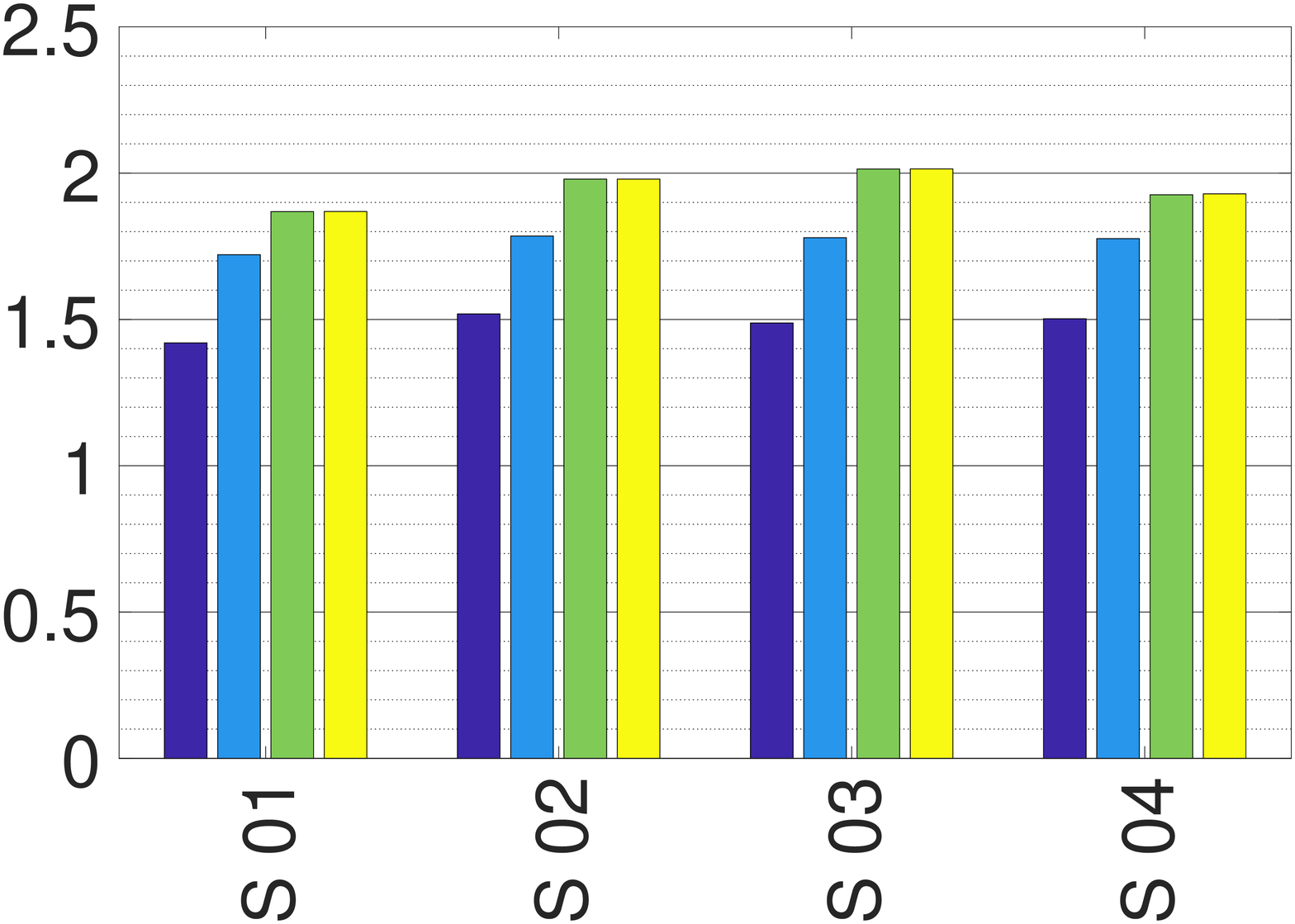}}
  \centerline{{\fontsize{9}{10}\selectfont(a) Room A ($L = 4$)}}\medskip
\end{minipage}
\begin{minipage}[b]{0.325\linewidth}
  \centering
  \centerline{\includegraphics[width=\linewidth]{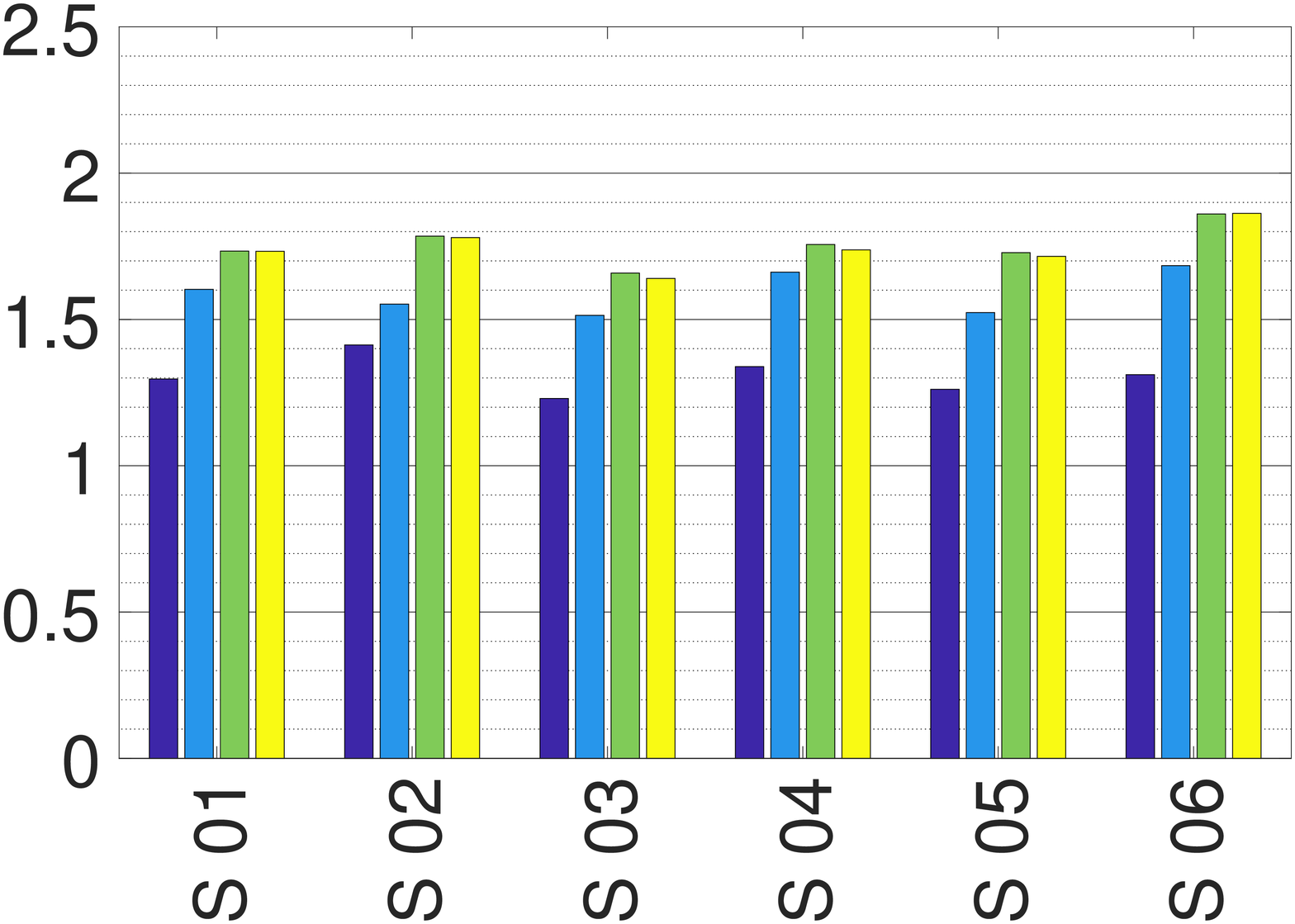}}
  \centerline{{\fontsize{9}{10}\selectfont(b) Room A ($L = 6$)}}\medskip
\end{minipage}
\begin{minipage}[b]{0.325\linewidth}
  \centering
  \centerline{\includegraphics[width=\linewidth]{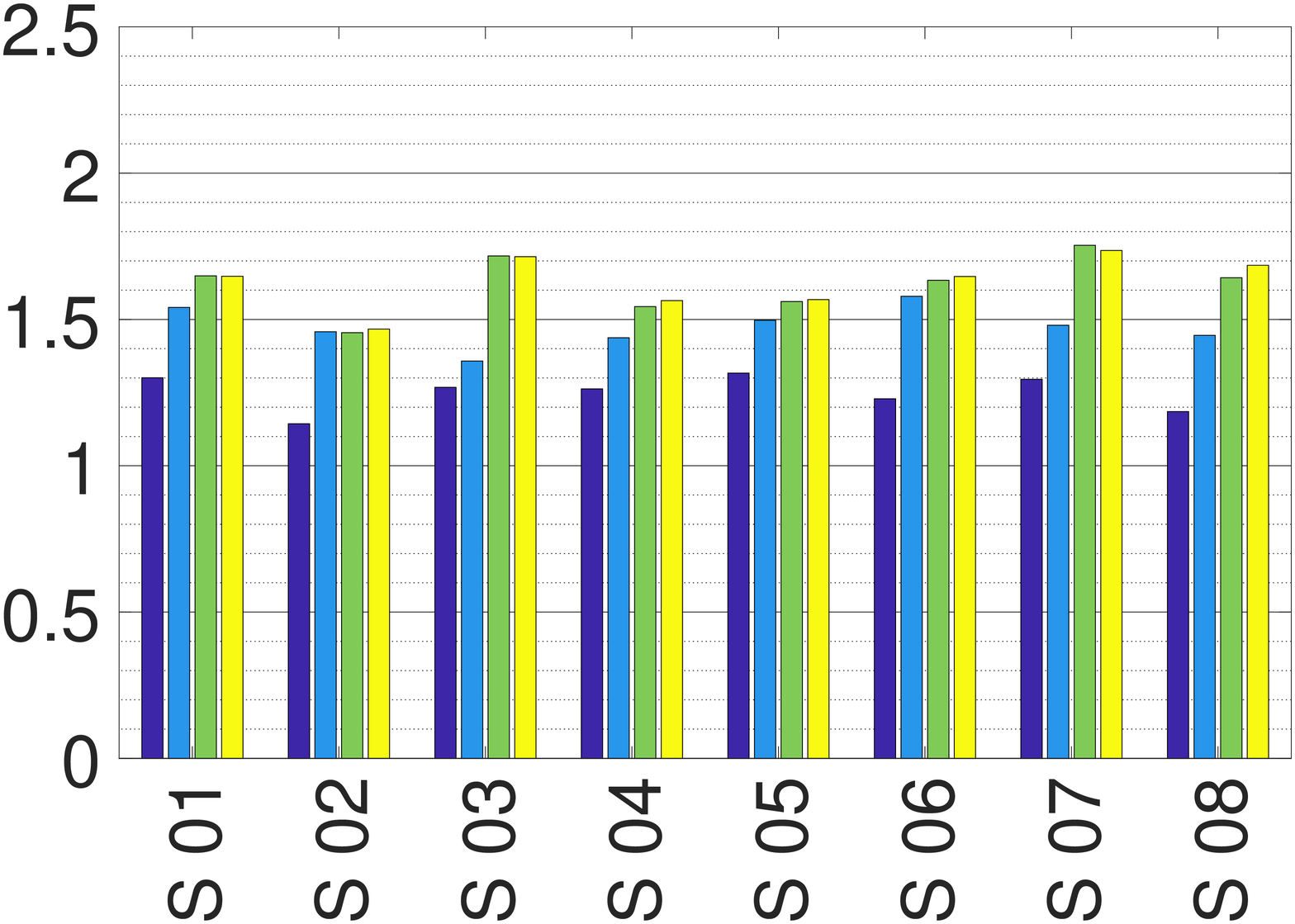}}
  \centerline{{\fontsize{9}{10}\selectfont(c) Room A ($L = 8$)}}\medskip
\end{minipage}\\
\begin{minipage}[b]{0.325\linewidth}
  \centering
  \centerline{\includegraphics[width=\linewidth]{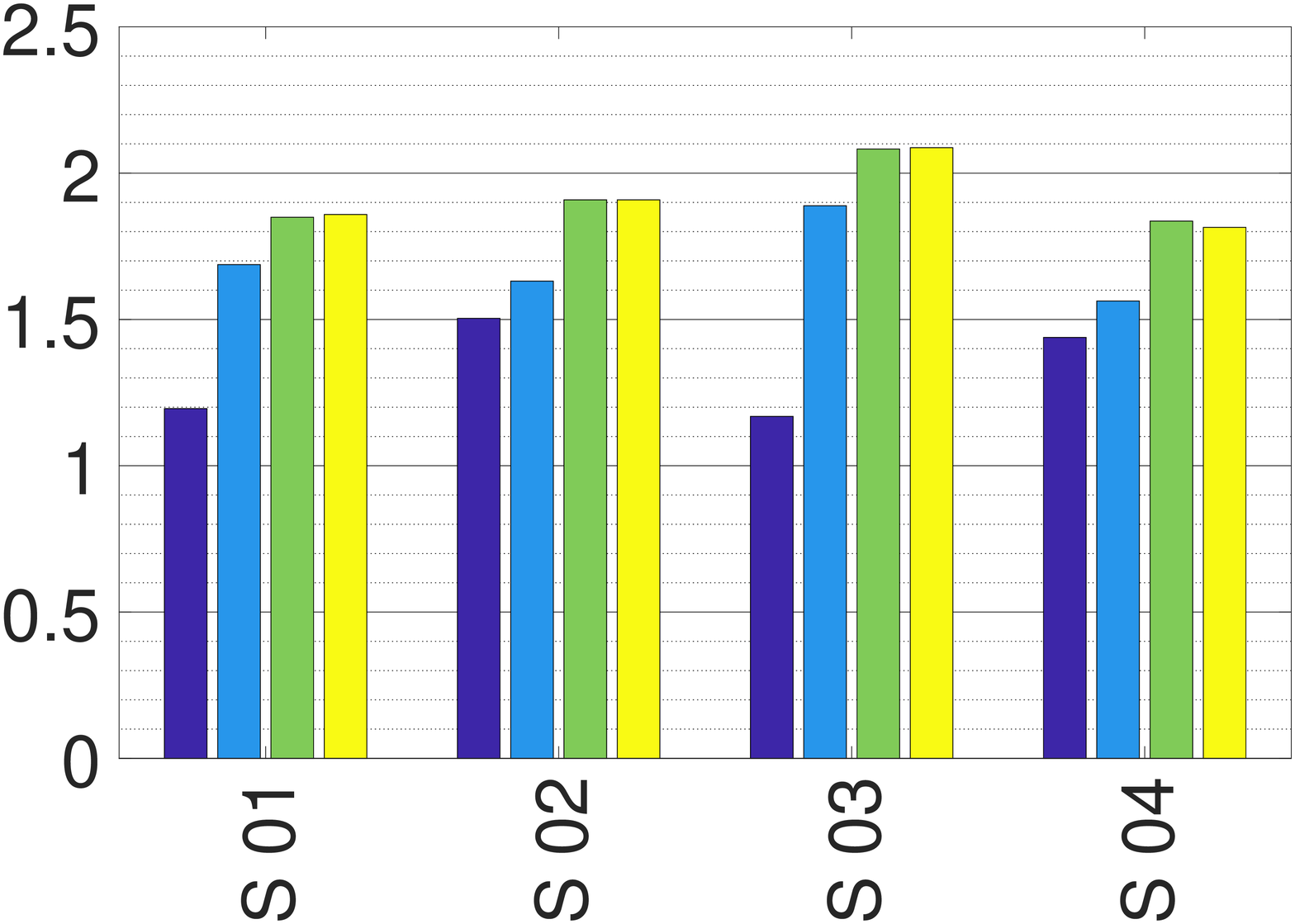}}
  \centerline{{\fontsize{9}{10}\selectfont(d) Room B ($L = 4$)}}\medskip
\end{minipage}
\begin{minipage}[b]{0.325\linewidth}
  \centering
  \centerline{\includegraphics[width=\linewidth]{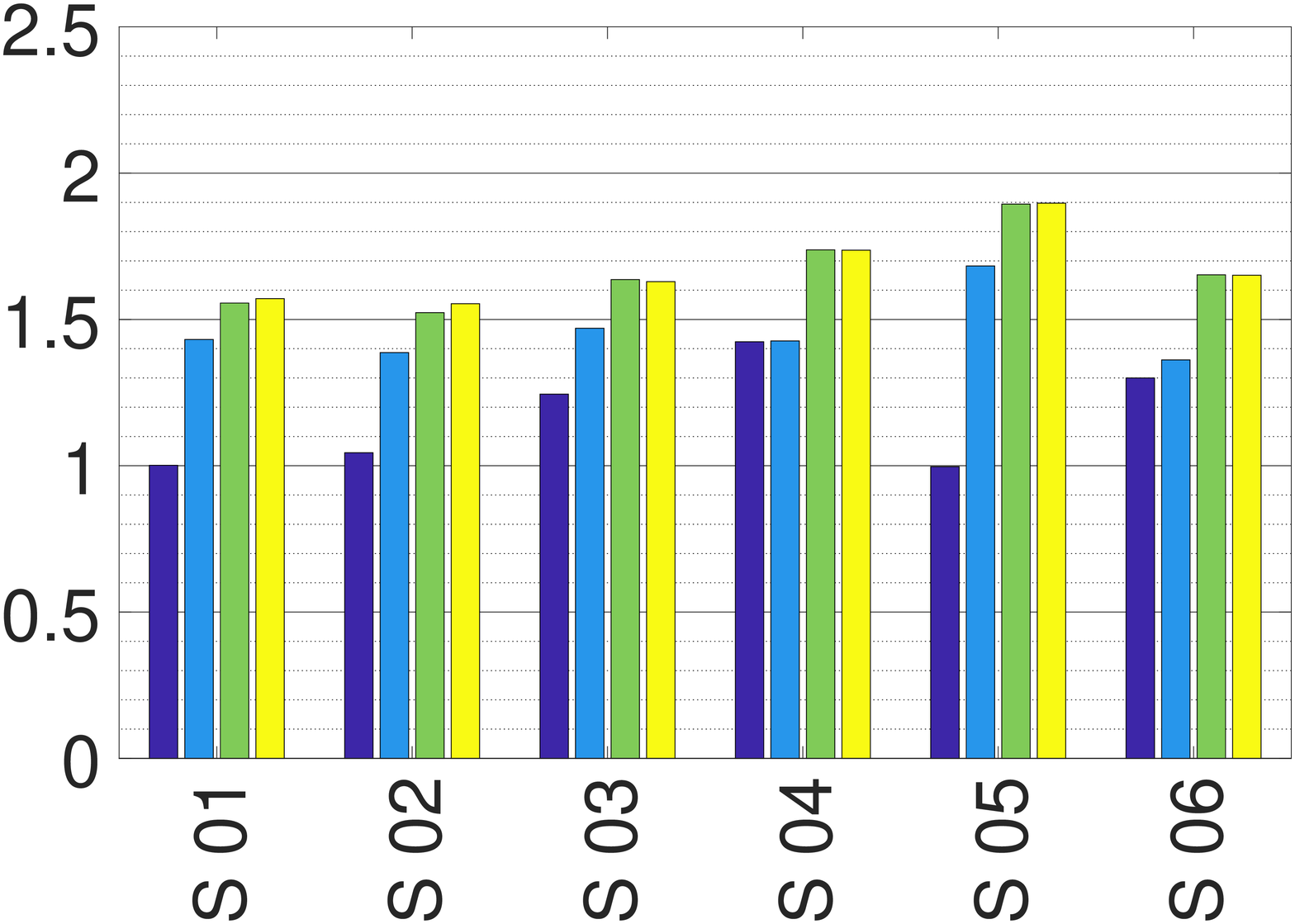}}
  \centerline{{\fontsize{9}{10}\selectfont(e) Room B ($L = 6$)}}\medskip
\end{minipage}
\hfill \\
\begin{minipage}[b]{0.325\linewidth}
  \centering
  \centerline{\includegraphics[width=\linewidth]{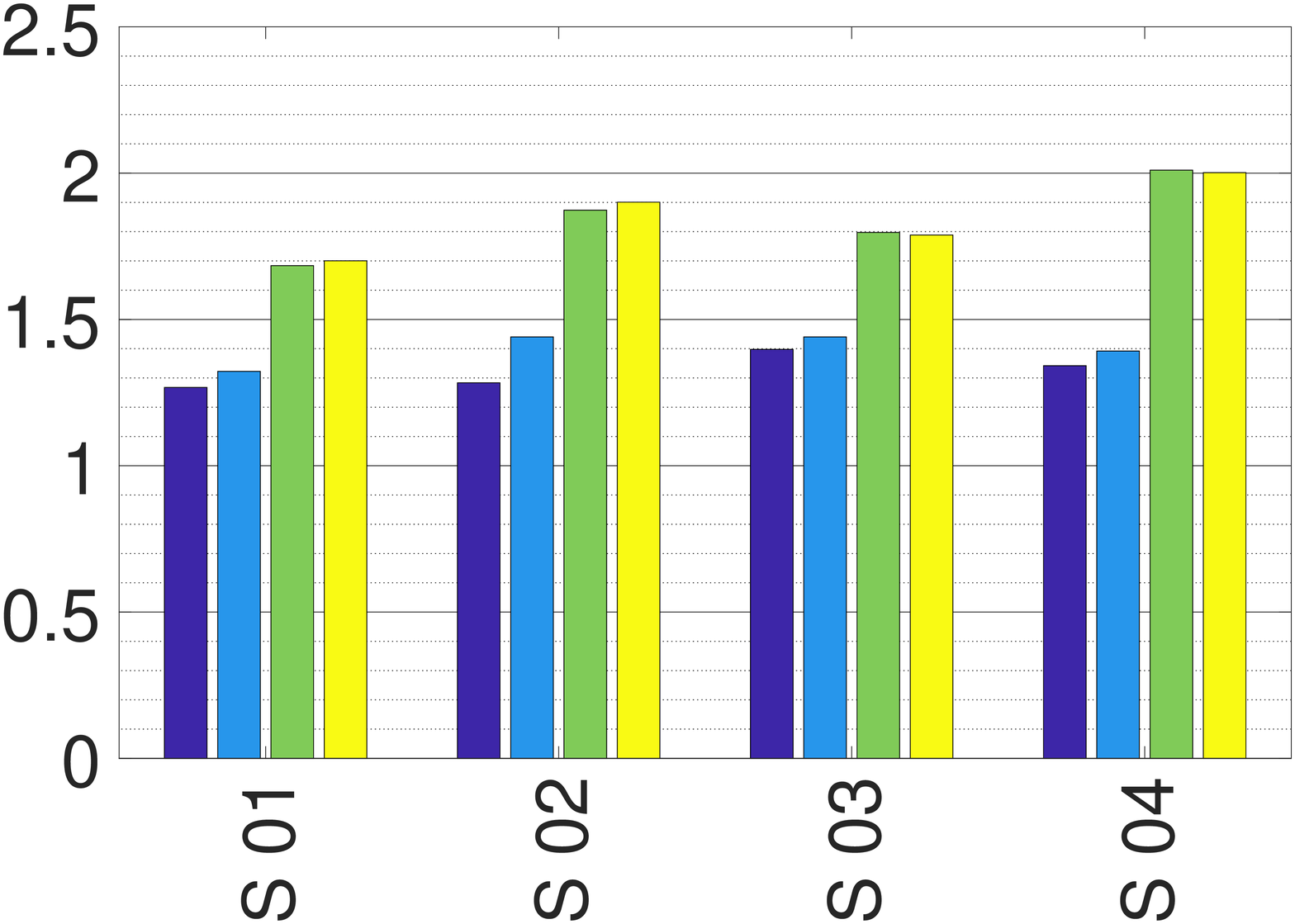}}
  \centerline{{\fontsize{9}{10}\selectfont(f) Room C ($L = 4$)}}\medskip
\end{minipage}
\begin{minipage}[b]{0.325\linewidth}
  \centering
  \centerline{\includegraphics[width=\linewidth]{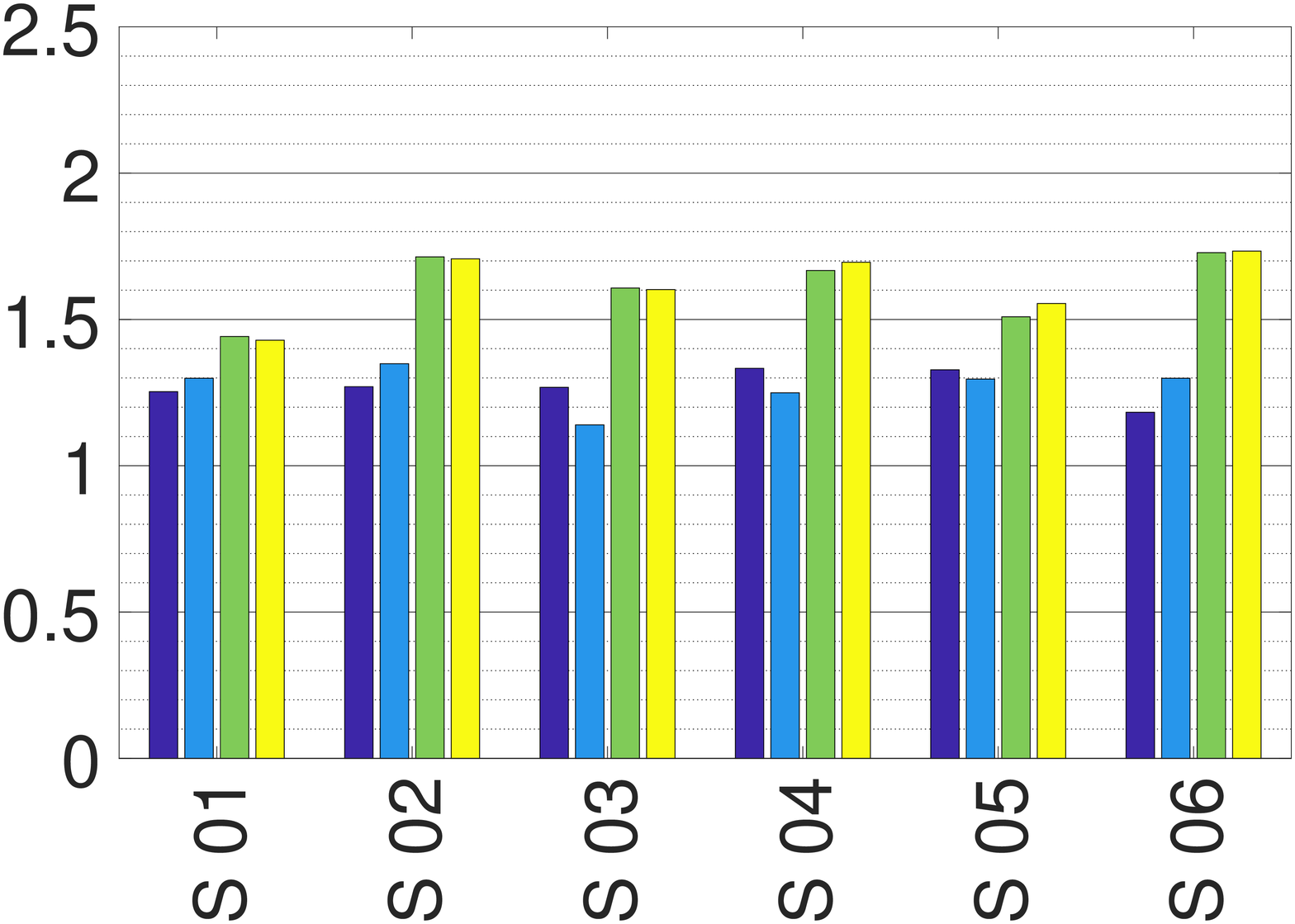}}
  \centerline{{\fontsize{9}{10}\selectfont(g) Room C ($L = 6$)}}\medskip
\end{minipage}
\hfill
\caption{PESQ in $3$ distinct reverberant environments (Table \ref{table:experimen-cases}) for different number of sources.}
\label{fig:pesq}
\end{figure}

We also plot FWSegSNR and PESQ for all $7$ scenarios in Fig. \ref{fig:snr} and \ref{fig:pesq}, respectively. The proposed method consistently exhibits a better performance in terms of FWSegSNR and PESQ compared to the competing methods. The results of the performance metrics agrees the observation made with the PSD estimation error in Section \ref{sec:psd-estimation-accuracy}. This suggests that the error level in Fig. \ref{fig:mse} is within an acceptable range for source separation application.\par
It is worth noting that, as our primarily goal was to estimate PSDs of the signal components, we used the source separation only as an example to demonstrate an application, and hence, we did not analyze beamformer design and musical noise reduction techniques within the scope of the current work.
\subsection{Number of microphones}
\begin{figure}[!t]
\centering
\begin{minipage}[b]{0.3\linewidth}
  \centering
  \centerline{\includegraphics[width=\linewidth]{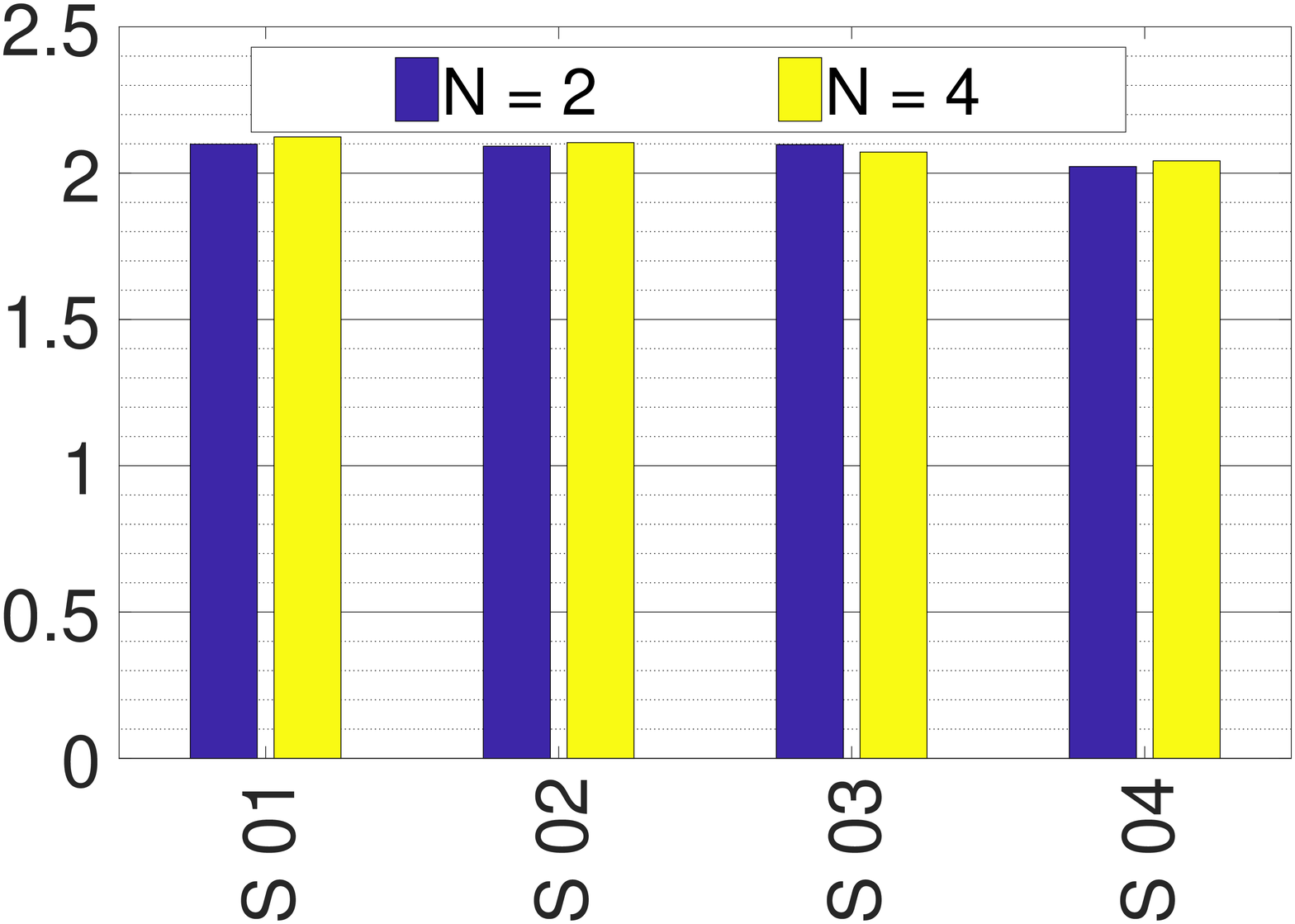}}
  \centerline{(a) 4-speaker case}\medskip
\end{minipage}
\begin{minipage}[b]{0.3\linewidth}
  \centering
  \centerline{\includegraphics[width=\linewidth]{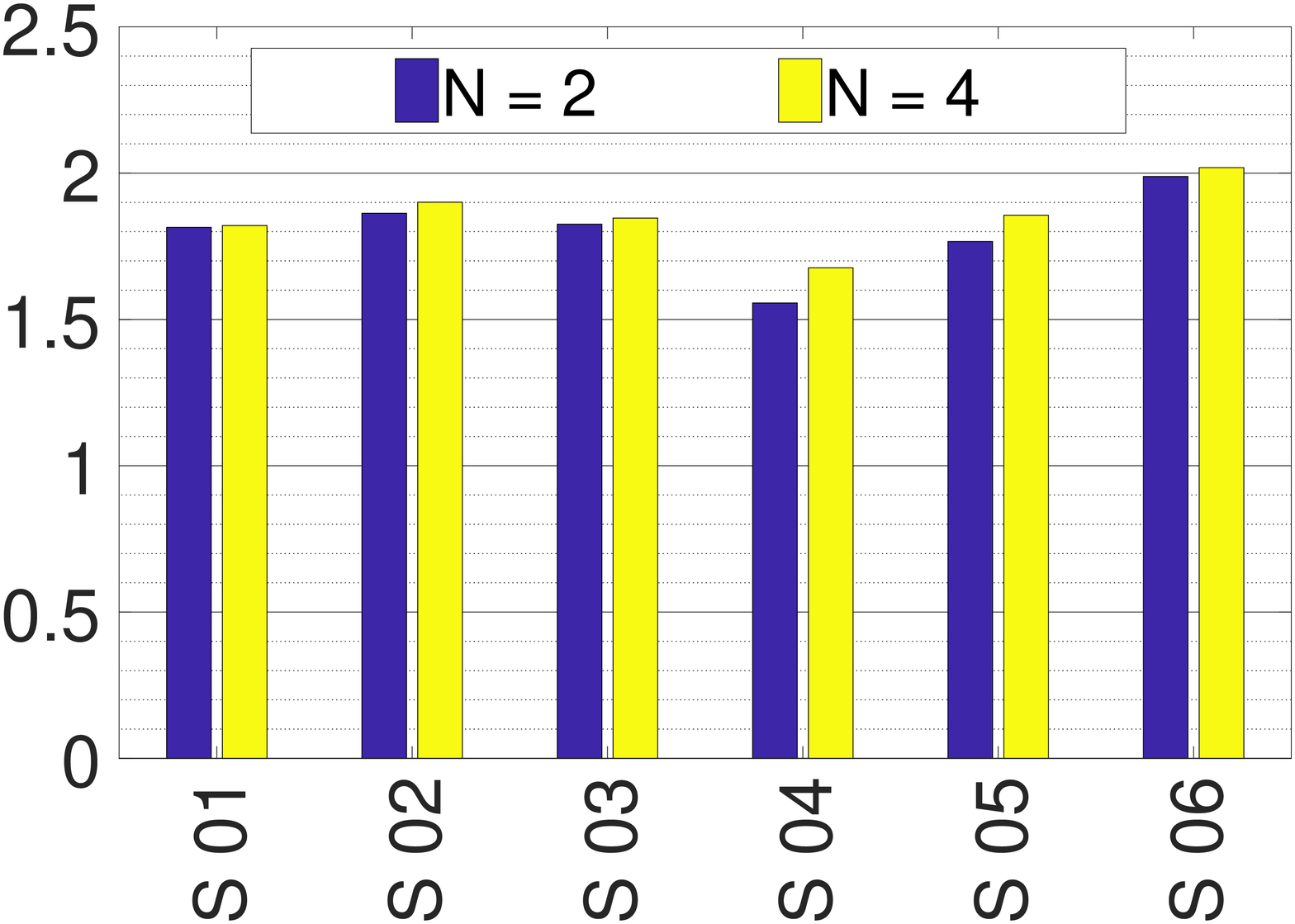}}
  \centerline{(b) 6-speaker case}\medskip
\end{minipage}
\begin{minipage}[b]{0.3\linewidth}
  \centering
  \centerline{\includegraphics[width=\linewidth]{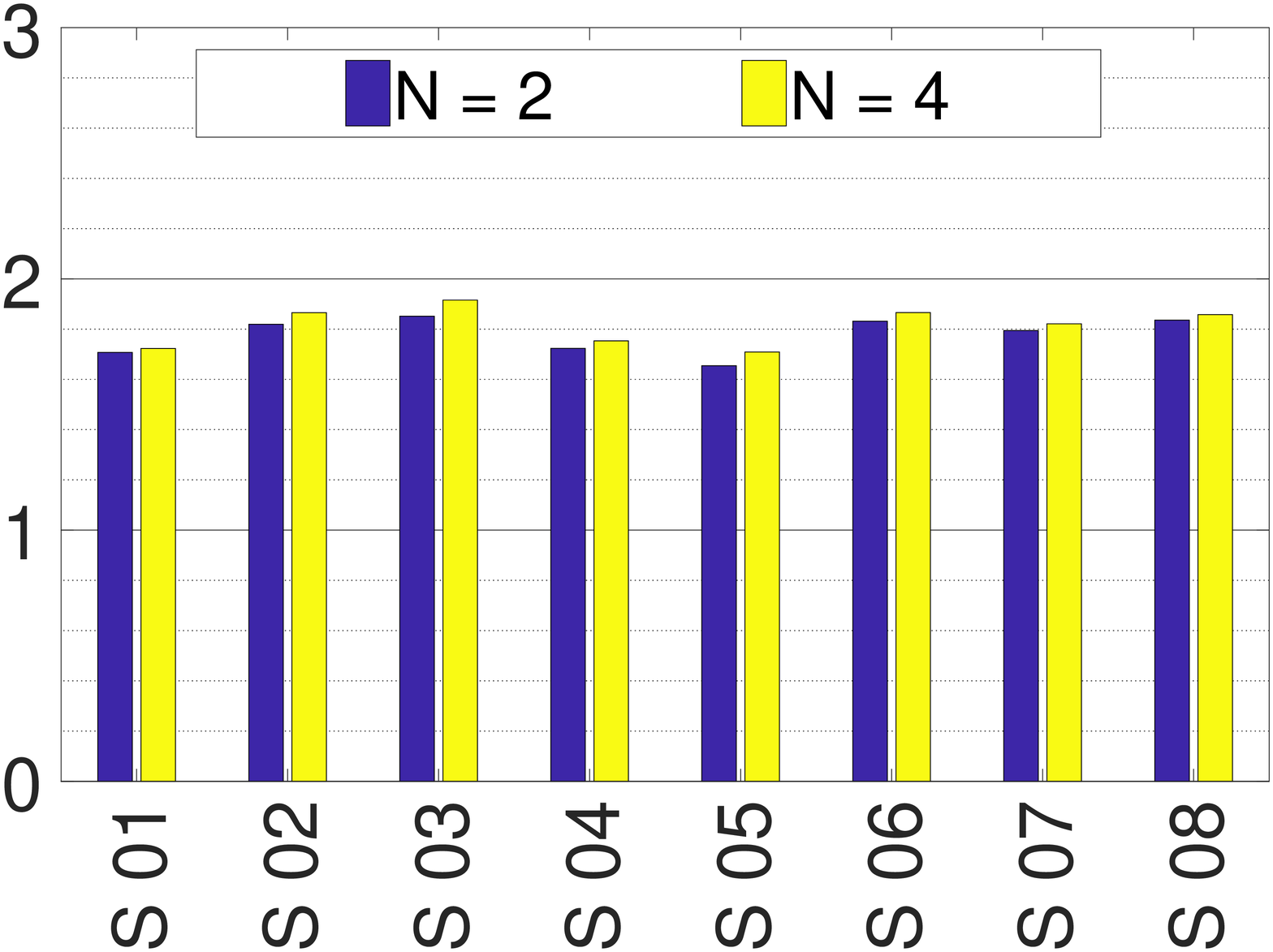}}
  \centerline{(b) 8-speaker case}\medskip
\end{minipage}
\caption{PESQ in Room A for estimated source signals with $N=2$ and $4$.}
\label{fig:pesq-N}
\end{figure}

One of the major challenges in most spherical harmonics-based solutions is the required number of microphones to calculate the required sound field coefficients. The required number of microphones is directly related to the maximum sound field order $N$. Theoretically, to calculate the spherical harmonics coefficients of an $N^{th}$-order sound field, we generally require at least $(N+1)^2$ microphones. So far, we have used $N=4$ in our experiments; however, reducing it to $N=2$ does not have any significant adverse impact on the performance, as shown in Fig. \ref{fig:pesq-N}. Hence, for the demonstrated examples, it is possible to utilize a lower-order microphone array \cite{samarasinghe2017planar} without having a major performance degradation.\par
\begin{figure}[!t]
\centering
\begin{minipage}[b]{0.47\linewidth}
  \centering
  \centerline{\includegraphics[width=\linewidth]{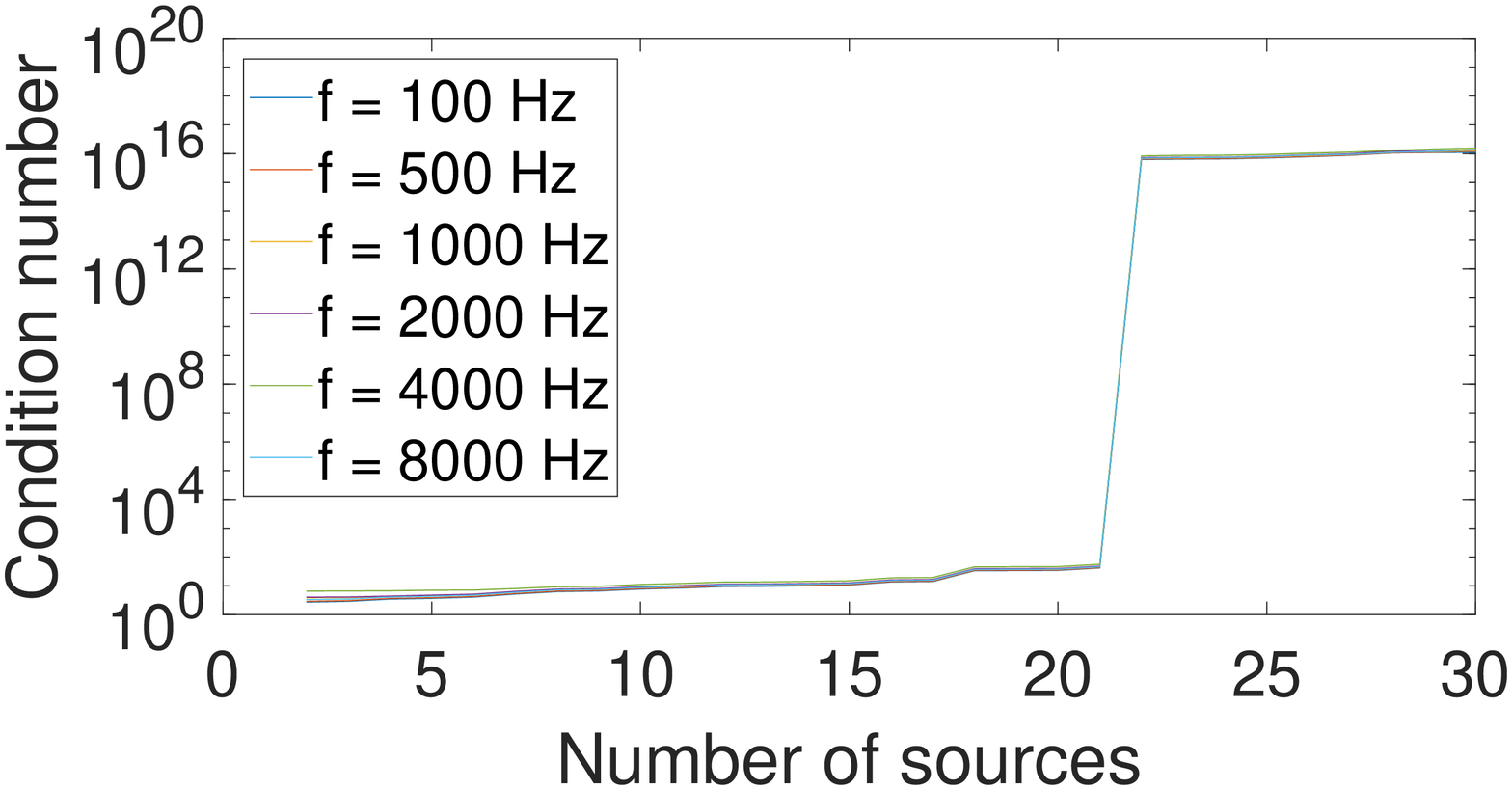}}
  \centerline{(a) $N=2$}\medskip
\end{minipage}
\begin{minipage}[b]{0.47\linewidth}
  \centering
  \centerline{\includegraphics[width=\linewidth]{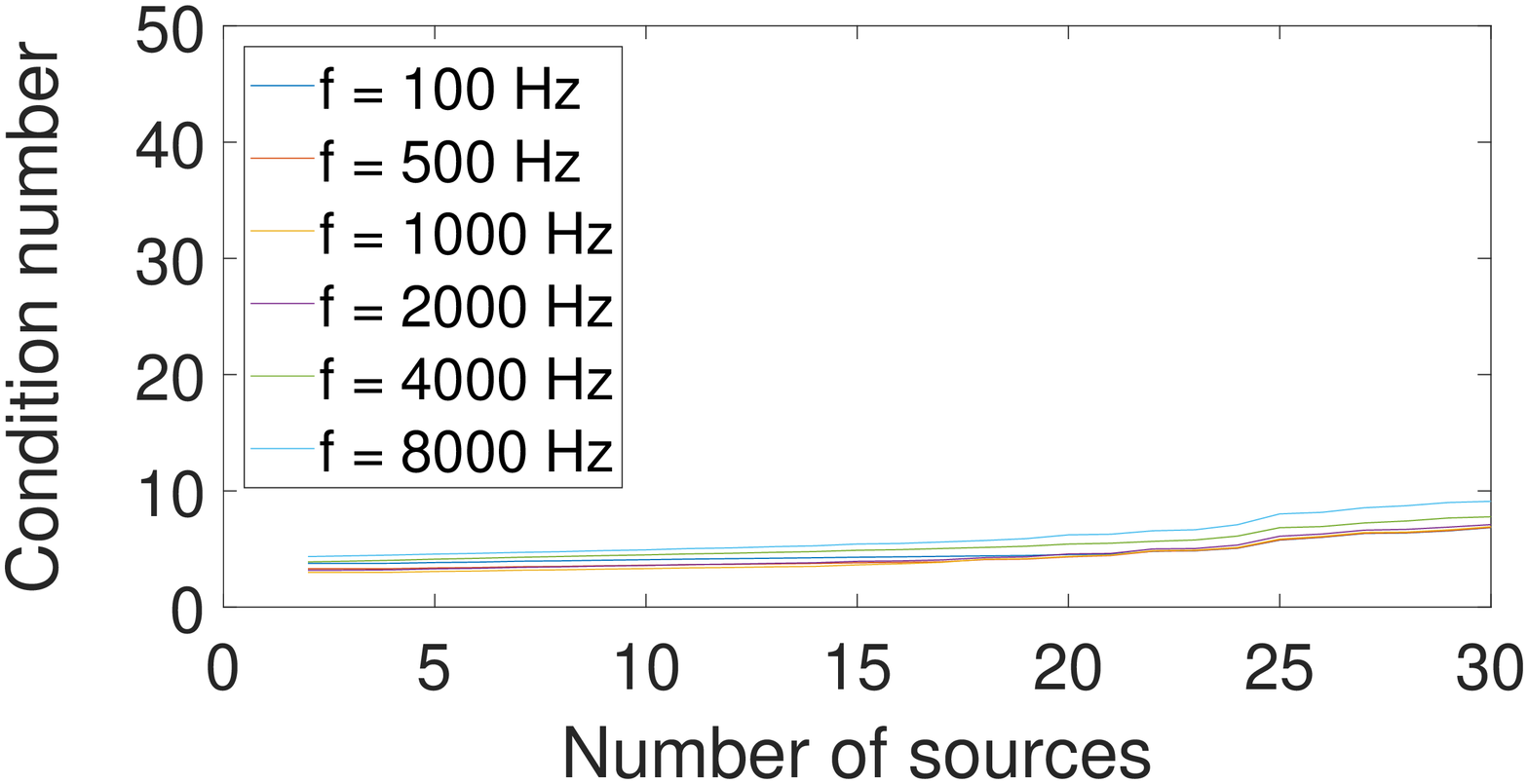}}
  \centerline{(b) $N=4$}\medskip
\end{minipage}
\caption{Condition number of the transfer matrix $\boldsymbol{T}$ with $N=2$ and $4$.}
\label{fig:condition-number}
\end{figure}

However, if the number of sources increases, it is expected that we would require the higher order sound field coefficients for a better estimation accuracy. The improved performance is achieved with a larger $N$ due to a better knowledge of the spatial sound field energy distribution with higher order modes. Furthermore, the higher order modes help to avoid ill-conditioning in matrix $\boldsymbol{T}$ for a large number of sources. This is evident from Fig. \ref{fig:condition-number} which plots the condition number of $\boldsymbol{T}$ for $N=[2, 4]$ with $V = 1$ against different number of sources and frequencies. The sources considered in Fig. \ref{fig:condition-number} were uniformly distributed on the surface of a $1$ m sphere at 5 different azimuth planes. For the case of $N=2$, the condition number of $\boldsymbol{T}$ remains low up to $21$ sources, but increases exponentially beyond that. On the contrary, the condition number never exhibits an issue for $N=4$ within the experimental limit of $30$ sources. Significantly, the behavior is almost identical over the whole frequency range. This is due to the fact that only the noise terms at the last column of $\boldsymbol{T}$ are frequency dependent, when the far-field assumption is made.
\section{Conclusions}
The objective of this work was to estimate individual PSD components in a multi-source noisy and reverberant environment which can be used in different speech enhancement techniques such as source separation, dereverberation or noise suppression. The use of the spherical harmonics coefficients of the sound field and their cross-correlation allowed us to address a larger number of sources in a mixed sound scene compared to the conventional beamformer-based solutions without facing an ill-posed problem. We measured PSD estimation accuracy and its application in source separation through different objective metrics in practical environments with a commercial microphone array. The relative comparison revealed that the proposed method outperformed other contemporary techniques in different acoustic environments.\par
In many spherical harmonics-based approaches for sound field processing, a major performance issue occurs at the nulls of the Bessel functions. Through our work, we analyzed and investigated the impact of the Bessel-zero issue and offered an engineering solution to the problem. Though the solution can introduce a new kind of estimation error, this proved to be a good trade-off, as the newly introduced error was found to be insignificant compared to the initial Bessel-zero issue.\par
Though we demonstrated the source separation application in this paper, the proposed method is also useful in applications such as DRR or SNR estimation, and selective denoising and dereverberation due to its ability to extract the reverberation and noise PSDs separately. For future work, we intend to investigate the accuracy of the estimated reverberant PSD and evaluate the proposed method in applications such as DRR measurement. We also plan to measure its performance with a simpler lower-order microphone array \cite{samarasinghe2017planar} for an easier commercial implementation.
\appendices
\section{The Definition of $W_{v, n, n'}^{u, m, m'}$}
\label{app:integral-sh}
The integral property of the spherical harmonics is defined over a sphere as
\begin{multline} \label{eq:integral-property-sh-app1}
\int_{\boldsymbol{\hat{y}}} Y_{vu}(\boldsymbol{\hat{y}}) Y_{nm}(\boldsymbol{\hat{y}}) Y_{n'm'}(\boldsymbol{\hat{y}}) d\boldsymbol{\hat{y}} = \\
\sqrt[]{\frac{(2v+1)(2n+1)(2n'+1)}{4 \pi}}  \text{ }\bar{W}_{m}
\end{multline}
where $\bar{W}_{m}$ represents a multiplication between two Wigner-3j symbols \cite{olver2010nist} as
\begin{equation} \label{eq:w12-app1}
\bar{W}_{m} = \left(\begin{array}{clcr}
v & n & n'\\
0 & 0 & 0  \end{array}\right) \text{ }
\left(\begin{array}{clcr}
v & n & n'\\
u & m & m'  \end{array}\right).
\end{equation}
Furthermore, a conjugated spherical harmonics coefficient follows the following property
\begin{equation} \label{eq:conjugate-sh-app1}
Y_{nm}^*(\boldsymbol{\hat{y}}) = (-1)^m \text{ } Y_{n(-m)}(\boldsymbol{\hat{y}}).
\end{equation}
From \eqref{eq:integral-property-sh-app1} and \eqref{eq:conjugate-sh-app1}, we obtain
\begin{align} \label{eq:integral-property-sh-1-app1}
W_{v, n, n'}^{u, m, m'} & = \int_{\boldsymbol{\hat{y}}} Y_{vu}(\boldsymbol{\hat{y}}) Y^*_{nm}(\boldsymbol{\hat{y}}) Y_{n'm'}(\boldsymbol{\hat{y}}) d\boldsymbol{\hat{y}} \nonumber \\
& = (-1)^m \int_{\boldsymbol{\hat{y}}} Y_{vu}(\boldsymbol{\hat{y}}) Y_{n(-m)}(\boldsymbol{\hat{y}}) Y_{n'm'}(\boldsymbol{\hat{y}}) d\boldsymbol{\hat{y}} \nonumber \\
& = (-1)^m \text{ } \sqrt[]{\frac{(2v+1)(2n+1)(2n'+1)}{4 \pi}} \text{ } \bar{W}_{-m}
\end{align}
where the notation $W_{v, n, n'}^{u, m, m'}$ is chosen for brevity.
\section{Closed-form expression of \eqref{eq:final-model-noise-coeff}}
\label{app:closed-form-noise}
Defining $\boldsymbol{x}'' = (\boldsymbol{x} - \boldsymbol{x}')$ where $\boldsymbol{x}'' = (r'', \boldsymbol{\hat{x}}'')$, we obtain \eqref{eq:app2-addition-bessel} from the addition theorem for spherical Bessel functions in a similar manner as in \cite[pp. 592-593]{chew1995waves}
\begin{multline}\label{eq:app2-addition-bessel}
Y_{a''b''}(\boldsymbol{\hat{x}}'') j_{a''}(k \norm{\boldsymbol{x}''}) = 4 \pi \sum \limits_{ab}^{\infty} \sum \limits_{a'b'}^{\infty} i^{(a-a'-a''+2b)} \text{ } Y_{ab}(\boldsymbol{\hat{x}}) \\
j_{a}(k \norm{\boldsymbol{x}}) \text{ } Y_{a'b'}(\boldsymbol{\hat{x}}') j_{a'}(k \norm{\boldsymbol{x}'}) \text{ } W_{a, a', a''}^{-b, b', b''}.
\end{multline}
As $Y_{00}(\cdot) = 1/\sqrt[]{4 \pi}$, we obtain \eqref{eq:app2-addition-bessel-zero} by letting $a''=b''=0$ in \eqref{eq:app2-addition-bessel}
\begin{multline}\label{eq:app2-addition-bessel-zero}
j_0(k \norm{\boldsymbol{x}''}) = (4 \pi)^{\frac{3}{2}} \sum \limits_{ab}^{\infty} \sum \limits_{a'b'}^{\infty} i^{(a-a'+2b)} \text{ } Y_{ab}(\boldsymbol{\hat{x}}) j_{a}(k \norm{\boldsymbol{x}}) \\
Y_{a'b'}(\boldsymbol{\hat{x}}') j_{a'}(k \norm{\boldsymbol{x}'}) \text{ } W_{a, a', 0}^{-b, b', 0}.
\end{multline}
Hence, using \eqref{eq:app2-addition-bessel-zero} in \eqref{eq:final-model-noise-coeff}, we obtain
\begin{multline}\label{eq:app2-final-model-noise-coeff}
\Omega_{nm}^{n'm'}(k) = \frac{1}{\lvert b_n(kr) \rvert ^2} \int_{\boldsymbol{\hat{x}}} \int_{\boldsymbol{\hat{x}}'} \Bigg((4 \pi)^{\frac{3}{2}} \sum \limits_{ab}^{\infty} \sum \limits_{a'b'}^{\infty} i^{(a-a'+2b)} \times \\
Y_{ab}(\boldsymbol{\hat{x}}) j_{a}(k \norm{\boldsymbol{x}}) \text{ } Y_{a'b'}(\boldsymbol{\hat{x}}') j_{a'}(k \norm{\boldsymbol{x}'}) \text{ } W_{a, a', 0}^{-b, b', 0} \Bigg) \times \\
Y_{nm}^*(\boldsymbol{\hat{x}}) Y_{n'm'}(\boldsymbol{\hat{x}}') \text{ } d\boldsymbol{\hat{x}} \text{ } d\boldsymbol{\hat{x}}'.
\end{multline}
Using the conjugate property of the spherical harmonics from \eqref{eq:conjugate-sh-app1} and rearranging \eqref{eq:app2-final-model-noise-coeff}, we obtain
\begin{multline} \label{eq:app2-final-model-noise-coeff2}
\!\!\!\! \Omega_{nm}^{n'm'}(k) = \frac{(4 \pi)^{\frac{3}{2}}}{\lvert b_n(kr) \rvert ^2} \sum \limits_{ab}^{\infty} \sum \limits_{a'b'}^{\infty} i^{(a-a'+2b)} j_{a}(k \norm{\boldsymbol{x}}) j_{a'}(k \norm{\boldsymbol{x}'}) \\
W_{a, a', 0}^{-b, b', 0} \Bigg( \int_{\boldsymbol{\hat{x}}} Y_{ab}(\boldsymbol{\hat{x}}) Y_{nm}^*(\boldsymbol{\hat{x}}) \text{ } d\boldsymbol{\hat{x}} \Bigg) \times \\
\Bigg( (-1)^{m'} \int_{\boldsymbol{\hat{x}}'} Y_{a'b'}(\boldsymbol{\hat{x}}') Y_{n'(-m')}^*(\boldsymbol{\hat{x}}') \text{ } d\boldsymbol{\hat{x}}' \Bigg).
\end{multline}
Finally, using they orthonormal property of the spherical harmonics from \eqref{eq: spherical-harmonics-orthonormal}, we obtain
\begin{equation} \label{eq:app2-final-model-noise-coeff3}
\Omega_{nm}^{n'm'} \! (k) \! = \! \frac{(4 \pi)^{\frac{3}{2}} i^{(n-n'+2m+2m')} j_{n}(k r) j_{n'}(k r) \text{ } W_{n, n', 0}^{-m, -m', 0}}{\lvert b_n(kr) \rvert ^2}
\end{equation}
as $(-1)^{m'} = i^{2m'}$ and $\norm{\boldsymbol{x}} = \norm{\boldsymbol{x}'} = r$, where $r$ is the radius of the spherical array.
\section{Source directions}
\label{app:source-directions}
Table \ref{table:doa} shows true $(\theta, \phi)$ and estimated $(\hat{\theta}, \hat{\phi})$ DOAs for $L = 4, 6$.
\begin{table}[!t]
\caption{Source directions in radian.}
\label{table:doa}
\renewcommand{\arraystretch}{1.2}
\resizebox{\linewidth}{!}{%
\begin{tabular}{| c || c | c || c | c || c | c |}
\hline
 & \multicolumn{2}{c||}{Room A} & \multicolumn{2}{c||}{Room B} & \multicolumn{2}{c|}{Room C} \\
\hhline{~------}
Source & $\theta, \phi$ & $\hat{\theta}, \hat{\phi}$ & $\theta, \phi$ & $\hat{\theta}, \hat{\phi}$ & $\theta, \phi$ & $\hat{\theta}, \hat{\phi}$\\
\hhline{|=======|}
\multicolumn{7}{|c|}{$4$-speaker case} \\
\hline
S-$01$ & $1.6, 5.81$ & $1.6, 5.8$ & $1.5, 0.75$ & $1.5, 0.75$ & $1.67, 0.77$ & $1.64, 0.73$ \\
\hline
S-$02$ & $1.58, 4.53$ & $1.58, 4.52$ & $1.51, 2.31$ & $1.51, 2.30$ & $1.89, 2.33$ & $1.93, 2.3$ \\
\hline
S-$03$ & $1.59, 3.19$ & $1.59, 3.18$ & $1.07, 4.01$ & $1.13, 4.04$ & $1.66, 3.87$ & $1.7, 3.9$ \\
\hline
S-$04$ & $1.57, 1.93$ & $1.57, 1.93$ & $1.51, 5.4$ & $1.52, 5.42$ & $1.86, 5.44$ & $1.89, 5.5$ \\
\hline
\multicolumn{7}{|c|}{$6$-speaker case} \\
\hline
S-$01$ & $0.54, 5.56$ & $0.52, 0.59$ & $1.45, 6.23$ & $1.46, 6.21$ & $1.66, 6.23$ & $1.64, 6.19$ \\
\hline
S-$02$ & $1.01, 3.51$ & $1.02, 3.54$ & $1.50, 0.74$ & $1.51, 0.76$ & $1.88, 1.01$ & $1.89, 1.03$ \\
\hline
S-$03$ & $1.58, 5.17$ & $1.55, 5.13$ & $1.70, 1.46$ & $1.72, 1.45$ & $1.68, 2.04$ & $1.69, 2.09$ \\
\hline
S-$04$ & $1.58, 1.32$ & $1.55, 1.37$ & $1.51, 2.31$ & $1.52, 2.32$ & $1.89, 3.11$ & $1.83, 3.10$ \\
\hline
S-$05$ & $2.13, 2.85$ & $2.15, 2.88$ & $1.07, 4.02$ & $1.09, 4.01$ & $1.66, 4.25$ & $1.62, 4.21$ \\
\hline
S-$06$ & $2.57, 6.18$ & $2.56, 6.11$ & $1.51, 5.41$ & $1.52, 5.41$ & $1.85, 5.19$ & $1.84, 5.17$ \\
\hline
\end{tabular}
}
\end{table}
%
%
% \section*{ACKNOWLEDGMENT}
\bibliography{abdfahim}
\bibliographystyle{IEEEtran}

\end{document}